\documentclass[aps,prstab,twocolumn,superscriptaddress,nofootinbib]{revtex4-2}

\pdfoutput=1
\makeatletter
\renewcommand\paragraph{\@startsection{paragraph}{4}{\z@}%
            {-2.5ex\@plus -1ex \@minus -.25ex}%
            {1.25ex \@plus .25ex}%
            {\normalfont\normalsize\bfseries}}
\makeatother

\usepackage{graphicx}
\usepackage{subcaption}
\usepackage[export]{adjustbox}
\usepackage{siunitx}
\usepackage{amsmath}
\usepackage{comment}
\usepackage{booktabs}

\usepackage[english]{babel}
\usepackage{hyperref}
\usepackage{tikz}
\usetikzlibrary{matrix, fit}
\usetikzlibrary{arrows.meta}
\tikzset{mymatrixstyle/.style={matrix of math nodes,
        ampersand replacement=\&,
        left delimiter=(,
        right delimiter=),
        inner sep=0pt,
        outer sep=0pt,
        row sep=2pt, column sep=10pt},
        myarrow/.style={-{Straight Barb[angle=60:3pt 3]}}}
\usepackage{mathtools}

\usepackage{graphicx}
\usepackage{sidecap}
\usepackage{caption, subcaption}

\mathtoolsset{showonlyrefs}

\DeclareMathOperator{\Tr}{Tr}
\begin{document}


\title{Review of coupled betatron motion parametrizations and applications to strongly coupled lattices}

\author{Marion Vanwelde}
\email[Email address: ]{marion.vanwelde@ulb.be}
\affiliation{Service de M{\'e}trologie Nucl{\'e}aire (CP165/84), Universit{\'e} libre de Bruxelles, Avenue Franklin Roosevelt 50, 1050 Brussels, Belgium}

\author{C{\'e}dric Hernalsteens}
\email[Email address: ]{cedric.hernalsteens@cern.ch}
\affiliation{CERN, European Organization for Nuclear Research, Esplanade des Particules 1, 1211 Meyrin, Switzerland}
\affiliation{Service de M{\'e}trologie Nucl{\'e}aire (CP165/84), Universit{\'e} libre de Bruxelles, Avenue Franklin Roosevelt 50, 1050 Brussels, Belgium}

\author{S. Alex Bogacz}
\affiliation{Thomas Jefferson National Accelerator Facility, Newport News, Virginia, U.S.A.}

\author{Shinji Machida}
\affiliation{STFC Rutherford Appleton Laboratory, Harwell Campus, Didcot, OX11 0QX, United Kingdom}

\author{Nicolas Pauly}
\affiliation{Service de M{\'e}trologie Nucl{\'e}aire (CP165/84), Universit{\'e} libre de Bruxelles, Avenue Franklin Roosevelt 50, 1050 Brussels, Belgium}

\date{\today}

\begin{abstract}
The coupling of transverse motion is a natural occurrence in particle accelerators, either in the form of a residual coupling arising from imperfections or originating by design from strong systematic coupling fields. While the first can be treated perturbatively, the latter requires a robust approach adapted to strongly coupled optics and a parametrization of the linear optics must be performed to explore beam dynamics in such peculiar lattices. This paper reviews the main concepts commonly put forth to describe coupled optics and clarifies the proposed parametrization formalisms. The links between the generalized Twiss parameters used by the different approaches are formally proven, and their physical interpretations are highlighted. The analytical methods have been implemented in a reference Python package and connected with a ray-tracing code to explore strongly coupled lattices featuring complex 3D fields. Multiple examples are discussed in detail to highlight the key physical interpretations of the parametrizations and characteristics of the lattices.
\end{abstract}


\maketitle

\section{Introduction}
The motion of charged particles in a particle accelerator is typically studied using the linear and uncoupled theory of betatron motion. The Courant-Snyder theory \cite{Courant_Snyder} allows the study of unidimensional and uncoupled motion by having an elegant parametrization whose optical parameters have a clear physical meaning. However, in many machines, coupling between the two transverse degrees of freedom is present. The coupling of the particle transverse motion has long been considered an undesirable effect. Coupling was first studied mainly because of imperfections (quadrupole tilt, vertical displacements of sextupoles \cite{Willeke}). This residual coupling, if not well controlled, can cause undesirable effects such as vertical emittance increase or impact linear and nonlinear observables such as amplitude detuning \cite{hoferEffectLocalLinear2020}. To take into account the effect of residual coupling, it is possible to start from the uncoupled theory and consider the coupling as a perturbation. This perturbation theory is no longer applicable as soon as the coupling arises from strong systematic coupling fields. In this case, the machine design contains elements that introduce coupling on purpose. In colliders, it is the case for interaction regions where large solenoidal fields and compensation elements are present. Atypical optics schemes based on strong coupling insertions have also been proposed to improve the performance of lepton and hadron colliders, such as the ``M\"obius accelerator'' \cite{Talman}, planar-to-circular beam adapters for circular modes operation \cite{burovCircularModesFlat2013} and round beam operation for lepton storage rings \cite{Chongchong_StudiesBeamHEPS2020}.

Recently, vertical excursion fixed field accelerators (vFFAs) were revived\footnote{Although first introduced in 1955 by Tihiro Ohkawa as electron cyclotrons, vFFAs received a lot of interest only from 2013 onwards, following Ref.\,\cite{Brooks}}, featuring coupling by design. The detailed linear and nonlinear study of vFFAs constitutes the main motivation for the present work. In what follows, $x$ is the horizontal coordinate, $y$ is the vertical coordinate, and $z$ is the longitudinal coordinate. In conventional, horizontal excursion, FFAs, the nonlinear magnetic field respects a scaling condition that allows having a constant tune for all energies \cite{Ohkawa, symonFixedFieldAlternatingGradientParticle1956, kolomenski} and higher momentum particles move to orbits of larger radius. By contrast, vFFA fields fulfill another scaling condition:
\begin{equation}
B = B_0 e^{k(y-y_0)}\text{,}
\end{equation}
where $k = \frac{1}{B}\frac{\partial B}{\partial y}$ is the normalized field gradient, $y_0$ is the reference vertical position and $B_0$ is the bending field at the reference position. The bending field increases exponentially in the vertical direction leading higher energy particle orbits to have the same radius but to shift vertically. The median plane of vFFA elements is the plane at $x = 0$ (vertical plane). Assuming $y_0 = 0$, we can write the three magnetic field components with an out-of-plane expansion \cite{machidaOpticsDesignVertical2021a}:
\begin{eqnarray*}
    B_x(x,y,z) &=& B_0 e^{ky} \sum_{i=0}^N b_{xi}(z) x^i \text{, }\\
    B_y(x,y,z) &=& B_0 e^{ky} \sum_{i=0}^N b_{yi}(z) x^i \text{, }\\
    B_z(x,y,z) &=& B_0 e^{ky} \sum_{i=0}^N b_{zi}(z) x^i \text{, }
\end{eqnarray*}
where, by taking into account the fringe field function $g(z)$, the coefficients of these equations are given by the following recurrence relations \cite{machidaOpticsDesignVertical2021a}:
\begin{align}
    b_{x0}(z) &= 0 \text{, }\quad  b_{x,i+1}(z) = -\frac{1}{i+1}(k b_{yi} + \frac{db_{zi}}{dz})\text{, }\\
    b_{y0}(z) &= g(z) \text{, }\quad  b_{y,i+2}(z) = \frac{k}{i+2} b_{x,i+1}\text{, } \\
    b_{z0}(z) &= \frac{1}{k} \frac{dg}{dz} \text{, }\quad  b_{z,i+2}(z) = \frac{1}{i+2} \frac{db_{x,i+1}}{dz} \text{.}
\end{align}

In the (vertical) median plane, the three field components are:
\begin{align}
\label{eq_B_vFFA_3D}
B_{x0}(0, y, z )&= 0\text{, }\\
B_{y0}(0, y, z )&= B_0 e^{ky} g(z)\text{, } \\
B_{z0}(0, y, z )&= \frac{B_0}{k}e^{ky} \frac{dg}{dz}\text{.}
\end{align}

The vFFAs thus present a non-zero longitudinal field component, which arises due to the fringe fields at the vFFA ends. It is especially important as the magnet construction, respecting the scaling law, will induce important fringe fields. If we look at the field in the element body, by neglecting the fringe field ($g(z) = const.$, $g^\prime(z) = 0$, $B_z = 0$), the transverse field components can be expressed as multipolar expansions by rewriting the exponential in terms of its Taylor series. It is readily seen that the first-order terms of this expansion correspond to skew quadrupolar components:
\begin{eqnarray*}
B_x (x,y,z) &\simeq& -B_0 (kx + \frac{k^2}{2!}(2xy)+ O(x^3))\\
        &\simeq& - B_0 k x \text{,}\\
B_y (x,y,z) & = & B_0 (1 + ky + \frac{k^2}{2!}(y^2 -x^2)+ O(x^3)) \\
        & \sim & B_0 + B_0 k y \text{.}
\end{eqnarray*}

Because of the longitudinal and skew quadrupolar field components, which are the main sources of transverse motion coupling, vFFAs feature strongly coupled optics. It is therefore necessary to study vFFA lattices with a model adapted to strongly coupled optics. The choice of a given parametrization for such a machine, suitable for the design, optimization, and operation phases, is key to a thorough understanding of the peculiar beam dynamics. All the methods and analyses presented are applicable to other coupled lattices in full generality and are relevant for snake \cite{Snake_ref} and spin rotator designs.

Several parametrizations attempt to describe coupled optics as elegantly as the Courant-Snyder theory for uncoupled motion. The most widely known parametrizations are those of Edwards and Teng (ET) \cite{Edwards_Teng} and of Mais and Ripken (MR) \cite{Mais_Ripken}. In addition, these parametrizations were extended and revisited in several works: Sagan and Rubin \cite{Sagan_Rubin}, Parzen \cite{Parzen}, Wolski \cite{wolskiAlternativeApproachGeneral2006, wolskiSimpleWayCharacterize2004, wolskiNormalFormAnalysis2004} and Lebedev and Bogacz (LB) in \cite{Lebedev}. The exact formalisms and notations used by these authors differ, and slightly different parametrization choices lead to an apparently inhomogeneous theory. To clarify the situation so that a clear picture can be obtained, we provide interpretations of these parameters and explicit links between them for the different parametrizations.

The general theory and formalism for the study of linear beam dynamics in 4D transverse phase-space are presented in Section~\ref{theoretical_background} and the peculiarities of coupled motion are highlighted.
In Section~\ref{parametrisation_review}, a review of the coupling parametrizations from ET, MR, and their extensions are detailed using unified approaches and notations. Physical interpretations regarding lattice functions and clarifications of the relationships between the quantities appearing in the different parametrizations are provided. The links between the ET and MR parametrization categories are discussed in Section~\ref{Lien_ET_MR}. The methods are implemented using the Zgoubidoo Python interface \cite{Zgoubidoo} for the Zgoubi code \cite{Zgoubi} and discussion in Section~\ref{implementation} where applications are presented for example lattices and for realistic examples of snakes and spin rotators. The implementations have been validated by comparing the generalized lattice functions computed by Zgoubidoo with those obtained by MAD-X \cite{madx} and PTC \cite{PTC}. Conclusions and recommendations for the study of vFFA lattices are provided in Section~\ref{section_conclusion}.

\section{Theory of coupled linear betatron motion \label{theoretical_background}}

\subsection{Notations}
Lowercase bold letters are used to indicate vectors of geometric coordinates, where prime denotes the differentiation with respect to the independent $s$ coordinate: $( )'  = \frac{d( )}{ds}$, $\mathbf{x} \equiv \left( x, x', y, y'\right)^T$. The vectors of canonical coordinates will be designated as:
\begin{equation}
    \mathbf{\hat{x}} \equiv \begin{pmatrix}
    x \\
    p_x \\
    y \\
    p_y \\
    \end{pmatrix} \text{.}
\end{equation}

Bold uppercase letters indicate matrices (for example, $\mathbf{M}$ will denote a transfer matrix), and a hat is added when it comes to the transfer matrice over a full period, (``one-turn transfer matrices'' $\mathbf{\hat{M}}$). No difference is made in the notation to denote the transfer matrices expressed in geometric variables or canonical variables. However, the identification of the variables for each of the matrices will be made clear from the context. Moreover, a bar is added on top to indicate symplectic conjugate matrices: the symplectic conjugate matrix of $\mathbf{M}$ will be denoted $\mathbf{\bar{M}}$. The symplectic conjugate of a symplectic matrix $\mathbf{M}$ is defined as $\mathbf{\bar{M}} = -\mathbf{S}\mathbf{M}^T \mathbf{S} = \mathbf{M}^{-1}$ \cite{Courant_Snyder, Laurent_Deniau}, where $\mathbf{S}$ is the unit symplectic matrix
\begin{equation}
    \mathbf{S} = \begin{pmatrix}
    0 & 1 & 0 & 0 \\
    -1 & 0 & 0 & 0 \\
    0 & 0 & 0 & 1 \\
    0 & 0 & -1 & 0
    \end{pmatrix}\text{,}
\end{equation}
with $\mathbf{S}^T\mathbf{S}= \mathbf{I}$, $\mathbf{S}\mathbf{S} = -\mathbf{I}$, and $\mathbf{S}^T = -\mathbf{S}$. The horizontal ($x$) and vertical ($y$) directions are referred to as ``physical directions'' or ``physical space'' as opposed to the ``eigen-directions'' related to the directions of the decoupled motion.

\subsection{Geometric coordinates and canonical coordinates \label{geom_canon}}
The relation between geometric coordinates and canonical variables reads:
\begin{align}
    x' &= p_x - \frac{e}{p_0} A_x \label{link_geom_canon_1} \text{, }\\
    y' &= p_y - \frac{e}{p_0} A_y \label{link_geom_canon_2} \text{,}
\end{align}
where the vector potential $\mathbf{A}$ components are related to the magnetic field by $\mathbf{B} = \nabla \times \mathbf{A}$. The components of the vector potential $\mathbf{A}$ can often be expressed as a series expansion. When studying the linear motion, this series can be approximated by its first-order terms in $x$, $p_x$, $y$, and $p_y$, which allows having only quadratic terms in the expression of the Hamiltonian. For example, for the longitudinal field produced by a solenoid, one obtains the components $A_x$ and $A_y$ of the vector potential as follows \cite{Willeke, Lebedev}:
\begin{align}
    \frac{e}{p_0} A_x &= - \frac{1}{2} R_1(s) y + O(y^3) \text{,}\\
    \frac{e}{p_0} A_y &= \frac{1}{2} R_2(s) x + O(x^3)\text{,}
\end{align}
where $R_1(s) = R_2(s) = \frac{e}{p_0} B_s(0,0,s)$ are constants proportional to the longitudinal component of the magnetic field.

In the case of a scaling vFFA field, the three vector potential components can be written:
\begin{align}
    A_x(x,y,z) &= B_0 e^{ky} \sum_{i=0}^N a_{xi}(z) x^i \text{, }\\
    A_y(x,y,z) &= B_0 e^{ky} \sum_{i=0}^N a_{yi}(z) x^i \text{, }\\
    A_z(x,y,z) &= B_0 e^{ky} \sum_{i=0}^N a_{zi}(z) x^i \text{, }
\end{align}
where the coefficients are given by the following recurrence relations:
\begin{eqnarray*}
    a_{x0}(z) &= 0 \text{, }\quad & a_{x1}(z) = 0 \text{, }\\
    a_{y0}(z) &= 0 \text{, }\quad & a_{y1}(z) = \frac{1}{k} \frac{dg}{dz}\text{,} \\
    a_{z0}(z) &= 0 \text{, }\quad & a_{z1}(z) = -g(z) \text{,}
\end{eqnarray*}
\begin{eqnarray*}
    a_{x,i+1}(z) &=& -\frac{k}{i+1} a_{yi}\\
    a_{y,i+2}(z) &=& -\frac{1}{(i+2)(i+1)} \biggl[ -k \frac{d a_{zi}}{dz} + \frac{d^2 a_{yi}}{dz ^2} \biggr] \\
                 & & + \frac{1}{i+2} k a_{x,i+1}\\
    a_{z,i+2}(z) &=& \frac{k}{(i+2)(i+1)} \biggl[ -k a_{zi} + \frac{d a_{yi}}{dz} \biggr] \\
                 & & +\frac{1}{i+2} \frac{d a_{x,i+1}}{dz} \text{.}
\end{eqnarray*}

By truncating the series in the first order, the transverse components $A_x$ and $A_y$ become

\begin{align}
    \frac{e}{p_0} A_x &= 0+ O(x^2) \text{, }\\
    \frac{e}{p_0} A_y &= \frac{1}{2} R_2(s) x + O(x^3)\text{,}
\end{align}
where, in this case, $R_1(s) = 0$ and $R_2(s) = 2 \frac{e}{p_0} B_z(0,y,s)$.

The expressions \eqref{link_geom_canon_1} and \eqref{link_geom_canon_2} for the transform between geometric and canonical variables can be rewritten using $R_{1,2}$:
\begin{align}
    x' &= p_x + \frac{1}{2} R_1 y \text{,}\\
    y' &= p_y - \frac{1}{2} R_2 x \text{.}
\end{align}

In the absence of a longitudinal field component, the canonical variables are equal to the geometric variables: $\mathbf{\hat{x}} = \mathbf{x}$. However, when there is a longitudinal field component, it must be considered. In a matrix form this reads $\mathbf{\hat{x}} = \mathbf{U} \mathbf{x}$, where
\begin{equation}
    \mathbf{U} = \begin{pmatrix}
    1 & 0 & 0 & 0 \\
    0 & 1 & -\frac{R_1}{2} & 0 \\
    0 & 0 & 1 & 0 \\
    \frac{R_2}{2} & 0 & 0 & 1 \\
    \end{pmatrix}\text{.}
\end{equation}

The matrix $\mathbf{U}$ can also be used to transform from a transfer matrix expressed in geometric variables to a transfer matrix expressed in canonical variables:
\begin{equation}
    \mathbf{M}_{s_0 \rightarrow s, \text{canon.}} = \mathbf{U}(s) \mathbf{M}_{s_0 \rightarrow s, \text{geom.}} \mathbf{U}^{-1}(s_0)\text{.}
\end{equation}

\subsection{Coupling sources}
The linear coupling between the two transverse directions originates from two types of field components: longitudinal or skew quadrupolar. Solenoids induce a rotation at a frequency equal to the Larmor frequency $\dot{\theta} = -\frac{q B_s}{2\gamma m} = -\frac{\Omega_c}{2} = -\omega_{Larmor}$ (where $B_s$ is the longitudinal field component and $\Omega_c$ is the cyclotron frequency) \cite{Wiedemann}. This rotation introduces a coupling between the vertical and horizontal motions of the particle. It can be shown that there is a $s$-dependent rotation that transforms the coordinate system into a frame where the motion is decoupled (the so-called ``Larmor'' frame) \cite{Wiedemann}. The rotation angle that decouples the motion is proportional to the integral of the longitudinal field along the trajectory of the particle. In the Larmor frame, the solenoid is a magnetic element that focuses in the two transverse directions \cite{Willeke}. The second magnetic field which induces linear coupling is the field produced by a skew quadrupole. A particle with a horizontal (resp. vertical) displacement will be affected by a horizontal (resp. vertical) magnetic field and will be subject to a vertical (resp. horizontal) force inducing a vertical displacement. The vertical motion ultimately becomes horizontal again. There is an energy exchange between the two transverse directions, and the motion is coupled \cite{wilsonBeamDynamics2020}.

\subsection{Equations and invariant of motion, symplecticity and stability}
We assume linearized transverse equations of motion expressed in the moving Frenet-Serret reference frame ($x$, $y$, and $s$) attached to the reference trajectory. In the geometric coordinates ($\mathbf{x} = \begin{pmatrix}x, x', y, y'\end{pmatrix}^T)$, the 2D coupled linear equations of motion can be written \cite{Willeke, Lebedev}:
\begin{align}
    x'' + (\kappa_x^2 + K) x + (N-\frac{1}{2}R_1') y - \frac{1}{2}(R_1 + R_2)y' &= 0 \label{equations_mouvement_x} \text{, }\\
    y'' + (\kappa_y^2 - K) y + (N+\frac{1}{2}R_2') x + \frac{1}{2}(R_1 + R_2)x' &= 0 \text{,} \label{equations_mouvement_y}
\end{align}\\
where the coefficients $\kappa_x$, $\kappa_y$, $K$, and $N$  are defined as follows:
\begin{align}
    \kappa_x &=  \frac{e B_y(0,0,s)}{p_0} \\
    \kappa_y &= - \frac{e B_x(0,0,s)}{p_0} \\
    K &= \frac{1}{B\rho}(\frac{\partial B_y}{\partial x})_{x=y=0} \\
    N &= \frac{1}{2 B\rho}(\frac{\partial B_y}{\partial y} - \frac{\partial B_x}{\partial x})_{x=y=0}\text{.}
\end{align}
$B_x$, $B_y$, $B_s$ are the field components along the closed orbit. $\kappa_x$ and $\kappa_y$ are the curvature of the design orbit in the horizontal and vertical directions, $K$ is related to the normal component of the field gradient, while $N$ is linked to the skew component of the field gradient \cite{Lebedev}. Finally, $R_{1}$ and $R_{2}$ (see Section \ref{geom_canon}) are related to the longitudinal field component. In the equations of motion, only the last two terms of the left-hand side reflect the coupling between the two transverse directions. Without these terms, the equations of motion are Hill's equations without coupling (see Appendix~\ref{appendix_decoupled}). The equations of motion are obtained from the Hamiltonian for a charged particle of charge $e$ and mass $m$ in an electromagnetic field expressed in Cartesian coordinates \cite{stupakovClassicalMechanicsElectromagnetism2018}:
\begin{equation}
   H(\mathbf{r}, \mathbf{\pi}, t) = c \sqrt{m^2c^2 + (\mathbf{\pi}- e \mathbf{A}(\mathbf{r}, t))^2)} + e \phi(\mathbf{r}, t)\text{, }
\end{equation}
where $\mathbf{r} = (x,y,z)$, $\mathbf{\pi} = (P_x, P_y, P_z)$ contains the three canonical conjugate momentum for the coordinates ($x,y,z$), $\mathbf{A}$ is the vector potential and $\phi$ is the scalar potential. To derive the equations of motion expressed in canonical variables, ensuring conservative solutions \cite{Willeke}, a transformation to the coordinates in the Frenet-Serret frame is performed and a change of independent variable from time $t$ to path length along the reference trajectory $s$ is performed. The Hamiltonian becomes
\begin{widetext}
\begin{equation}
    H(x,p_x, y, p_y ; s) = -(1+\kappa_x x + \kappa_y y) \biggl[\sqrt{1 - \biggl(p_x -(1-\delta)\frac{e}{p_0} A_x\biggr)^2 - \biggl(p_y - (1-\delta)\frac{e}{p_0} A_y\biggr)^2} + (1-\delta)\frac{e}{p_0}As\biggr]\text{,}
\end{equation}
\end{widetext}
where $\delta = \frac{P-P_0}{P_0}$ and $p_x$, $p_y$ are the canonical momentum normalized by the total reference momentum $P_0$. Moreover, because the transverse momentum components are much smaller than the total reference momentum $P_0$, it is possible to expend the Hamiltonian in a power series. For linear motion, the Hamiltonian is truncated to a quadratic form. Considering in addition the nominal energy only ($\delta = 0$) and expressing the vector potential components in terms of $\kappa_x$, $\kappa_y$, $K$, $N$, $R_1$ and $R_2$, the Hamiltonian becomes \cite{Lebedev}:
\begin{widetext}
\begin{equation}
    H = \frac{p_x^2 +p_y^2}{2} + (\kappa_x^2 + K + \frac{R_2^2}{4}) \frac{x^2}{2} + (\kappa_y^2 - K + \frac{R_1^2}{4}) \frac{y^2}{2} + N xy + \frac{1}{2}(R_1 yp_x - R_2 xp_y)\text{.}
    \label{hamiltonian}
\end{equation}
\end{widetext}
The coupling terms are readily apparent, with skew quadrupolar fields gradient $N$ coupling the motion through the $xy$ term and longitudinal fields coupling the motion through the $y p_x$ and $x p_y$ terms. The Hamiltonian equations of motion can be written in a $4\times4$ matrix formalism using the bilinear form \cite{Teng, Laurent_Deniau, Edwards_Teng, Lebedev}
\begin{equation}
    H = \frac{1}{2}\mathbf{\hat{x}^T} \mathbf{H} \mathbf{\hat{x}}\text{, }
\end{equation}
where $\mathbf{H}$ is a real and symmetric matrix:
\begin{equation}
    \mathbf{H} = \begin{pmatrix}
    \kappa_x^2 + K + \frac{R_2^2}{4} & 0 & N & -\frac{R_2}{2} \\
    0   & 1 & \frac{R_1}{2} & 0\\
    N & \frac{R_1}{2} & \kappa_y^2 - K + \frac{R_1^2}{4} & 0 \\
    - \frac{R_2}{2} & 0 & 0 & 1
    \end{pmatrix}\text{.}
\end{equation}

In a matrix form the equations of motion become
\begin{align}
    \mathbf{\hat{x}'} &= \mathbf{S} \mathbf{H} \mathbf{\hat{x}} \label{equation_mouvement_matricielle} = \mathbf{A}(s) \mathbf{\hat{x}}\text{.}
\end{align}

From Equation \eqref{equation_mouvement_matricielle}, we can show that, for any solutions $\mathbf{\hat{x}_1}$ and $\mathbf{\hat{x}_2}$, the quantity $\mathbf{\hat{x}_2^T} \mathbf{S} \mathbf{\hat{x}_1}$ is a constant of motion \cite{Courant_Snyder}; the so-called Lagrange invariant. Indeed, if $\mathbf{\hat{x}_1}$ and $\mathbf{\hat{x}_2}$ are solutions of equation \eqref{equation_mouvement_matricielle}, then
\begin{align}
    \frac{d}{ds}(\mathbf{\hat{x}_2^T} \mathbf{S} \mathbf{\hat{x}_1}) &= \frac{d\mathbf{\hat{x}_2^T}}{ds} \mathbf{S} \mathbf{\hat{x}_1} + \mathbf{\hat{x}_2^T} \mathbf{S} \frac{d \mathbf{\hat{x}_1}}{ds}\\
                &= \mathbf{\hat{x}_2^T} \mathbf{H}^T \mathbf{S}^T \mathbf{S} \mathbf{\hat{x}_1} + \mathbf{\hat{x}_2^T} \mathbf{S} \mathbf{S} \mathbf{H} \mathbf{\hat{x}_1} \\
                &= 0 \\
                & \Rightarrow  \mathbf{\hat{x}_2^T} \mathbf{S} \mathbf{\hat{x}_1} = \text{constant} \label{invariant_Lagrange}\text{.}
\end{align}

The linear motion described by the quadratic Hamiltonian (Eq.~\eqref{hamiltonian}) is a succession of linear canonical transformations represented by the transfer matrices $\mathbf{M}$. The solution of the equations of motion can therefore be written in the form $\mathbf{\hat{x}}(s) = \mathbf{M_{s_0 \rightarrow s}} \mathbf{\hat{x}}(s_0)$, where $\mathbf{M_{s_0 \rightarrow s}}$ is the transfer matrix allowing to propagate the coordinates from $s_0$ to $s$. The transfer matrix must satisfy the following conditions \cite{Willeke}:
\begin{eqnarray*}
    \frac{d}{ds} \mathbf{M_{s_0\rightarrow s}} & =&  \mathbf{A}(s) \mathbf{M_{s_0 \rightarrow s}}\text{, }\\
    \mathbf{M_{s_0 \rightarrow s_0}} &=& I \text{.}
\end{eqnarray*}

Expressing the particle trajectory with these transfer matrices is equivalent to integrating the linear differential equations over a finite distance. The particle motion can thus be described either by the equations of motion derived from the Hamiltonian (continuous formalism) or by transfer matrices (discrete formalism) \cite{NF94}. One of the advantages of using matrix formalism is that we can obtain the one-turn transfer matrix $\mathbf{\hat{M}}$ by multiplying the transfer matrices of the sections contained in the period. The one-turn transfer matrix computed at $s$ can be obtained from the one-turn transfer matrix computed at $s_0$:
\begin{equation}
    \mathbf{\hat{M}}(s) = \mathbf{M_{s_0\rightarrow s}} \mathbf{\hat{M}}(s_0) \mathbf{M^{-1}_{s_0\rightarrow s}}\text{.}
\end{equation}

The Jacobian matrix of a canonical transformation is symplectic \cite{Teng}. In the case of linear motion, the transfer matrix is equal to the Jacobian matrix, and therefore $\mathbf{M_{s_i \rightarrow s_j}}$ (expressed in terms of canonical variables) is also symplectic:
\begin{equation}
    \mathbf{M}^T \mathbf{S} \mathbf{M} = \mathbf{S}\text{.}
    \label{symplectic_matrix}
\end{equation}
It is possible to find this symplecticity condition from the Lagrange invariant \cite{USPAS}. $\mathbf{\hat{x}_2^T} \mathbf{S} \mathbf{\hat{x}_1}$ being an invariant of the motion, we know that $\mathbf{\hat{x}_2^T}(s) \mathbf{S} \mathbf{\hat{x}_1}(s) = \mathbf{\hat{x}_2^T}(s_0) \mathbf{S} \mathbf{\hat{x}_1}(s_0)$. By expressing the coordinates as functions of $s$ using the transfer matrix and the initial coordinates at $s_0$, we get:
\begin{eqnarray*}
    \mathbf{\hat{x}_2^T}(s) \mathbf{S} \mathbf{\hat{x}_1}(s) &=& \mathbf{\hat{x}_2^T}(s_0) \mathbf{M}^T \mathbf{S} \mathbf{M} \mathbf{\hat{x}_1}(s_0)\\
                &=& \mathbf{\hat{x}_2^T}(s_0) \mathbf{S} \mathbf{\hat{x}_1}(s_0) \\
                &\Rightarrow& \mathbf{M}^T \mathbf{S} \mathbf{M} = \mathbf{S}\text{.}
\end{eqnarray*}
The matrix $\mathbf{S}$ being anti-symmetric, the symplecticity condition on the transfer matrix (Eq.~\eqref{symplectic_matrix}) gives $\frac{(n^2-n)}{2}$ scalar conditions. The transfer matrix $\mathbf{M}$ will therefore contain $n^2 - \frac{(n^2-n)}{2} = \frac{n}{2}(n+1)$ independent elements \cite{Teng, Edwards_Teng, Lebedev, Courant_Snyder}. For a two-dimensional motion, at least 10 independent parameters are needed to parameterize the matrix.

It is insightful to study the eigenvalues and eigenvectors of the transfer matrix $\mathbf{\hat{M}}$. The expression of the Lagrange invariant allows to find conditions on these eigenvalues and eigenvectors. For a $4\times4$ transfer matrix, there are 4 eigenvectors $\mathbf{\hat{v}_j}$ corresponding to the eigenvalues $\lambda_j$: $\mathbf{\hat{M}} \mathbf{\hat{v}_j} = \lambda_j \mathbf{\hat{v}_j}$. By expressing the Lagrange invariant for two eigenvectors $\mathbf{\hat{v}_i}$ et $\mathbf{\hat{v}_j}$ of the transfer matrix, we obtain \cite{Willeke}:
\begin{eqnarray*}
    \mathbf{\hat{v}_i^T}(s) \mathbf{S} \mathbf{\hat{v}_j}(s) &=& (\mathbf{\hat{M}}\mathbf{\hat{v}_i}(s_0))^{\mathbf{T}} \mathbf{S} \mathbf{\mathbf{\hat{M}}\hat{v}_j}(s_0) \\
                &=& \lambda_i \lambda_j \mathbf{\hat{v}_i^T}(s_0) \mathbf{S} \mathbf{\hat{v}_j}(s_0)\text{.}
\end{eqnarray*}

One obtains:
\begin{equation}
	\left\lbrace
		\begin{aligned}
		\lambda_i \lambda_j &= 1 \Rightarrow \mathbf{\hat{v}_i^T} \mathbf{S} \mathbf{\hat{v}_j} \ne 0 \\
		\lambda_i \lambda_j &\ne 1 \Rightarrow \mathbf{\hat{v}_i^T} \mathbf{S} \mathbf{\hat{v}_j} = 0
		\end{aligned}
	\right.
\end{equation}

\begin{figure}[h]
\includegraphics[width=0.8\linewidth]{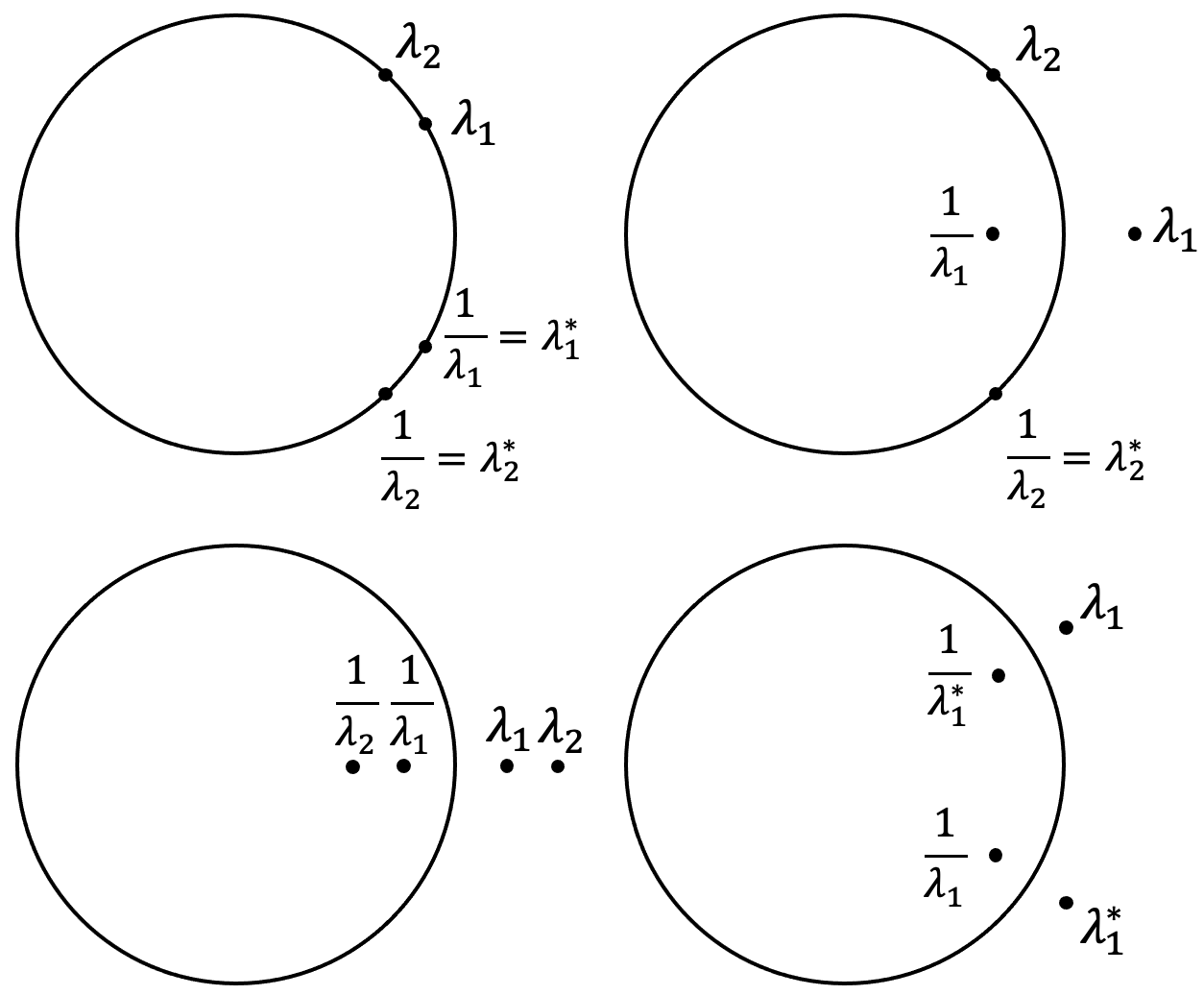}
\caption{Different situations for the four eigenvalues of the transfer matrix $\mathbf{\hat{M}}$; only the first case is stable - Reproduced from \cite{Courant_Snyder}.}
\label{valeurs_propres_schema}
\end{figure}

The eigenvalues of the transfer matrix thus appear in reciprocal pairs ($\lambda_{j}, \frac{1}{\lambda_j}$) \cite{Courant_Snyder}. Moreover, the matrix $\mathbf{\hat{M}}$ being real, the eigenvalues appear in complex conjugate pairs. As illustrated in Fig.~\ref{valeurs_propres_schema}, there are four possibilities respecting these two conditions: (i) the four eigenvalues are complex with $|\lambda| = 1 $, (ii) one of the pairs of eigenvalues is real, the other is complex and lies on the unit circle, (iii) both pairs of reciprocal eigenvalues are real, (iv) one of the eigenvalues is complex but is not on the unit circle $|\lambda_1| \ne 1$. The other eigenvalues are then $\lambda_2 = \frac{1}{\lambda_1}$, $\lambda_3 = \lambda_1^*$, $\lambda_4 = \frac{1}{\lambda_1^*}$ \cite{Courant_Snyder}.

To guarantee stable motion, $|\lambda| = 1$ is necessary (case (i) above). The eigenvectors of the transfer matrix will then be complex conjugate $\mathbf{\hat{v}_{-j}} = \mathbf{\hat{v}^*_{j}}$ (with $j = 1,2$) and will correspond to the eigenvalues $\lambda_{\pm j} = e^{\pm i 2\pi Q_j}$, where $Q_j$ are the linear tunes. Since the eigenvectors appear in complex conjugate pairs, the Lagrange invariant $\mathbf{\hat{v}_j^{+}} \mathbf{S} \mathbf{\hat{v}_j}$ (where $\mathbf{\hat{v}_j^{+}} = \mathbf{\hat{v}_j^{*T}}$) will be imaginary. We can therefore normalize the eigenvectors of the matrix $\mathbf{\hat{M}}$ as follows:
\begin{equation}
    \label{normalisation_eigenvectors_}
	\left\lbrace
		\begin{aligned}
            \mathbf{\hat{v}_i^+}\  \mathbf{S}\  \mathbf{\hat{v}_j} &= \pm i \quad & \text{if} \quad \delta_{ij} = 1\\
            \mathbf{\hat{v}_i^+}\  \mathbf{S}\  \mathbf{\hat{v}_j} &= 0 \quad &\text{if} \quad \delta_{ij} = 0 \text{.}
		\end{aligned}
	\right.
\end{equation}

\section{Review and comparison of $x-y$ coupled motion parametrizations\label{parametrisation_review}}
The study of two-dimensional uncoupled transverse motion can be reduced to the study of one-dimensional motions in the horizontal and vertical directions. The parameters used to describe this transverse motion have a clear physical meaning and are related to the optical properties of the lattice (see Appendix \ref{appendix_decoupled} for more details). For the linear coupled transverse motion, several parametrizations attempt to describe the coupled optics and characterize the coupling in an elegant fashion. Among these parametrizations, the most widely known are the parametrization from Edwards and Teng \cite{Edwards_Teng} (``ET'' parametrization) and the parametrization from Mais and Ripken \cite{Mais_Ripken} (``MR'' parametrization). Other variants can be linked to one of these two classes. These two parametrization categories differ in their way of describing coupled optics; their lattice parameters are different and have different interpretations.

The ET parametrization transforms the transfer matrix in a decoupled block-diagonal matrix using a symplectic rotation. The lattice functions are then defined for each block of the block-diagonal matrix; each block is parameterized as a Twiss matrix, as shown in Appendix~\ref{appendix_decoupled}, Eq.~\eqref{Twiss_matrix} with three lattice parameters $\alpha$, $\beta$ and $\mu$. The lattice functions are thus connected to the eigenmodes of oscillation and not to the physical directions of the transverse plane. The coupling between the transverse motions is characterized by the parameters of the symplectic rotation.

The MR parametrization is based on the coupled transfer matrix in physical space. It consists in parameterizing the normalization matrix\textemdash the matrix that transforms the transfer matrix into its normal form\textemdash with lattice functions or, in an equivalent way, parameterizing the eigenvectors of the coupled transfer matrix. The resulting lattice parameters represent the effect of the two eigenmodes of oscillation on each of the physical transverse directions. These optical parameters are therefore linked not only to the oscillation eigenmode but also to the physical directions of the transverse plane, which allows for interpreting them in relation to the physical beam sizes.

The link between these parametrizations was clarified by Lebedev and Bogacz in Ref.~\cite{Lebedev}. In addition, several authors have revisited and extended these two parametrizations: Sagan and Rubin \cite{Sagan_Rubin}, Parzen \cite{Parzen}, and Wolski \cite{wolskiAlternativeApproachGeneral2006, wolskiSimpleWayCharacterize2004, wolskiNormalFormAnalysis2004}. However, although these approaches are based on the same principles, different notations and slightly different parameters are used, or the derivations proceed differently. We review the different parametrizations using the same formalism to be able to compare and highlight the links, similarities, and fundamental differences in the lattice parameters. In this section, the ET and the MR parametrizations are detailed individually, together with their variants, and we make explicit links between the various parameters. The links between the parametrizations belonging to the ET or MR categories are highlighted in Section~\ref{Lien_ET_MR}.

\subsection{Edwards and Teng (ET) parametrization}\label{ET_parametrization}
The ET parametrization initially aimed to generalize in a straightforward manner the Courant-Snyder parameters to coupled motion. To that end, 10 independent parameters are used to parameterize the 2D coupled motion. These parameters include the usual $\alpha$, $\beta$, and $\mu$ functions for each eigenmode of oscillation. These functions are defined as the Twiss parameters of the $2\times2$ matrices representing the decoupled motion in the eigen-directions. In addition, the ET parametrization also includes four parameters that represent the coupling strength and structure in the lattice. These parameters originate from the decoupling matrix.

To obtain this parametrization, one starts from a very general transfer matrix:
\begin{equation}
    \mathbf{M_{s_1 \rightarrow s_2}} = \begin{pmatrix}
    \mathbf{A} & \mathbf{B} \\
    \mathbf{C} & \mathbf{D} \\
    \end{pmatrix}\text{,}
    \label{T}
\end{equation}
where $\mathbf{A}$, $\mathbf{B}$, $\mathbf{C}$ and $\mathbf{D}$ are $2\times2$ matrices. This transfer matrix propagates the physical coordinates $(x, p_x, y, p_y)$ from $s_1$ to $s_2$: $\mathbf{\hat{x}}(s_2) = \mathbf{M} \mathbf{\hat{x}}(s_1)$. If the lattice does not introduce coupling, the horizontal and vertical motions do not mix and the transfer matrix $\mathbf{M}$ is block-diagonal ($\mathbf{B}$ and $\mathbf{C}$ are zeros, and the unimodular $2\times2$ matrices $\mathbf{A}$ and $\mathbf{D}$ respectively describe the horizontal and vertical motions). If the transverse motion is coupled, the transfer matrix is no longer block-diagonal and couples the horizontal and vertical motions. The motion in the coupled physical space is described by a transfer matrix $\mathbf{M}$ where none of the elements is \textit{a priori} zero. A decoupled space appears in which the motion along two so-called ``eigen-directions'' can be described independently. The transfer matrix in the decoupled frame $\mathbf{P}$ propagates the decoupled coordinates $(u, p_u, v, p_v)$ from one point to another in the accelerator: $\mathbf{\hat{u}}(s_2) = \mathbf{P} \mathbf{\hat{u}}(s_1)$. This transfer matrix $\mathbf{P}$ is block-diagonal:
\begin{equation}
    \mathbf{P_{s_1 \rightarrow s_2}} = \begin{pmatrix}
    \mathbf{E} & 0 \\
    0 & \mathbf{F} \\
    \end{pmatrix}\text{.}
\end{equation}

It is possible to go from the coupled physical space to the decoupled space using a linear similarity transformation. The decoupling matrix $\mathbf{\widetilde{R}}$ transforms the transfer matrix $\mathbf{M}$ into the block-diagonal matrix $\mathbf{P}$ \cite{Parzen}:
\begin{align}
    \mathbf{\hat{x}} &= \mathbf{\widetilde{R}\hat{u}} \text{, }\\
    \mathbf{P}_{s_0 \rightarrow s} &= \mathbf{\widetilde{R}}^{-1}(s) \mathbf{M}_{s_0 \rightarrow s} \mathbf{\widetilde{R}}(s_0)\text{.}
\end{align}

This transformation is also valid for the one-turn matrices in the coupled and decoupled spaces $\mathbf{\hat{M}}(s)$ and $\mathbf{\hat{P}}(s)$:
\begin{equation}
  \mathbf{\hat{P}}(s) = \mathbf{\widetilde{R}}^{-1}(s) \mathbf{\hat{M}}(s) \mathbf{\widetilde{R}}(s)\text{.} \label{RTR}
\end{equation}

The most general form of the similarity transformation which block-diagonalizes $\mathbf{\hat{M}}$ can be written as \cite{Parzen}:
\begin{equation}
  \mathbf{\widetilde{R}} = \begin{pmatrix}
  q_1 \mathbf{I} & \mathbf{R_{12}}\\
  \mathbf{R_{21}} &  q_2 \mathbf{I}
  \end{pmatrix} \text{,}
\end{equation}
where $q_1$, $q_2$ are scalar quantities, $\mathbf{R_{12}}$, $\mathbf{R_{21}}$ are $2\times2$ matrices and $\mathbf{I}$ is the unit matrix. The matrix $\mathbf{\widetilde{R}}$ being a $4\times4$ symplectic matrix, there are 6 symplecticity conditions. These symplecticity conditions on $\mathbf{\widetilde{R}}$ can be written as follows:
\begin{align}
q_1^2 +|\mathbf{R_{12}}| &= 1 \text{, }\\
|\mathbf{R_{21}}| + q_2^2 &= 1 \text{, }\\
q_1 \overline{\mathbf{R_{21}}} + \mathbf{R_{12}} q_2 &= 0 \label{third_condition} \text{.}
\end{align}

From these conditions, we can simplify the expression of $\mathbf{\widetilde{R}}$ by imposing that $q_1 = q_2$ (we will call this constant $\gamma$ in what follows) and that $\mathbf{R_{21}} =  - \overline{\mathbf{R_{12}}}$. To simplify the expression of $\mathbf{\widetilde{R}}$, we write $\mathbf{R_{12}} = \mathbf{\mathcal{C}}$ and $\mathbf{R_{21}}
= -\mathbf{\mathcal{\overline{C}}}$ \cite{Sagan_Rubin}:
\begin{equation}
  \mathbf{\widetilde{R}} = \begin{pmatrix}
  \gamma \mathbf{I} & \mathbf{\mathcal{C}} \\
  -\mathbf{\mathcal{\overline{C}}} &\gamma \mathbf{I}
  \end{pmatrix} \text{.}
\end{equation}

To take into account the last remaining symplecticity condition on $\mathbf{\widetilde{R}}$ ($\gamma^2 + |\mathbf{\mathcal{C}}| = 1$), the ET parametrization goes one step further and describes the matrix $\mathbf{\widetilde{R}}$ as a symplectic rotation, which imposes $\gamma = \cos{\phi}$ and $\mathbf{\mathcal{C}} = \mathbf{\mathcal{D}}^{-1}\sin{(\phi)}$, where $\mathbf{\mathcal{D}}$ is a symplectic $2\times2$ matrix \cite{Edwards_Teng}:
\begin{equation}
    \mathbf{\widetilde{R}}= \begin{pmatrix}
    \mathbf{I} \cos{(\phi)} & \mathbf{\mathcal{D}}^{-1}\sin{(\phi)} \\
    -\mathbf{\mathcal{D}} \sin{(\phi)} & \mathbf{I} \cos{(\phi)}
    \end{pmatrix} \text{,}
    \label{R_matrix}
\end{equation}
\begin{equation}
    \mathbf{\mathcal{D}}= \begin{pmatrix}
    a & b \\
    c & d
    \end{pmatrix} \text{.}
\end{equation}

All the symplecticity conditions on $\mathbf{\widetilde{R}}$ are taken into account in this last expression. One can see that the matrix $\mathbf{\widetilde{R}}$ has four independent elements: the rotation angle $\phi$ and the three independent elements of the symplectic matrix $\mathbf{\mathcal{D}}$. The parameter $\gamma = \cos{(\phi)}$ represents the coupling strength, while the matrix $\mathbf{\mathcal{D}}$ represents the coupling structure \cite{Edwards_Teng, Laurent_Deniau, Teng}. This manner of parametrizing the matrix $\mathbf{\widetilde{R}}$ was first presented in Ref.~\cite{Teng}. Multiple ways to write the similarity matrix $\mathbf{\widetilde{R}}$ in terms of a symplectic rotation exist in order to block-diagonalize the symplectic transfer matrix $\mathbf{\hat{M}}$. The symplectic rotation is a four-dimensional rotation of the $x-p_x$ and $y-p_y$ planes, which gives the orientation of the normal modes compared to the axes of the physical system. Once the motion is decoupled, each of the blocks of the matrix $\mathbf{\hat{P}}$ corresponds to an eigenmode of oscillation and can be rewritten as a Twiss matrix (Eq.~\eqref{Twiss_matrix}):
\begin{equation*}
    \mathbf{A_i} = \begin{pmatrix}
    \cos{(\mu_i)} + \alpha_i \sin{(\mu_i)} &  \beta_i \sin{(\mu_i)} \\
    -\gamma \sin{(\mu_i)} & \cos{(\mu_i)} - \alpha_i \sin{(\mu_i)}
    \end{pmatrix}\text{, }
\end{equation*}
where $\mathbf{A_i}$ represents each of the blocks of the diagonal of $\mathbf{\hat{P}}$, with $i = 1,2$ indicating the considered eigenmode.

It should be noted that forcing $\mathbf{\widetilde{R}}$ to be symplectic forces the matrix $\mathbf{\hat{P}}$ to be symplectic as well. This allows the $\mathbf{E}$ and $\mathbf{F}$ matrices to be parameterized with only 3 independent parameters, as in the case of uncoupled motion. The Twiss parameters $\alpha_i$, $\beta_i$, and $\mu_i$ characterize the eigenmode motion and are not related to the physical axes. As a result, the $\beta$-functions are not directly related to the beam size in the physical plane, and the interpretation of these Twiss parameters is more complicated than in the case of an uncoupled motion. The generalized Twiss parameters of the ET parametrization thus describe the beam dynamics in the decoupled axes but are not related to the measurable parameters of the beam.

In order to calculate the elements of the matrix $\mathbf{\widetilde{R}}$ as well as the lattice parameters for each oscillation mode, we can use two different methods of calculation. The first method (see Section~\ref{subsection_first_method}) is based on the analytical solution of a system of equations, which allows us to express $\mathbf{\widetilde{R}}$, $\mathbf{E}$ and $\mathbf{F}$ from the elements of the matrix $\mathbf{\hat{M}}$. This method is used by Edwards and Teng in \cite{Edwards_Teng} and was extended by Sagan and Rubin in \cite{Sagan_Rubin} for strong coupling. This method was also used to implement the ET parametrization in MAD-X \cite{Laurent_Deniau, madx}. The second method (see Section~\ref{subsection_second_method}), explained by Parzen in Ref.~\cite{Parzen}, is based on the eigenvectors of $\mathbf{\hat{M}}$ and on the link between these eigenvectors and the eigenvectors of $\mathbf{\hat{P}}$. These are easily parameterized with the optical functions $\alpha$, $\beta$, and $\mu$ of each eigenmode. This method is advantageous because it allows studying the coupled motion in a phase space of greater dimension: for example, the ET parametrization is extended by Parzen for coupled motion in 6 degrees of freedom \cite{Parzen}.

The following sections detail these two different methods, which we use to calculate the elements of $\mathbf{\widetilde{R}}$, as well as the generalized Twiss parameters. The two procedures are based on the transfer matrix in coupled space and it is possible to find all the parameters of the ET parametrization using a tracking code that provides this transfer matrix.

\subsubsection{ET parameters from an explicit analytical solution\label{subsection_first_method}}
$\mathbf{\widetilde{R}}$ is a symplectic matrix allowing to obtain the decoupled block-diagonal matrix $\mathbf{\hat{P}}$ starting from $\mathbf{\hat{M}}$. Using Eq.~\eqref{RTR} and the symplecticity condition on $\mathbf{\widetilde{R}}$, all the elements of $\mathbf{\hat{P}}$ and $\mathbf{\widetilde{R}}$ can be found in term of the elements of $\mathbf{\hat{M}}$ (in the form of Eq.~\eqref{T}) \cite{Edwards_Teng, Sagan_Rubin, Laurent_Deniau}:
\begin{align}
    \gamma &= \sqrt{\frac{1}{2} + \frac{1}{2} \sqrt{\frac{(\Tr{\mathbf{A}}-\Tr{\mathbf{D}})^2}{(\Tr{\mathbf{A}}-\Tr{\mathbf{D}})^2 + 4 \times |\mathbf{B}+\overline{\mathbf{C}}|}}}\text{,}\label{solution_ET}\\
    \mathbf{\mathcal{C}} &= \frac{-(\mathbf{B}+\overline{\mathbf{C}}) \times \text{sign}(\Tr{\mathbf{A}}-\Tr{\mathbf{D}})}{\gamma \sqrt{(\Tr{\mathbf{A}}-\Tr{\mathbf{D}})^2 + 4\times |\mathbf{B}+\overline{\mathbf{C}}|}}\text{,} \label{solution_ET_1}\\
    \mathbf{E} &= \mathbf{A} - \frac{\mathbf{\mathcal{C}} \mathbf{C}}{\gamma} \text{, } \label{solution_ET_2}\\
    \mathbf{F} &= \mathbf{D} + \frac{\mathbf{\mathcal{\overline{C}}} \mathbf{B}}{\gamma}\text{.}
    \label{solution_EF}
\end{align}
\noeqref{solution_ET_1, solution_ET_2}

The solutions written in this form directly provide a stability condition: having a real solution imposes $|\mathbf{B}+\overline{\mathbf{C}}|> - \frac{1}{4}(\Tr{\mathbf{A}}-\Tr{\mathbf{D}})^2$. This stability condition is equivalent to a constraint on the eigenvalues (see section \ref{theoretical_background}). If this condition is not met, the sum of the two eigenvalues within a pair is complex, which corresponds to the case where all the eigenvalues are complex but do not lie on the unit circle. A discussion of this stability condition in case of weak coupling is done in \cite{Courant_Snyder}, concluding that sum resonances can induce instability while difference resonances cannot.

Once the matrices $\mathbf{E}$ and $\mathbf{F}$ are calculated, it is possible to express them in the form of Twiss matrices to find the lattice parameters of the two eigenmodes. The solution presented above (Eqn. \eqref{solution_ET} and \eqref{solution_EF}) is the solution presented in Ref.~\cite{Edwards_Teng} and corresponds to the solution used in the case of weak coupling. However, when $|\mathbf{B}+\mathbf{\bar{C}}| > 0$, a second solution exists for the parameters of $\mathbf{\widetilde{R}}$ \cite{Sagan_Rubin}:
\begin{align}
\gamma &= \sqrt{\frac{1}{2} - \frac{1}{2} \sqrt{\frac{(\Tr{\mathbf{A}}-\Tr{\mathbf{D}})^2}{(\Tr{\mathbf{A}}-\Tr{\mathbf{D}})^2 + 4\times |\mathbf{B}+\overline{\mathbf{C}}|}}}\text{, }\\
\mathbf{\mathcal{C}} &= \frac{(\mathbf{B}+\overline{\mathbf{C}}) \times \text{sign}(\Tr{\mathbf{A}}-\Tr{\mathbf{D}})}{\gamma \sqrt{(\Tr{\mathbf{A}}-\Tr{\mathbf{D}})^2 + 4\times |\mathbf{B}+\overline{\mathbf{C}}|}}\text{.}
\end{align}

This second solution corresponds to another symplectic rotation matrix $\mathbf{\widetilde{R}}$. Depending on the chosen analytical solution and therefore depending on the matrix $\mathbf{\widetilde{R}}$ used for the decoupling, one obtains a different block-diagonal matrix $\mathbf{\hat{P}}$, and the blocks of this matrix will be associated differently with the eigenmodes. The $2\times2$ matrix $\mathbf{E}$ can thus be associated with the eigenmode I using one of the two solutions but associated with the eigenmode II using the other solution. As the decoupled matrix is different depending on the solution used, the Twiss parameters will also have different values in one case or the other \cite{Sagan_Rubin}.

In a weakly coupled lattice, the horizontal and vertical oscillations are nearly unchanged, and the eigenmode oscillations can be associated with the horizontal and vertical motions. The decoupling matrix must be close to the unit matrix so that the eigen-axes are close to the horizontal and vertical directions. Therefore, only the first solution will be chosen and used ($\gamma \approx 1$ and $\mathbf{\mathcal{C}} \approx 0$). With $\gamma = \cos{\phi}$ and developing $\cos{\phi}$ in terms of $\cos{2\phi}$, we get:
$$\gamma = \cos{\phi} = \sqrt{\frac{1}{2}+ \frac{1}{2} \cos{2\phi}} \text{.}$$ Imposing the choice of the first solution is equivalent to imposing a condition on the angle of rotation $\phi$: $$-\frac{\pi}{4}\leq \phi \leq \frac{\pi}{4}.$$ It is the condition originally imposed by Edwards and Teng in Ref.~\cite{Edwards_Teng}.

However, in a strongly coupled lattice, it is more complicated to associate the eigenmodes with the $2\times2$ matrices that lie on the diagonal of $\mathbf{\hat{P}}$. At some locations of the lattice (where $|\mathbf{B}+ \mathbf{\bar{C}}| < 0$), only the first solution may exist, which forces the identification of the modes. The eigenmode I can thus be associated with the matrix $\mathbf{E}$ at a given location and associated with the matrix $\mathbf{F} $ at others. Therefore, it might not be possible to keep the identification of a mode with one of the two matrices $\mathbf{E}$ or $\mathbf{F}$. The change in mode identification at different locations of the lattice is referred to as \textit{mode flipping} and only occurs in elements that introduce a strong coupling between horizontal and vertical motions. The number of mode flips must be even in order to find the same eigen-axes at the exit of the element. In strongly coupled lattices, the Twiss parameters can thus be different depending on the chosen mode identification. The knowledge of the Twiss parameters alone is not sufficient to compare lattices; the identification of the eigen-axes is required as well.

The above optical function and decoupling matrix calculations are based on the one-turn transfer matrix $\mathbf{\hat{M}}$ at a specific location in the accelerator. To propagate the decoupled matrix from one point to another of the lattice, one can calculate the transfer matrix ($\mathbf{W_{12}}$ = $\mathbf{W_{s_1 \rightarrow s_2}}$) in the decoupled space \cite{Sagan_Rubin}:
\begin{equation}
    \mathbf{W_{12}} = \mathbf{\widetilde{R}^{-1}_2} \mathbf{M_{12}} \mathbf{\widetilde{R}_1}\text{, }
\end{equation}
where $\mathbf{M_{12}} = \mathbf{M_{s_1 \rightarrow s_2}}$ is the coupled transfer matrix between $s_1$ and $s_2$, $\mathbf{\widetilde{R}_2} = \mathbf{\widetilde{R}}(s_2)$ and $\mathbf{\widetilde{R}_1} = \mathbf{\widetilde{R}}(s_1)$. From $\mathbf{W_{12}}$, we can compute the decoupled one-turn transfer matrix at $s_2$ \cite{Laurent_Deniau}:
\begin{equation}
\mathbf{\hat{P}}(s_2) = \mathbf{W_{12}}\mathbf{\hat{P}}(s_1)\mathbf{W^{-1}_{12}}\text{.}
\end{equation}

It is possible to see if there is a mode flip between $s_1$ and $s_2$ using the trace of the matrix $\mathbf{M_ {12}}$, or \textit{via} the structure of the matrix $\mathbf{W_{12}}$. Because the oscillation eigenmodes are independent of each other, the propagation matrix $\mathbf{W_{12}}$ can only be block-diagonal or anti-block-diagonal \cite{Sagan_Rubin, Laurent_Deniau}. It will be block-diagonal if there is no mode flip and anti-block-diagonal if there is a mode flip\footnote{Refs. \cite{Laurent_Deniau, Sagan_Rubin} provide more information on the propagation of normal modes and initial Twiss parameters.}. The sign change of $|\mathbf{B}+ \mathbf{\bar{C}}|$ characterizes a forced mode flip. When the determinant of the matrix is equal to 0, only one solution remains. Thus, if there is a forced mode flip in the lattice, $\gamma \rightarrow 0$. The Twiss parameters are then associated with different modes so that the lattice functions are discontinuous at this location, and the $\beta$-functions can diverge. Because the $\beta$-functions can become infinite or negative, it is not possible to preserve their physical interpretation in terms of envelope functions.

\subsubsection{ET parameters from the eigenvectors of $\mathbf{\hat{M}}$\label{subsection_second_method}}
The second method allows to express the lattice parameters and the decoupling matrix in terms of the components of the one-turn transfer matrix eigenvectors. The transfer matrix in the decoupled space $\mathbf{\hat{P}}$ is a block-diagonal matrix whose blocks can be expressed with the linear and periodic optical parameters $\alpha_i$, $\beta_i$ and $\mu_i$ for $i = 1,2$. It is therefore possible to express the eigenvectors of $\mathbf{\hat{P}}$\textemdash $\mathbf{\hat{u}_1}$, $\mathbf{\hat{u}_2}$, $\mathbf{\hat{u}_3}$, $\mathbf{\hat{u}_4}$ (where $\mathbf{\hat{u}_2} = \mathbf{\hat{u}_1}^*$ and $\mathbf{\hat{u}_4} = \mathbf{\hat{u}_3}^*$)\textemdash in terms of $\alpha_i$, $\beta_i$ and $\mu_i$:

\begin{equation}
    \mathbf{\hat{u}_1} = \begin{pmatrix}
    \beta_1^{\frac{1}{2}}\\
    \beta_1^{-\frac{1}{2}}(-\alpha_1+i)\\
    0\\
    0
    \end{pmatrix} e^{i \mu_1}\text{, }
\end{equation}

\begin{equation}
    \mathbf{\hat{u}_3} = \begin{pmatrix}
    0\\
    0\\
    \beta_2^{\frac{1}{2}}\\
    \beta_2^{-\frac{1}{2}}(-\alpha_2+i)\\
    \end{pmatrix} e^{i \mu_2}\text{.}
\end{equation}\\

In the coupled physical space, the eigenvectors of the one-turn matrix $\mathbf{\hat{M}}$\textemdash $\mathbf{\hat{x}_1}$, $\mathbf{\hat{x}_2}$, $\mathbf{\hat{x}_3}$, $\mathbf{\hat{x}_4}$ (where $\mathbf{\hat{x}_2} = \mathbf{\hat{x}_1}^*$ and $\mathbf{\hat{x}_4} = \mathbf{\hat{x}_3}^*$)\textemdash can be calculated. The eigenvalues corresponding to these eigenvectors can be grouped in pairs $\lambda_1 = e^{i 2\pi Q_1}$,  $\lambda_2 = e^{-i 2\pi Q_1} = \lambda_1^*$, $\lambda_3 = e^{i 2\pi Q_2}$,  $\lambda_4 = e^{-i 2\pi Q_2} = \lambda_3^*$\footnote{It should be noted that, in this second method, the eigenvectors are ordered: $\mathbf{\hat{x}_1}$ corresponds to a positive phase and $\mathbf{\hat{x}_2}$ to a negative phase. Also, as the transfer matrices $\mathbf{\hat{M}}$ and $\mathbf{\hat{P}}$ are related by a similarity transformation, they have the same eigenvalues \cite{Edwards_Teng}.}

The eigenvectors of $\mathbf{\hat{P}}$ and $\mathbf{\hat{M}}$ are normalized as follows (see also Eq.~\eqref{normalisation_eigenvectors_}):
\begin{equation}
    \label{normalisation_eigenvectors}
	\left\lbrace
		\begin{aligned}
            \mathbf{\hat{e}_i^+}\  \mathbf{S}\  \mathbf{\hat{e}_j} &= \pm i \quad &if \quad \delta_{ij} = 1\\
            \mathbf{\hat{e}_i^+}\  \mathbf{S}\  \mathbf{\hat{e}_j} &= 0 \quad &if \quad \delta_{ij} = 0\\
		\end{aligned}
	\right.
\end{equation}
where $\mathbf{\hat{e}}$ refers either to the $\mathbf{\hat{u}}$ eigenvectors or to the $\mathbf{\hat{x}}$ eigenvectors, and $\mathbf{\hat{e}_i^+} = \mathbf{\hat{e}_i^{*T}}$. The eigenvectors in the coupled space are related to the eigenvectors in the decoupled space via $\mathbf{\widetilde{R}}$:
\begin{equation}
    \mathbf{\hat{x}_1} = \mathbf{\widetilde{R}} \mathbf{\hat{u}_1}\text{.}
\end{equation}

It is thus possible to express the lattice parameters in terms of the $\mathbf{\hat{M}}$-eigenvector components \cite{Parzen}:
\begin{equation}
	\left\lbrace
		\begin{aligned}
		&x_1 = q_1 \sqrt{\beta_1} e^{i \mu_1} \\
		&p_{x1} = q_1 \frac{(-\alpha_1+i)}{\sqrt{\beta_1}} e^{i \mu_1}\\
		&\frac{p_{x1}}{x_1} = \frac{(-\alpha_1+i)}{\beta_1}
		\end{aligned}
	\right.	 \Rightarrow
	\left\lbrace
		\begin{aligned}
		\beta_1 &= \frac{1}{\text{Im }(\frac{p_{x1}}{x_1})} \\
		\alpha_1 &= -\beta_1 \text{Re }(\frac{p_{x1}}{x_1})\\
		\mu_1 &= \arg(x_1)
		\end{aligned}
	\right.
\end{equation}

Once the optical parameters are known, the eigenvectors of $\mathbf{\hat{P}}$ and $\mathbf{\hat{M}}$ are known explicitly. It is then possible to calculate the decoupling matrix $\mathbf{\widetilde{R}}$ from these two eigenvector sets. First, two matrices based on these eigenvectors are built:
\begin{align}
    \mathbf{U} &= \frac{1}{\sqrt{-i}}[\mathbf{\hat{u}_1}\  \mathbf{\hat{u}_2}\  \mathbf{\hat{u}_3}\  \mathbf{\hat{u}_4}] \label{U_matrix}\text{, }\\
    \mathbf{X} &= \frac{1}{\sqrt{-i}}[\mathbf{\hat{x}_1}\  \mathbf{\hat{x}_2}\  \mathbf{\hat{x}_3}\  \mathbf{\hat{x}_4}]\text{.}
    \label{X_matrix}
\end{align}
By taking into account the eigenvector normalization (Eq.~\eqref{normalisation_eigenvectors}), we see that the factor $\sqrt{-i}$ in the above expressions ensures the symplecticity of the matrices $\mathbf{U}$ and $\mathbf{X}$. The matrices $\mathbf{U}$ and $\mathbf{X}$ contain the eigenvectors of the coupled and decoupled spaces, and we can thus connect them by the matrix $\mathbf{\widetilde{R}}$: $\mathbf{X} = \mathbf{\widetilde{R}} \mathbf{U}$. By inverting this relation and by taking into account that $\mathbf{U}$ is symplectic, and thus $\mathbf{U}^{-1} = \mathbf{\overline{U}}$, the decoupling matrix is obtained:
\begin{equation}
    \mathbf{\widetilde{R}} = \mathbf{X} \mathbf{\overline{U}}\text{.}
\end{equation}

The symplecticity of the matrices $\mathbf{U}$ and $\mathbf{X}$ ensures the symplecticity of the matrix $\mathbf{\widetilde{R}}$. Using Floquet's theorem (Eq.~\eqref{Floquet_theorem}), the eigenvectors are expressed as the product of a periodic function and a harmonic factor. Each column of the matrices $\mathbf{U}$ and $\mathbf{X}$ will then contain a periodic function and a harmonic factor ($e^{\pm i\mu(s)}$). The harmonic factors cancel each other out when one calculates the product of $\mathbf{X}$ and $\mathbf{\overline{U}}$ so that only the periodic functions remain in the matrix $\mathbf{R}$, which is therefore periodic.

\subsubsection{Interpretation and advantages of the ET parametrization}
The 10 parameters of the ET parametrization for the coupled transverse motion are the 2 $\alpha$-functions, 2 $\beta$-functions, 2 phase advance $\mu$ and 4 periodic functions which describe the decoupling matrix $\mathbf{\widetilde{R}}$. The functions $\alpha$, $\beta$, and $\mu$ characterize the two eigenmodes and, thus, the oscillations in the decoupled space, while the parameters of the matrix $\mathbf{\widetilde{R}}$ describe the coupling between the two transverse motions (strength and structure). The functions $\alpha$, $\beta$, and $\mu$ of each eigenmode are defined in the same way as in the Courant-Snyder theory for uncoupled motion. However, these parameters are defined with respect to eigen-axes that no longer correspond to the physical axes. These Twiss parameters thus no longer have their usual physical interpretation, and some commonly used relations are no longer valid. In particular, the $\beta$-functions can become negative or infinite. In addition, the relations between the functions $\alpha$, $\beta$, and $\mu$ of the Courant-Snyder theory are no longer valid in decoupled space and must be generalized \cite{Parzen, parzenLinearOrbitParameters}. Finally, as mentioned above, mode identification can be tedious.

The phase advance $\mu_i$ is directly connected to the oscillation in the physical direction, which is the principal direction associated with the eigenmode $i$. As we will see in section \ref{Lien_ET_MR}, these phase advances are identical in the ET and  MR parametrizations.

The interpretation of the parameters of the decoupling matrix is detailed in Refs. \cite{Sagan_Rubin, bagleyCorrectionTransverseCoupling1989}. Notably, $\mathbf{\mathcal{C}}$, normalized by the $\beta$ functions, characterizes the coupling strength and can be used in coupling correction algorithms \cite{Sagan_Rubin}. The elements of $\mathbf{\mathcal{C}}$ are associated with the ellipse formed in the physical plane ($x-y$) when only one of the eigenmodes is excited\footnote{A more detailed interpretation can be found in \cite{bagleyCorrectionTransverseCoupling1989}.}. In addition, the parameters of the decoupling matrix ($\gamma$ and $\mathbf{\mathcal{C}}$) can be linked to the parameters of the difference coupling resonances obtained from the perturbative approach for weak coupling \cite{Desforges}. By making this link, we can see that the $\gamma$ parameter provides the coupling strength and indicates if the system is close to a coupling resonance and the type of this resonance. Finally, in the ET parametrization, the linear invariants are easily expressed in terms of the eigenmode lattice functions $\alpha$, $\beta$, and $\mu$ and have the same expression as the usual Courant-Snyder invariants (Eq.~\eqref{Courant_Snyder_invariant}).

\subsection{Mais and Ripken (MR) parametrization}\label{MR_parametrization}
The MR parametrization does not focus on the motion in the decoupled eigen-axes but provides lattice functions that depend on the oscillation modes and physical directions along which the beam envelope can be measured. The physical interpretation of these lattice functions is similar to the usual Twiss interpretation of the $\Sigma$ matrix of the second-order moments in the physical laboratory axes. For example, the $\beta$ functions (always positive and finite) characterize the amplitude of the betatron oscillations and can be used to obtain the beam sizes. At least 10 parameters are required for a $4\times4$ symplectic transfer matrix and the parameter set typically includes two main phase advances, four main lattice functions $\beta$, $\alpha$, or $\gamma$ (which reflect the motion of an oscillation mode in its principal transverse direction), and parameters reflecting the coupling. The chosen set may differ between authors and may include more than 10 parameters \cite{wolskiAlternativeApproachGeneral2006}. In particular, variants exist for the coupling parameters describing the off-diagonal part of the normalization matrix. These parameters are described either by non-principal $\beta$, $\alpha$, and phase advances as in Ref.~\cite{Willeke, Lebedev} or by complex parameters which combine these non-principal functions into a single quantity as in Ref.~\cite{wolskiNormalFormAnalysis2004}. Considering additional parameters allows having similar expressions for all the optical functions, as well as elegant expressions for measurable beam parameters.

This description of the coupled motion can be performed using two distinct but related approaches. The first uses transfer matrices as the basis of the description. In the uncoupled case, a normalization transformation casts the transfer matrix into a rotation matrix - $\mathbf{T^{-1} \hat{M} T = R}$. Analogously, the MR approach parameterizes this normalization matrix for coupled motion. The second approach starts from the phase space trajectories with the generating vectors defining the curve defined by the turn-by-turn coordinates. These generating vectors can be parameterized by the lattice functions (similarly to the uncoupled case, see Eqs. \eqref{uncoupled_generating_vectors_1}, \eqref{uncoupled_generating_vectors}).

Starting from the eigenvectors of the one-turn transfer matrix of the canonical coordinates $\mathbf{\hat{v}_1}$, $\mathbf{\hat{v}_1^*}$, $\mathbf{\hat{v}_2}$, $\mathbf{\hat{v}_2}^*$, the eigenvectors $\mathbf{\hat{v}_1}$ and $\mathbf{\hat{v}_1^*}$ are associated with the eigenvalues $e^{\pm i 2 \pi Q_1} = e^{\pm i \mu_1}$ while the eigenvectors $\mathbf{\hat{v}_2}$ and $\mathbf{\hat{v}_2}^*$ are associated with the eigenvalues $e^{\pm i 2\pi Q_2} = e^{\pm i \mu_2}$, where $Q_1$ and $Q_2$ are the eigen-tunes of the machine. These eigenvectors can be expressed by their real and imaginary parts:
\begin{align}
    \mathbf{\hat{v}_1} &= \frac{1}{\sqrt{2}}(\mathbf{\hat{z}_1}+ i \mathbf{\hat{z}_2}) \text{,}\label{Re_Im_v1}\\
    \mathbf{\hat{v}_2} &= \frac{1}{\sqrt{2}}(\mathbf{\hat{z}_3}+ i \mathbf{\hat{z}_4})\text{.}\label{Re_Im_v2}
\end{align}

The matrix $\mathbf{E}$, with columns corresponding to the one-turn transfer matrix eigenvectors, diagonalizes the transfer matrix:
\begin{equation}
    \mathbf{E^{-1}} \mathbf{\hat{M}} \mathbf{E} = \mathbf{\Lambda} \text{,}
\end{equation}
where $\mathbf{\Lambda} = \begin{pmatrix}
    e^{i\mu_1} & 0 & 0 & 0\\
    0 & e^{-i\mu_1} & 0 & 0\\
    0 & 0 & e^{i\mu_2} & 0\\
    0 & 0 & 0 & e^{-i\mu_2}\\
\end{pmatrix}$, and
\begin{eqnarray*}
    \mathbf{E} &=& [\mathbf{\hat{v}_1} \quad \mathbf{\hat{v}_1^*} \quad \mathbf{\hat{v}_2} \quad \mathbf{\hat{v}_2^*}]\\
               &=& \frac{1}{\sqrt{2}} [\mathbf{\hat{z}_1}+ i \mathbf{\hat{z}_2} \quad \mathbf{\hat{z}_1}- i \mathbf{\hat{z}_2} \quad \mathbf{\hat{z}_3}+ i \mathbf{\hat{z}_4} \quad \mathbf{\hat{z}_3}- i \mathbf{\hat{z}_4}] \text{.}
\end{eqnarray*}

In addition, there is a normalization transformation that transforms the transfer matrix into its normal form (\textit{i.e.} a rotation matrix): $\mathbf{N^{-1} \hat{M} N} = \mathbf{R}(\mu_1, \mu_2)$. The rotation matrix $\mathbf{R}(\mu_1, \mu_2)$ can also be diagonalized by its eigenvectors:
\begin{equation}
\mathbf{K^{-1}} \mathbf{R} \mathbf{K} = \mathbf{\Lambda} \text{,}
\end{equation}
with
\begin{equation*}
    \mathbf{K} = \frac{1}{\sqrt{2}}\begin{pmatrix}
        1 & 1 & 0 & 0\\
        i & -i & 0 &0 \\
        0 & 0 & 1 & 1 \\
        0 & 0 & i & -i
    \end{pmatrix} \text{.}
\end{equation*}
We can therefore link the normalization matrix $\mathbf{N}$ and the diagonalization matrix $\mathbf{E}$ containing the eigenvectors of the one-turn transfer matrix:
\begin{eqnarray*}
    \mathbf{\hat{M}} &=& \mathbf{N} \mathbf{R} \mathbf{N}^{-1} \\
               &=& \mathbf{N} \mathbf{K} \mathbf{\Lambda} \mathbf{K}^{-1} \mathbf{N}^{-1}\\
               &=& \mathbf{E}\mathbf{\Lambda} \mathbf{E^{-1}}\text{.}
\end{eqnarray*}
The diagonalization transformation $\mathbf{E}$ combines a normalization transformation $\mathbf{N}$ that contains the lattice functions reflecting the oscillation amplitudes and a transformation $\mathbf{K}$ that diagonalizes the rotation matrix. The matrix $\mathbf{K}$ transforms the Courant-Snyder coordinates into the complex Courant-Snyder coordinates \cite{NF94}. The normalization transformation thus contains the real and imaginary parts of the eigenvectors, and the $\mathbf{K}$ transformation combines these real and imaginary parts to form a single complex eigenvector. We can therefore write the normalization matrix in terms of the real and imaginary parts of the one-turn transfer matrix eigenvectors:
\begin{align}
     \mathbf{N} &= \sqrt{2} [\text{Re}(\mathbf{\hat{v}_1}) \quad \text{Im}(\mathbf{\hat{v}_1}) \quad \text{Re}(\mathbf{\hat{v}_2}) \quad \text{Im}(\mathbf{\hat{v}_2})] \label{N_vecteurs_propres} \\
                &= [\mathbf{\hat{z}_1} \quad \mathbf{\hat{z}_2} \quad \mathbf{\hat{z}_3} \quad \mathbf{\hat{z}_4}]\text{.}
\end{align}

By parameterizing the normalization matrix, the eigenvectors of the one-turn transfer matrix in the coupled physical space are parameterized. Now that we have the relation between the normalization matrix and the one-turn transfer matrix eigenvectors, we want to establish the link between these eigenvectors and the generating vectors of the surface supporting the motion in phase space. Since we consider the four-dimensional $x-x'-y-y'$ phase space, the transfer matrix and vectors are here written in terms of geometric coordinates. The eigenvectors of the transfer matrix fully describe the motion and the trajectory of the particle can be written as a linear combination of these four eigenvectors. These eigenvectors form pairs of conjugated eigenvectors. The particle trajectory in phase space is expressed as a linear combination of two eigenvectors weighted by complex constants:
\begin{equation}
    \mathbf{z}(s) = \text{Re}(\sqrt{2 \epsilon_{I}} \mathbf{v_1}(s)e^{i\phi_{I,0}} + \sqrt{2 \epsilon_{II}} \mathbf{v_2}(s)e^{i\phi_{II,0}})\text{.}
\end{equation}

The eigenvectors $\mathbf{v_1}(s)$ and $\mathbf{v_2}(s)$ correspond to the initial eigenvectors $\mathbf{v_1}(s_0)$ and $\mathbf{v_2}(s_0)$ propagated by the transfer matrix $\mathbf{M}_{s_0 \rightarrow s}$:
\begin{align}
    \mathbf{v_1}(s) &= \mathbf{M}_{s_0 \rightarrow s} \mathbf{v_1}(s_0) \text{, }\\
    \mathbf{v_2}(s) &= \mathbf{M}_{s_0 \rightarrow s} \mathbf{v_2}(s_0)\text{.}
\end{align}
Each of these eigenvectors can also be written as the product of a harmonic factor and a periodic function (see Appendix \ref{appendeix_floquet} on Floquet's theorem):
\begin{align}
    \mathbf{v_1}(s) &= e^{i\mu_1(s)} \mathbf{v_1} \label{Floquet_v1}\text{,}\\
    \mathbf{v_2}(s) &= e^{i\mu_2(s)} \mathbf{v_2} \label{Floquet_v2}\text{.}
\end{align}
The periodic functions $\mathbf{v_1}$ and $\mathbf{v_2}$ correspond to the eigenvectors of the one-turn transfer matrix at $s$, $$\mathbf{\hat{M}}(s) = \mathbf{M}_{s_0 \rightarrow s} \mathbf{\hat{M}} \mathbf{M}_{s_0 \rightarrow s}^{-1} \text{,}$$ while the harmonic factors are phase factors in which the phase advance functions $\mu_1(s)$ and $\mu_2(s)$ appear. In what follows, we describe the particle trajectory in phase space with $\mathbf{v_1}(s)$ and $\mathbf{v_2}(s)$. These vectors thus include a phase factor. The eigenvectors that we used to build the normalization matrix are the one-turn transfer matrix eigenvectors and therefore do not contain this phase factor.

With the real and imaginary parts of $\mathbf{v_1}(s)$ and $\mathbf{v_2}(s)$, we can express any point on the particle's trajectory as follows:
\begin{equation}
  \label{particle_trajectory}
	\begin{aligned}
        \mathbf{z}(s) = & \sqrt{\epsilon_{I}}[\mathbf{z_1}(s) \cos{\phi_{I,0}} - \mathbf{z_2}(s) \sin{\phi_{I,0}}] \\
           & + \sqrt{\epsilon_{II}} [\mathbf{z_3}(s) \cos{\phi_{II,0}} - \mathbf{z_4}(s) \sin{\phi_{II,0}}]\text{.}
	\end{aligned}
\end{equation}
This expression describes trajectories in the four-dimensional phase space. The motion remains confined on a surface. In the coupled case, it is a toroidal surface whose projections in the $x-x^\prime$ plane and in the $y-y^\prime$ plane no longer correspond to an ellipse. The vectors $\mathbf{z_1}(s)$, $\mathbf{z_2}(s)$, $\mathbf{z_3}(s)$ and $\mathbf{z_4(s)}$ are four independent generating vectors of this toroidal surface. It clearly appears that the generating vectors of the phase space surface are linked to the real and imaginary parts of the one-turn transfer matrix eigenvectors. The parametrization of these generating vectors, in turn, parameterizes the eigenvectors of the transfer matrix in the coupled physical space.

Two variants are in use for the parametrization method. The first one uses the generating vectors, as for Willeke and Ripken (WR) in Ref.~\cite{Willeke}. The second one uses the normalization matrix, as for Lebedev and Bogacz (LB) in Ref.~\cite{Lebedev}, and Wolski in Refs. \cite{wolskiNormalFormAnalysis2004, wolskiSimpleWayCharacterize2004}. Both methods ultimately parameterize the eigenvectors of the one-turn transfer matrix.

\subsubsection{Parametrization of generating vectors}\label{generating_vectors_parametrization}
Willeke and Ripken \cite{Willeke} proceed as follows:
\begin{equation*}
    \mathbf{z_1}(s) = \begin{pmatrix}
        \sqrt{\beta_{xI}} \cos{\phi_{xI}} \\
        \sqrt{\gamma_{xI}} \cos{\overset{\sim}\phi_{xI}} \\
        \sqrt{\beta_{yI}} \cos{\phi_{yI}} \\
        \sqrt{\gamma_{yI}} \cos{\overset{\sim}\phi_{yI}}
    \end{pmatrix} \text{, }\quad
    \mathbf{z_2}(s) = \begin{pmatrix}
        \sqrt{\beta_{xI}} \sin{\phi_{xI}} \\
        \sqrt{\gamma_{xI}} \sin{\overset{\sim}\phi_{xI}} \\
        \sqrt{\beta_{yI}} \sin{\phi_{yI}} \\
        \sqrt{\gamma_{yI}} \sin{\overset{\sim}\phi_{yI}}
    \end{pmatrix}\text{, }
\end{equation*}
\begin{equation*}
    \mathbf{z_3}(s) = \begin{pmatrix}
        \sqrt{\beta_{xII}} \cos{\phi_{xII}} \\
        \sqrt{\gamma_{xII}} \cos{\overset{\sim}\phi_{xII}} \\
        \sqrt{\beta_{yII}} \cos{\phi_{yII}} \\
        \sqrt{\gamma_{yII}} \cos{\overset{\sim}\phi_{yII}}
    \end{pmatrix} \text{, }\quad
    \mathbf{z_4}(s) = \begin{pmatrix}
        \sqrt{\beta_{xII}} \sin{\phi_{xII}} \\
        \sqrt{\gamma_{xII}} \sin{\overset{\sim}\phi_{xII}} \\
        \sqrt{\beta_{yII}} \sin{\phi_{yII}} \\
        \sqrt{\gamma_{yII}} \sin{\overset{\sim}\phi_{yII}}
    \end{pmatrix}\text{. }
\end{equation*}
In these expressions, the generating vectors $\mathbf{z_1}$(s) and $\mathbf{z_2}$(s) correspond to the oscillation mode I, projected in the $x$ and $y$ directions. These vectors form the eigenvector $\mathbf{v_1}$(s), associated with the eigenvalue $e^{i 2\pi Q_1}$. The generating vectors $\mathbf{z_3}$(s) and $\mathbf{z_4}$(s) correspond to the oscillation mode II, and are associated with the eigenvector $\mathbf{v_2}$(s) whose eigenvalue is  $e^{i 2\pi Q_2}$. The normalization of the eigenvectors implies that the generating vectors are normalized as follows:
\begin{align}
 \mathbf{\hat{z}_1}^T \ \mathbf{S} \  \mathbf{\hat{z}_2} &= 1\text{,} \\
 \mathbf{\hat{z}_3}^T \ \mathbf{S} \  \mathbf{\hat{z}_4} &= 1\text{.}
\end{align}
The normalization conditions apply to the generating vectors expressed in canonical coordinates ($x$, $p_x$, $y$, $p_y$) while the parametrization presented above applies to generating vectors expressed in geometric coordinates ($x$, $x^\prime$, $y$, $y^\prime$); the matrix $\mathbf{U}$ transforms geometric coordinates into canonical coordinates:
\begin{equation*}
\mathbf{\hat{z}} = \mathbf{U}\mathbf{z}\text{.}
\end{equation*}

The parametrization of the generating vectors $\mathbf{\hat{z}_1}(s)$, $\mathbf{\hat{z}_2}(s)$, $\mathbf{\hat{z}_3}(s)$ and $\mathbf{\hat{z}_4}(s)$ is given by:
\begin{equation*}
    \mathbf{\hat{z}_1}(s) = \begin{pmatrix}
        \sqrt{\beta_{xI}} \cos{\phi_{xI}} \\
        \sqrt{\gamma_{xI}} \cos{\overset{\sim}\phi_{xI}} - \frac{R_1}{2} \sqrt{\beta_{yI}} \cos{\phi_{yI}}\\
        \sqrt{\beta_{yI}} \cos{\phi_{yI}} \\
        \sqrt{\gamma_{yI}} \cos{\overset{\sim}\phi_{yI}} + \frac{R_2}{2} \sqrt{\beta_{xI}} \cos{\phi_{xI}}
    \end{pmatrix}\text{, }
\end{equation*}
\begin{equation*}
    \mathbf{\hat{z}_2}(s) = \begin{pmatrix}
        \sqrt{\beta_{xI}} \sin{\phi_{xI}} \\
        \sqrt{\gamma_{xI}} \sin{\overset{\sim}\phi_{xI}} -\frac{R_1}{2} \sqrt{\beta_{yI}} \sin{\phi_{yI}}\\
        \sqrt{\beta_{yI}} \sin{\phi_{yI}} \\
        \sqrt{\gamma_{yI}} \sin{\overset{\sim}\phi_{yI}} + \frac{R_2}{2} \sqrt{\beta_{xI}} \sin{\phi_{xI}}
    \end{pmatrix}\text{, }
\end{equation*}
\begin{equation*}
    \mathbf{\hat{z}_3}(s) = \begin{pmatrix}
        \sqrt{\beta_{xII}} \cos{\phi_{xII}} \\
        \sqrt{\gamma_{xII}} \cos{\overset{\sim}\phi_{xII}} -\frac{R_1}{2}\sqrt{\beta_{yII}} \cos{\phi_{yII}}\\
        \sqrt{\beta_{yII}} \cos{\phi_{yII}} \\
        \sqrt{\gamma_{yII}} \cos{\overset{\sim}\phi_{yII}}+\frac{R_2}{2}\sqrt{\beta_{xII}} \cos{\phi_{xII}}
    \end{pmatrix}\text{, }
\end{equation*}
\begin{equation*}
    \mathbf{\hat{z}_4}(s) = \begin{pmatrix}
        \sqrt{\beta_{xII}} \sin{\phi_{xII}} \\
        \sqrt{\gamma_{xII}} \sin{\overset{\sim}\phi_{xII}} -\frac{R_1}{2} \sqrt{\beta_{yII}} \sin{\phi_{yII}} \\
        \sqrt{\beta_{yII}} \sin{\phi_{yII}}\\
        \sqrt{\gamma_{yII}} \sin{\overset{\sim}\phi_{yII}} + \frac{R_2}{2}\sqrt{\beta_{xII}} \sin{\phi_{xII}}
    \end{pmatrix}\text{. }
\end{equation*}

The normalization conditions on the generating vectors allow finding the link between the lattice parameters of the parametrization. For each mode, the normalization condition imposes:
\begin{equation}
    \beta_x\phi_x'+ \beta_y\phi_y'+ \frac{1}{2} (R_1 + R_2) \sqrt{\beta_x \beta_y}\sin{(\phi_x - \phi_y)} = 1 \text{.}
    \label{normalisation_condition}
\end{equation}
In addition, some lattice parameters are related. The phase functions $\overset{\sim}\phi$ are related to the phase advance functions $\phi$ by the relation \cite{Wiedemann}:
\begin{equation}
\overset{\sim}\phi(s) = \phi(s) - \text{arctan}(\frac{\beta \phi'}{\alpha}) \text{,}
\end{equation}
and there is a relation between the lattice parameters $\alpha$, $\beta$, and $\gamma$ for each set of optical functions associated with a mode (I or II) and with a transverse direction ($x$ or $y$):
\begin{equation}
    \gamma = \frac{\beta^2\phi'^2 + \alpha^2}{\beta}\text{.}
\end{equation}

With these relations, we can express the generating vectors using only the lattice functions $\alpha$, $\beta$, and $\phi$ associated with each mode and each transverse direction:
\begin{widetext}
\begin{equation*}
    \mathbf{\hat{z}_1}(s) = \begin{pmatrix}
        \sqrt{\beta_{xI}} \cos{\phi_{xI}} \\
        -\frac{\alpha_{xI}}{\sqrt{\beta_{xI}}}\cos{\phi_{xI}} - \frac{\beta_{xI}\phi_{xI}'}{\sqrt{\beta_{xI}}}\sin{\phi_{xI}} - \frac{R_1}{2} \sqrt{\beta_{yI}} \cos{\phi_{yI}}\\
        \sqrt{\beta_{yI}} \cos{\phi_{yI}} \\
        -\frac{\alpha_{yI}}{\sqrt{\beta_{yI}}}\cos{\phi_{yI}} - \frac{\beta_{yI}\phi_{yI}'}{\sqrt{\beta_{yI}}}\sin{\phi_{yI}} + \frac{R_2}{2} \sqrt{\beta_{xI}} \cos{\phi_{xI}}
    \end{pmatrix}\text{, }
\end{equation*}
\begin{equation*}
    \mathbf{\hat{z}_2}(s) = \begin{pmatrix}
        \sqrt{\beta_{xI}} \sin{\phi_{xI}} \\
        -\frac{\alpha_{xI}}{\sqrt{\beta_{xI}}}\sin{\phi_{xI}} + \frac{\beta_{xI}\phi_{xI}'}{\sqrt{\beta_{xI}}}\cos{\phi_{xI}} -\frac{R_1}{2} \sqrt{\beta_{yI}} \sin{\phi_{yI}}\\
        \sqrt{\beta_{yI}} \sin{\phi_{yI}} \\
         -\frac{\alpha_{yI}}{\sqrt{\beta_{yI}}}\sin{\phi_{yI}} + \frac{\beta_{yI}\phi_{yI}'}{\sqrt{\beta_{yI}}}\cos{\phi_{yI}} + \frac{R_2}{2} \sqrt{\beta_{xI}} \sin{\phi_{xI}}
    \end{pmatrix}\text{, }
\end{equation*}
\begin{equation*}
    \mathbf{\hat{z}_3}(s) = \begin{pmatrix}
        \sqrt{\beta_{xII}} \cos{\phi_{xII}} \\
        -\frac{\alpha_{xII}}{\sqrt{\beta_{xII}}}\cos{\phi_{xII}} - \frac{\beta_{xII}\phi_{xII}'}{\sqrt{\beta_{xII}}}\sin{\phi_{xII}} -\frac{R_1}{2}\sqrt{\beta_{yII}} \cos{\phi_{yII}}\\
        \sqrt{\beta_{yII}} \cos{\phi_{yII}} \\
        -\frac{\alpha_{yII}}{\sqrt{\beta_{yII}}}\cos{\phi_{yII}} - \frac{\beta_{yII}\phi_{yII}'}{\sqrt{\beta_{yII}}}\sin{\phi_{yII}} + \frac{R_2}{2}\sqrt{\beta_{xII}} \cos{\phi_{xII}}
    \end{pmatrix}\text{, }
\end{equation*}
\begin{equation*}
    \mathbf{\hat{z}_4}(s) = \begin{pmatrix}
        \sqrt{\beta_{xII}} \sin{\phi_{xII}} \\
        -\frac{\alpha_{xII}}{\sqrt{\beta_{xII}}}\sin{\phi_{xII}} + \frac{\beta_{xII}\phi_{xII}'}{\sqrt{\beta_{xII}}}\cos{\phi_{xII}} - \frac{R_1}{2} \sqrt{\beta_{yII}} \sin{\phi_{yII}} \\
        \sqrt{\beta_{yII}} \sin{\phi_{yII}}\\
        -\frac{\alpha_{yII}}{\sqrt{\beta_{yII}}}\sin{\phi_{yII}} + \frac{\beta_{yII}\phi_{yII}'}{\sqrt{\beta_{yII}}}\cos{\phi_{yII}} + \frac{R_2}{2}\sqrt{\beta_{xII}} \sin{\phi_{xII}}
    \end{pmatrix}\text{.}
\end{equation*}
\end{widetext}

\begin{figure}[h]
\includegraphics[width=0.99\linewidth]{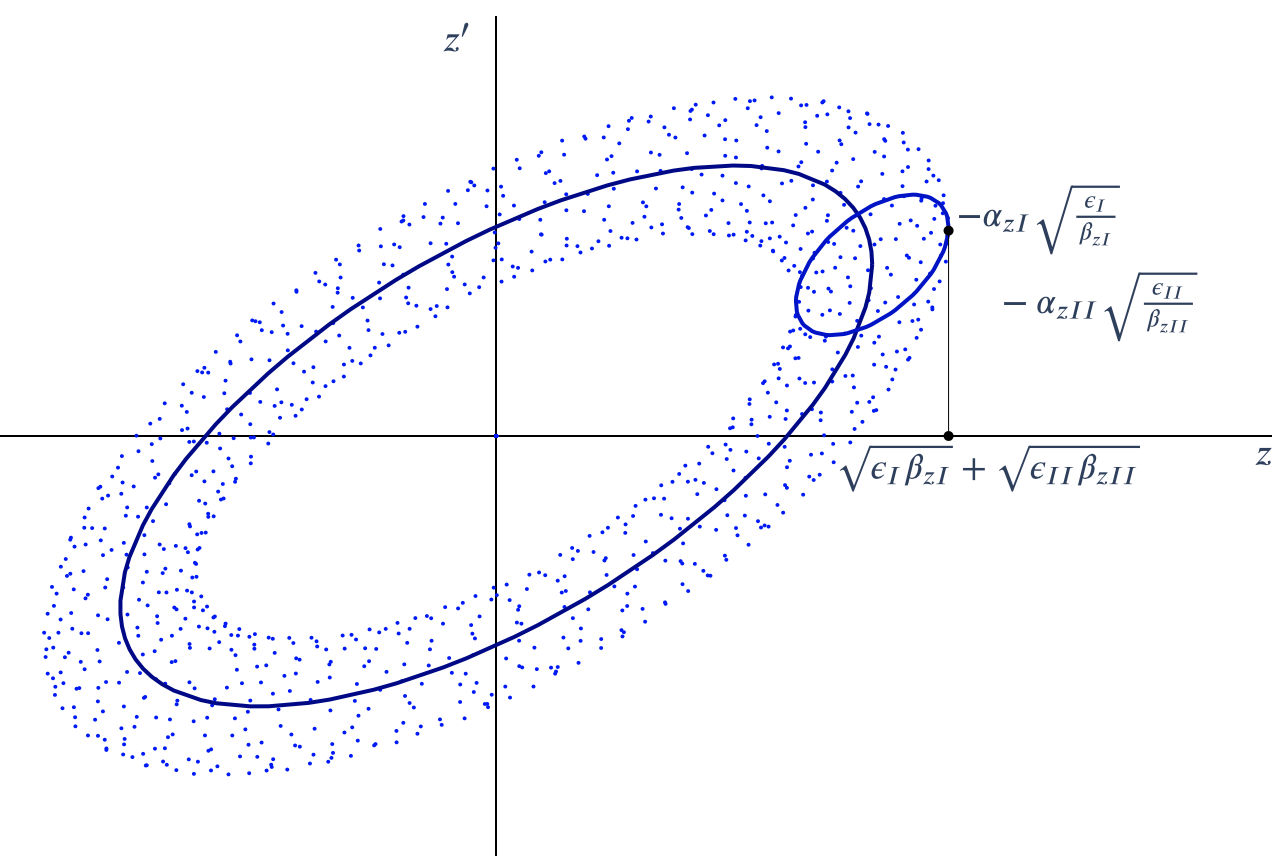}
\caption{Projection of the 4D torus on the $z-z'$ phase space ($z=x,y$): superposition of two ellipses described by the lattice functions associated with the plane and reflecting the projection of the 2 oscillation modes on the plane. Reproduced from Ref.~\cite{Willeke}.}
\label{ellipses_Willeke}
\end{figure}

We thus have a set of 20 parameters that are related to each other: ($\beta$, $\alpha$, $\gamma$, $\phi$, and $\overset{\sim}\phi$) for each mode and each transverse direction. It is possible to express the generating vectors with a subset of these parameters. The parametrization presented in Ref.~\cite{Willeke} uses the $\beta$, $\gamma$, $\phi$ and $\overset{\sim}\phi$ functions to highlight the meaning of these parameters and the parallel that can be made between position functions and angle functions: the $\beta$ and $\gamma$ functions are envelope functions for the position and angle coordinates while the $\phi$ and $\overset{\sim}\phi$ functions are phase functions for the position and angle coordinates. In what follows, we will rather express the generating vectors in terms of the $\alpha$, $\beta$, and $\phi$ functions to compare the parameters of this parametrization with others. The set of lattice parameters presented in Ref.~\cite{Willeke} characterizes the 4D phase space surface. When looking at the projections of this surface in the $x-x'$ and $y-y'$ phase planes, we get a set of points that can be characterized by the superposition of two ellipses, as shown in Fig.~\ref{ellipses_Willeke}. These two ellipses are characterized by the optical functions associated with the two oscillation modes. The areas of the ellipses corresponding to the two oscillation modes projected into one of the transverse phase planes ($z-z'$ where $z$ represents $x$ or $y$) can be calculated using the optical functions:
\begin{align}
    \Gamma_{zI} &= \pi \epsilon_1 \beta_{zI}|\phi_{zI}'| \label{area_1} \text{, }\\
    \Gamma_{zII} &= \pi \epsilon_2 \beta_{zII}|\phi_{zII}'| \label{area_2} \text{.}
\end{align}

Finally, the lattice functions can be easily related to the size of the beam in the horizontal and vertical directions. Expressing $\mathbf{z}$(s) as a weighted sum of the generating vectors, one clearly sees that the maximum horizontal oscillation amplitude is $\sqrt{\epsilon_1 \beta_{xI}} + \sqrt{\epsilon_2 \beta_{xII}}$, while the maximum vertical oscillation amplitude is $\sqrt{\epsilon_1 \beta_{yI}} + \sqrt{\epsilon_2 \beta_{yII}}$.

\subsubsection{Parametrization of the normalization matrix}
The second way to introduce the coupled lattice functions is to parameterize the normalization matrix that transforms the transfer matrix to a rotation matrix $\mathbf{R}(\mu_1, \mu_2)$ \textemdash a block-diagonal matrix where each diagonal block is a 2D rotation matrix of angles $\mu_1$ and $\mu_2$. This approach is followed in Refs.~\cite{wolskiNormalFormAnalysis2004, Lebedev}.

The $2\times2$ blocks of the $\mathbf{N}$ diagonal will be associated with principal lattice functions $\beta$ and $\alpha$. The off-diagonal blocks characterize the coupling between the two transverse oscillations and are described by coupling functions that represent the so-called ``non-principal oscillations'', \textit{i.e.} the motion in a transverse direction due to the eigenmode which is not related to the considered transverse direction in the limit of weak coupling. In general, none of the elements of the normalization matrix can \textit{a priori} be considered zero. The normalization matrix is therefore written in the most general way as follows:
\begin{equation}
    \mathbf{N} = \begin{pmatrix}
    n_{11} & n_{12} & n_{13} & n_{14} \\
    n_{21} & n_{22} & n_{23} & n_{24} \\
    n_{31} & n_{32} & n_{33} & n_{34} \\
    n_{41} & n_{42} & n_{43} & n_{44}
    \end{pmatrix} \text{.}\\
\end{equation}
As for the uncoupled case, there is some freedom in the choice of $\mathbf{N}$ as a matrix $\mathbf{\bar{N}} = \mathbf{N} \mathbf{R}(\theta_1, \theta_2)$ is also a normalization matrix. This degeneracy originates from the fact that only the phase differences are physically measurable and that the overall phases are not known. The angles $\theta_1$ and $\theta_2$ are usually chosen so as to cancel the elements $n_{12}$ and $n_{34}$ \cite{wolskiSimpleWayCharacterize2004}. The normalization matrix is a symplectic matrix and therefore has 10 independent parameters. By imposing $n_{12}=0$ and $n_{34} = 0$, 2 of these parameters are set. A minimal parametrization of the normalization matrix, therefore, requires 8 other independent parameters. The total transfer matrix will then be parameterized by these 8 independent parameters as well as by the two phase advances appearing in the rotation matrix $\mathbf{R}(\mu_1, \mu_2)$.

LB parameterize the normalization matrix as follows \cite{Lebedev}:
\begin{widetext}
\begin{equation}
    \mathbf{N} = \begin{pmatrix}
    \sqrt{\beta_{1x}}                      & 0  & \sqrt{\beta_{2x}}\cos{\nu_2} & -\sqrt{\beta_{2x}}\sin{\nu_2}\\
    - \frac{\alpha_{1x}}{\sqrt{\beta_{1x}}}  & \frac{1-u}{\sqrt{\beta_{1x}}} & \frac{u \sin {\nu_2} -\alpha_{2x}\cos{\nu_2}}{\sqrt{\beta_{2x}}} & \frac{u \cos {\nu_2} +\alpha_{2x} \sin{\nu_2}}{\sqrt{\beta_{2x}}}\\

    \sqrt{\beta_{1y}}\cos{\nu_1}              & - \sqrt{\beta_{1y}}\sin{\nu_1} & \sqrt{\beta_{2y}} &  0 \\
    \frac{u \sin {\nu_1} -\alpha_{1y}\cos{\nu_1}}{\sqrt{\beta_{1y}}} & \frac{u \cos{\nu_1} +\alpha_{1y} \sin{\nu_1}}{\sqrt{\beta_{1y}}} & - \frac{\alpha_{2y}}{\sqrt{\beta_{2y}}} & \frac{1-u}{\sqrt{\beta_{2y}}}\\
    \end{pmatrix} \text{.}
    \label{param_lebedev}
\end{equation}
\end{widetext}
This parametrization includes 10 independent parameters (four $\beta$ functions, four $\alpha$ functions and the two phase advances $\mu_1$ and $\mu_2$ appearing in the rotation matrix) and 3 additional real functions ($\nu_1$, $\nu_2$ and $u$). The interpretation of these coupling parameters is clarified in the next section.

A slightly different way to parameterize the normalization matrix $\mathbf{N}$ is proposed by Wolski in Refs.~\cite{wolskiSimpleWayCharacterize2004, wolskiNormalFormAnalysis2004}:
\begin{widetext}
\begin{align}
    \mathbf{N} &= \begin{pmatrix}
        n_{11} & 0 & n_{13} & n_{14} \\
        n_{21} & n_{22} & n_{23} & n_{24} \\
        n_{31} & n_{32} & n_{33} & 0 \\
        n_{41} & n_{42} & n_{43} & n_{44}
    \end{pmatrix}\\
     &= \begin{tikzpicture}[baseline]
        \matrix[mymatrixstyle] (mymatr) {
        \sqrt{\beta_x} \& 0 \& n_{13} \& n_{14} \&[-7pt]\\
        -\dfrac{\alpha_x}{\sqrt{\beta_x}} \& n_{22} \& n_{23} \& n_{24} \\
        n_{31} \& n_{32} \& \sqrt{\beta_y} \& 0 \\
        n_{41} \& n_{42} \& -\dfrac{\alpha_y}{\sqrt{\beta_y}} \& n_{44}\\};
        \node[draw, fit=(mymatr-3-1)(mymatr-3-2)] (n3132) {};
        \node[draw, fit=(mymatr-4-1)(mymatr-4-2)] (n4142) {};
        \node[draw, fit=(mymatr-1-3)(mymatr-1-4)] (n1314) {};
        \node[draw, fit=(mymatr-2-3)(mymatr-2-4)] (n2324) {};
        \draw[myarrow, overlay] (n3132) -- +(-1.8,0) node[anchor=east] {$\zeta_x = n_{31}+in_{32}$};
        \draw[myarrow, overlay] (n4142) -- +(-1.8,0) node[anchor=east] {$\tilde{\zeta}_y=n_{41}-in_{42}$};
        \draw[myarrow, overlay] (n1314) -- +(1.8,0) node[anchor=west] {$\zeta_y=n_{13}+ i n_{14}$};
        \draw[myarrow, overlay] (n2324) -- +(1.8,0) node[anchor=west] {$\tilde{\zeta}_x=n_{23}-in_{24}$};
    \end{tikzpicture}
\end{align}
\end{widetext}
This parametrization includes the main optical functions $\beta_x$, $\alpha_x$, $\beta_y$, $\alpha_y$, and functions reflecting the coupling $\zeta_x$, $\zeta_y$, $\overset{\sim}{\zeta_x}$, $\overset{\sim}{\zeta_y}$, which combines the different non-principal optical functions appearing in \cite{Willeke} and \cite{Lebedev}. These coupling functions correspond to the components of the eigenvectors that only appear when there is coupling between the two transverse directions.

In addition to these approaches which are restricted to a minimal number of parameters, it should be noted that parametrizations with more parameters have also been proposed. In Ref.~\cite{wolskiAlternativeApproachGeneral2006}, Wolski defines three matrices $\mathbf{B}^k$ ($k=$I, II, III), whose elements determine the lattice functions. Each of these matrices $\mathbf{B}^k$ is associated with an oscillation mode ($k=$I, II, III) and is defined as follows:
\begin{equation}
    \mathbf{B}^k = \mathbf{N}\mathbf{T}^k \mathbf{N}^T\text{,}
\end{equation}
where the matrix $\mathbf{N}$ is the normalization matrix, and the matrix $\mathbf{T}^k$ is a block-diagonal matrix where the block $k$ is the identity matrix and the other blocks are zero. It is equivalent to defining the lattice functions as follows:
\begin{eqnarray*}
    \beta_{ij}^I = n_{i1}n_{j1}+ n_{i2}n_{j2} \text{, }\\
    \beta_{ij}^{II} = n_{i3}n_{j3}+ n_{i4}n_{j4} \text{, }\\
    \beta_{ij}^{III} = n_{i5}n_{j5}+ n_{i6}n_{j6} \text{.}
\end{eqnarray*}

The matrices $\mathbf{B}^k$ always have the same structure. The blocks ($2\times2$ matrices) on the $\mathbf{B}^k$ diagonal contain the principal and non-principal lattice functions $\alpha$, $\beta$ and $\gamma$ associated with the eigenmode of oscillation $k$. Each of these blocks is associated with a physical direction. For example, in a 2D motion, the first block will contain the lattice functions associated with the $x$ direction, and the second one with the $y$ direction. The matrix elements outside the diagonal blocks will correspond to the factors weighing the invariant associated with the eigenmode ($\epsilon_k$) in the expressions of the correlation matrix elements. For example, in the case of a 2D motion, the matrix $\mathbf{B}^I$, which is associated with the oscillation eigenmode I, can be written as follows using the parameters of LB:
\begin{widetext}
\begin{align}
  \mathbf{B}^I &= \begin{pmatrix}
      \mathbf{B}^I_{11}     & \mathbf{B}^I_{12} \\
      \mathbf{B}^I_{21}     & \mathbf{B}^I_{22}
  \end{pmatrix}\text{,} \quad
  \mathbf{B}^I_{11} = \begin{pmatrix}
    \beta_{1x}      &         -\alpha_{1x} \\
     -\alpha_{1x}    & \frac{\alpha_{1x}^2 + (1-u)^2}{\beta_{1x}}
  \end{pmatrix}\text{,} \quad
  \mathbf{B}^I_{22} = \begin{pmatrix}
      \beta_{1y} & -\alpha_{1y} \\
      -\alpha_{1y} & \frac{\alpha^2_{1y} + u^2}{\beta_{1y}}
  \end{pmatrix}\text{,}\\
  \mathbf{B}^I_{12} &= \begin{pmatrix}
      \sqrt{\beta_{1x}\beta_{1y}}\cos{\nu_1}      & \sqrt{\frac{\beta_{1x}}{\beta_{1y}}}(u\sin{\nu_1}- \alpha_{1y}\cos{\nu_1})\\
      -\sqrt{\frac{\beta_{1y}}{\beta_{1x}}}(\alpha_{1x}\cos{\nu_1} + (1-u) \sin{\nu_1})                                                   &      \frac{ \sin{\nu_1}(\alpha_{1y} (1-u)-\alpha_{1x}u) + \cos{\nu_1}(\alpha_{1x}\alpha_{1y} + (1-u)u)}{\sqrt{\beta_{1y}\beta_{1x}}}
  \end{pmatrix}\text{,}\\
  \mathbf{B}^I_{21} &= \begin{pmatrix}
      \sqrt{\beta_{1x}\beta_{1y}}\cos{\nu_1} &  -\sqrt{\frac{\beta_{1y}}{\beta_{1x}}}(\alpha_{1x}\cos{\nu_1} + (1-u) \sin{\nu_1}) \\
      \sqrt{\frac{\beta_{1x}}{\beta_{1y}}}(u\sin{\nu_1}- \alpha_{1y}\cos{\nu_1}) & \frac{ \sin{\nu_1}(\alpha_{1y} (1-u)-\alpha_{1x}u) + \cos{\nu_1}(\alpha_{1x}\alpha_{1y} + (1-u)u)}{\sqrt{\beta_{1y}\beta_{1x}}}
  \end{pmatrix}\text{.}
\end{align}
\end{widetext}

Any element of the correlation matrix $<ij>$ can thus be expressed as the sum of the three invariants ($\epsilon_{I}$, $\epsilon_{II}$ and $\epsilon_{III}$ in the more general case of a 3D motion) weighted by the elements $\beta^k_{ij}$ of the matrices $\mathbf{B}^k$:
\begin{equation}
<ij> = \beta^I_{ij} \epsilon_{I} + \beta^{II}_{ij} \epsilon_{II}+ \beta^{III}_{ij} \epsilon_{III} \text{.}
\end{equation}
This parametrization allows having similar definitions for all the lattice functions and finding elegant expressions for the correlation matrix elements in terms of these generalized lattice functions. It is straightforward to generalize the lattice functions to higher dimensions. However, despite this elegant formalism, some of the lattice functions $\beta^k_{ij}$ of this parametrization are a combination of the parameters commonly used, \textit{i.e.} the amplitudes and the phase shifts of the principal and non-principal oscillations. It is, therefore, complicated to physically interpret the individual meaning of these additional lattice functions.

\subsubsection{Comparison between variants of the MR parametrization}
To compare the parameters of WR\cite{Willeke} to parameterize the generating vectors with the parameters of Wolski \cite{wolskiNormalFormAnalysis2004} and LB \cite{Lebedev} to parameterize the normalization matrix, we can write the normalization matrix in terms of the parameters of Willeke. Using equations \eqref{Floquet_v1} and \eqref{Floquet_v2}, we can express the columns of the normalization matrix ($\mathbf{\hat{z}_1}$, $\mathbf{\hat{z}_2}$, $\mathbf{\hat{z}_3}$, $\mathbf{\hat{z}_4}$) using the generating vectors at any point $s$ ($\mathbf{\hat{z}_1}(s)$, $\mathbf{\hat{z}_2}(s)$, $\mathbf{\hat{z}_3}(s)$, $\mathbf{\hat{z}_4}(s)$):
\begin{align}
    \mathbf{\hat{z}_1} &= \cos{(\mu_1(s))}\mathbf{\hat{z}_1}(s) + \sin{(\mu_1(s))}\mathbf{\hat{z}_2}(s)\text{, }\\
    \mathbf{\hat{z}_2} &= -\sin{(\mu_1(s))}\mathbf{\hat{z}_1}(s) + \cos{(\mu_1(s))}\mathbf{\hat{z}_2}(s) \text{, }\\
    \mathbf{\hat{z}_3} &= \cos{(\mu_2(s))}\mathbf{\hat{z}_3}(s) + \sin{(\mu_2(s))}\mathbf{\hat{z}_4}(s)\text{, }\\
    \mathbf{\hat{z}_4} &= -\sin{(\mu_2(s))}\mathbf{\hat{z}_3}(s) + \cos{(\mu_2(s))}\mathbf{\hat{z}_4}(s)\text{.}
\end{align}
By noticing that $\mu_1(s) = \phi_{xI}(s)$ and that $\mu_2(s)$ = $\phi_{yII}(s)$, we get:
\begin{widetext}
\begin{equation}
    \mathbf{\hat{z_1}} = \begin{pmatrix}
    \sqrt{\beta_{xI}} \\
    -\frac{\alpha_{xI}}{\sqrt{\beta_{xI}}} - \frac{R_1}{2} \frac{\sqrt{\beta_{xI}\beta_{yI}}}{\sqrt{\beta_{xI}}}\cos{(\mu_1 - \phi_{yI})} \\
    \sqrt{\beta_{yI}}\cos{(\mu_1 - \phi_{yI})}\\
    -\frac{\alpha_{yI}}{\sqrt{\beta_{yI}}}\cos{(\mu_1 - \phi_{yI})} +\frac{\beta_{yI}\phi_{yI}'}{\sqrt{\beta_{yI}}}\sin{(\mu_1 - \phi_{yI})} + \frac{R_2}{2} \frac{\sqrt{\beta_{xI}\beta_{yI}}}{\sqrt{\beta_{yI}}}
    \end{pmatrix}\text{, }
\end{equation}
\begin{equation}
    \mathbf{\hat{z_2}} = \begin{pmatrix}
    0\\
    \frac{\beta_{xI}\phi_{xI}'}{\sqrt{\beta_{xI}}} + \frac{R_1}{2} \frac{\sqrt{\beta_{xI}\beta_{yI}}}{\sqrt{\beta_{xI}}}\sin{(\mu_1 - \phi_{yI})}\\
    -\sqrt{\beta_{yI}}\sin{(\mu_1 - \phi_{yI})}\\
    \frac{\alpha_{yI}}{\sqrt{\beta_{yI}}}\sin{(\mu_1 - \phi_{yI})} +\frac{\beta_{yI}\phi_{yI}'}{\sqrt{\beta_{yI}}}\cos{(\mu_1 - \phi_{yI})}
    \end{pmatrix}\text{, }
\end{equation}
\begin{equation}
    \mathbf{\hat{z_3}} = \begin{pmatrix}
    \sqrt{\beta_{xII}}\cos{(\mu_2 - \phi_{xII})}\\
    -\frac{\alpha_{xII}}{\sqrt{\beta_{xII}}}\cos{(\mu_2 - \phi_{xII})} +\frac{\beta_{xII}\phi_{xII}'}{\sqrt{\beta_{xII}}}\sin{(\mu_2 - \phi_{xII})} - \frac{R_1}{2} \frac{\sqrt{\beta_{xII}\beta_{yII}}}{\sqrt{\beta_{xII}}}\\
    \sqrt{\beta_{yII}} \\
    -\frac{\alpha_{yII}}{\sqrt{\beta_{yII}}} + \frac{R_2}{2} \frac{\sqrt{\beta_{xII}\beta_{yII}}}{\sqrt{\beta_{yII}}}\cos{(\mu_2 - \phi_{xII})}
    \end{pmatrix}\text{, }
\end{equation}
\begin{equation}
    \mathbf{\hat{z_4}} = \begin{pmatrix}
    -\sqrt{\beta_{xII}}\sin{(\mu_2 - \phi_{xII})}\\
    \frac{\alpha_{xII}}{\sqrt{\beta_{xII}}}\sin{(\mu_2 - \phi_{xII})} +\frac{\beta_{xII}\phi_{xII}'}{\sqrt{\beta_{xII}}}\cos{(\mu_2 - \phi_{xII})}\\
    0 \\
    \frac{\beta_{yII}\phi_{yII}'}{\sqrt{\beta_{yII}}} - \frac{R_2}{2} \frac{\sqrt{\beta_{xII}\beta_{yII}}}{\sqrt{\beta_{yII}}}\sin{(\mu_2 - \phi_{xII})}
    \end{pmatrix}\text{.}
\end{equation}\\
\end{widetext}
By comparing these columns with the normalization matrix parameterized by LB (Eq.~\eqref{param_lebedev}), it clearly appears that the non-principal phase advances $\phi_{yI}$ and $\phi_{xII}$ of WR are related to the real functions $\nu_1$ and $\nu_2$ of LB:
\begin{align}
    \nu_1 &= \mu_1 - \phi_{yI}\text{, } \\
    \nu_2 &= \mu_2 - \phi_{xII}\text{.}
\end{align}
The function $\nu_1$ (resp. $\nu_2$) therefore represents the difference between the main phase advance due to mode I in the $x$-direction (resp. mode II in the $y$-direction) and the non-principal phase advance due to mode I in the $y$-direction (resp. mode II in the $x$-direction). These functions represent the phase shift of the non-principal oscillation with respect to the principal oscillation of the same oscillation eigenmode. Moreover, we see that the principal and non-principal optical functions are similar, except that the coupling due to the longitudinal field is directly taken into account in the $\alpha$-functions of the LB parametrization, while this is not the case in the WR parametrization. Table~\ref{compa_params} summarizes the link between the $\alpha$ and $\beta$-functions in the two approaches. Finally, LB also introduce a real function $u$ in their parametrization, which combines in a single expression the non-principal lattice functions $\beta_{xII}$ and $\beta_{yI}$, the non-principal phase advances $\phi_{xII}$ and $\phi_{yI} $, and the coupling parameters $R_1$ and $R_2$ that represent the coupling due to a longitudinal field. This real function quantifies the lattice coupling within a single parameter. If there is no coupling in the lattice, $u$ is zero.

\begin{align}
    1-u &= \beta_{xI}\phi_{xI}' +\frac{R_1}{2}\sqrt{\beta_{xI}\beta_{yI}}\sin{(\nu_1)}\label{expressions_u_3}\\
        &= \beta_{yII}\phi_{yII}' -\frac{R_2}{2}\sqrt{\beta_{xII}\beta_{yII}}\sin{(\nu_2)}\label{expressions_u_4}\text{, }\\
    u &= \beta_{yI}\phi_{yI}' +\frac{R_2}{2}\sqrt{\beta_{xI}\beta_{yI}}\sin{(\nu_1)}\label{expressions_u_1}\\
        &= \beta_{xII}\phi_{xII}' -\frac{R_1}{2}\sqrt{\beta_{xII}\beta_{yII}}\sin{(\nu_2)}\text{.}
    \label{expressions_u_2}
\end{align}
We can see that $u$ also highlights the normalization condition that appears explicitly in the work of WR (Ref.~\cite{ Willeke}) \textit{via} Eq.~\eqref{normalisation_condition}:
\begin{eqnarray*}
  1 &=& (1-u) + u \\
    &=& \beta_{xI}\phi_{xI}' + \beta_{yI}\phi_{yI}' + \frac{1}{2}(R_1 + R_2) \sqrt{\beta_{xI}\beta_{yI}}\sin{(\nu_1)} \\
    &=& \beta_{xII}\phi_{xII}' + \beta_{yII}\phi_{yII}' - \frac{1}{2}(R_1 + R_2) \sqrt{\beta_{xII}\beta_{yII}}\sin{(\nu_2)}\text{.}
\end{eqnarray*}
The expressions above also show that if the non-principal components of the eigenvector linked to a mode increase, the parameter $u$ increases. This parameter thus represents the relative importance of the $x$ and $y$ components of an eigenvector associated with a mode. In addition, the parameter $u$ can also be linked to the surfaces of the two ellipses due to a mode in the 2 phase planes ($x-x'$, $y-y'$). The relative importance of the $x$ and $y$ components of an eigenvector will also be reflected by the relative size of these ellipses. The higher the coupling, the more $u$ will increase, and the more the surface of these two ellipses will approach each other. The lower the coupling, the more the non-principal ellipse in a plane will shrink until disappearing in the limit of uncoupled motion corresponding to $u = 0$. Section~\ref{Lien_ET_MR} shows that the $u$ parameter is also related to the rotation angle of the ET parametrization, which confirms its interpretation as coupling strength.

By comparing the parametrization of LB to that of Wolski, we obtain:
\begin{eqnarray*}
    \zeta_x &=& \sqrt{\beta_{1y}} e^{-i\nu_1}\text{, }\\
    \zeta_y &=& \sqrt{\beta_{2x}} e^{-i\nu_2}\text{, }\\
    \overset{\sim}{\zeta_x} &=& -\frac{\alpha_{2x}}{\sqrt{\beta_{2x}}}e^{i\nu_2} - i\frac{u}{\sqrt{\beta_{2x}}}e^{i\nu_2}\text{,}\\
    \overset{\sim}{\zeta_y} &=&  -\frac{\alpha_{1y}}{\sqrt{\beta_{1y}}}e^{i\nu_1} - i\frac{u}{\sqrt{\beta_{1y}}}e^{i\nu_1}\text{.}
\end{eqnarray*}
We see that the parameters $\zeta_x$ and $\zeta_y$ are coupling parameters and are zero when there is no local coupling. The motion in the $x$-direction can be seen as the superposition of two quasi-harmonic motions. The first motion corresponds to the projection of the oscillation mode I in the $x$-direction and is characterized by the main lattice functions $\beta_x$ and $\alpha_x$, while the second motion corresponds to the projection of the oscillation mode II in the $x$-direction and is characterized by $\zeta_y$ and $\overset{\sim}{\zeta_x}$. The ``main'' motion can be seen as a quasi-harmonic oscillation whose amplitude is characterized by $\beta_x$ and which is not out of phase with respect to the oscillation in the eigen direction I. The ``non-principal'' motion is a quasi-harmonic oscillation whose amplitude is characterized by $|\zeta_y| = \sqrt{\beta_{2x}}$ and whose phase shift compared to the oscillation eigenmode II is $-\nu_2$. The parameters appearing in the different parametrizations are summarized in Table \ref{compa_params}.

\begin{table}[ht!]
\centering
\begin{tabular}{@{}lllll@{}}
  \toprule
  \multicolumn{5}{c}{Principal lattice functions}  \\
    Willeke \& Ripken  &  & Lebedev \& Bogacz &  & Wolski  \\
  \midrule
  $\beta_{xI}$ &  & $\beta_{1x}$ &  & $\beta_{x}$ \\
  $\beta_{yII}$ &  & $\beta_{2y}$ &  & $\beta_{y}$ \\
  $\alpha_{xI} + \frac{R_1}{2}\sqrt{\beta_{xI}\beta_{yI}} \cos{(\nu_1)}$ &  & $\alpha_{1x}$ &  & $\alpha_{x}$ \\
  $\alpha_{yII} -  \frac{R_2}{2}\sqrt{\beta_{xII}\beta_{yII}} \cos{(\nu_2)}$ &  & $\alpha_{2y}$ &  & $\alpha_{y}$ \\
  $\phi_{xI}$ &  & $\mu_1$ &  & $\mu_I$ \\
  $\phi_{yII}$ &  & $\mu_2$ &  & $\mu_{II}$ \\
  \toprule
  \multicolumn{5}{c}{Non-principal lattice functions}\\
  Willeke \& Ripken &  & Lebedev \& Bogacz &  & Wolski \\
  \midrule
  $\beta_{xII}$ &  & $\beta_{2x}$ &  & $|\zeta_y|^2$ \\
  $\beta_{yI}$ &  & $\beta_{1y}$ &  & $|\zeta_x|^2$ \\
  $\alpha_{xII} + \frac{R_1}{2}\sqrt{\beta_{xII}\beta_{yII}} \cos{(\nu_2)}$ &  & $\alpha_{2x}$ &  & -$Re(\zeta_y \overset{\sim}{\zeta_x})$ \\
  $\alpha_{yI} - \frac{R_2}{2}\sqrt{\beta_{xI}\beta_{yI}} \cos{(\nu_1)}$ &  & $\alpha_{1y}$ &  & -$Re(\zeta_x \overset{\sim}{\zeta_y})$\\
   $\phi_{xII}$ &  & $\mu_1 -\nu_1$ &  &  $\mu_I + ph(\zeta_x)$ \\
   $\phi_{yI}$ &  & $\mu_2 -\nu_2$ &  & $\mu_{II} + ph(\zeta_y)$ \\
    \bottomrule
    \end{tabular}
    \caption{Comparison of the parameters appearing in \cite{Willeke}, \cite{Lebedev} and \cite{wolskiNormalFormAnalysis2004}.}
    \label{compa_params}
\end{table}

The main differences between the parametrization of WR, that of LB, and that of Wolski are as follows. The parameters given by WR are each associated with an oscillation mode and a transverse direction. Each of the oscillations (principal and non-principal) in the transverse directions can be described by a set of distinct parameters. These parameters also characterize the two ellipses appearing in the phase space associated with a transverse direction. The normalization condition reflects the link between the principal oscillation and the non-principal oscillation due to one eigenmode in the two transverse directions. LB slightly reduce the number of parameters associated with a mode and a transverse direction. Instead, they introduce real functions which highlight the differences between the principal and non-principal oscillations linked to an oscillation eigenmode. In that respect, the parameters $\nu$ give the phase shift between these two oscillations, while the parameter $u$ appears in the normalization condition. This condition can be written as $u + (1-u) = 1 $ where $(1-u)$ and $u$ are related to the areas of the principal and non-principal ellipses associated with an oscillation mode. LB, and WR characterize a non-principal oscillation by giving several parameters (amplitudes and phases/phase shifts), while Wolski describes this non-principal oscillation as a single complex parameter that combines amplitude and phase shift. The coupling due to the longitudinal magnetic field in the lattice does not appear in the optical functions of the WR parametrization, while this coupling is directly taken into account in the lattice functions of LB and those of Wolski.

\subsubsection{Interpretation, advantages and disadvantages of the MR parametrization}
The parameters set of the MR parametrization generally includes 6 main optical functions as well as non-principal parameters that reflect coupling. The principal optical functions are 2 $\beta$-functions, 2 $\alpha$-functions, and 2 phase advances $\mu$, which describe the oscillation of a mode in its ``principal'' transverse direction. The non-principal optical parameters describe the non-principal oscillation due to a mode. By ``non-principal oscillation'', we denote the quasi-harmonic oscillation in the transverse direction that is not mainly associated with the mode eigendirection in the limit of weak coupling. This non-principal oscillation is described in a slightly different way depending on the exact choice of parameters. The first way is to describe it independently of the principal oscillation. The non-principal oscillation will then have its own parameters ($\alpha$-function, $\beta$-function, and a phase advance) \cite{Willeke}. The second way is to describe it in relation to the principal oscillation. Its amplitude will then be described again by the $\alpha$ and $\beta$ functions, but its phase will be characterized by its phase shift with respect to the main oscillation. Describing the non-principal oscillation in this manner, some authors explicitly give the amplitudes and phase shifts \cite{Lebedev}, while others describe the quasi-harmonic oscillation by a phasor \textemdash a complex number whose modulus describes the oscillation amplitude and whose argument represents the phase shift \cite{wolskiNormalFormAnalysis2004}. These non-principal optical functions are characteristics of the coupling. If there is no coupling, the non-principal $\beta$ and $\alpha$ functions (or equivalently the complex $\zeta$ functions) are zero: $\beta_{1y} = \beta_{2x} = \alpha_{1y} = \alpha_{2x} = \zeta_{y} = \zeta_{x} = u = 0$.

The advantage of using the MR parametrization instead of the ET parametrization is that the interpretation of the MR parameters is similar to the interpretation of the Twiss parameters in the Courant-Snyder theory. Indeed, the lattice parameters of this parametrization are related to the physical directions, and it is possible to associate them with the amplitudes of transverse betatron oscillations and to physical beam parameters that can be measured in the laboratory axes. The $\beta$-functions (and the modulus of $\zeta$) are positive and finite functions (unlike the $\beta$-functions of the ET parametrization) and are related to the horizontal and vertical beam sizes. The motion in each of the transverse directions will be characterized by a sum of two motions due to the two oscillation eigenmodes. The maximum beam sizes in each direction are given by the sum of the mode invariants weighted by $\beta$-functions. The $\beta$-functions of the MR parametrization thus allow to easily generalize the envelope expression of the uncoupled motion. The lattice functions give clear information on the focusing properties of the lattice: looking at their evolution, we have information on the amplitude of the oscillations in the transverse plane at any point of the lattice. The $\alpha$-functions also have the same meaning as in Courant-Snyder's theory if there is no longitudinal field that couples motion. Otherwise, the $\alpha$-functions of the WR parametrization will remain identical, while the $\alpha$ parameters of the other parametrizations will have an additional term that takes into account this coupling due to a longitudinal field. The MR parametrization allows computing the envelope parameters explicitly: it is possible to calculate the elements of the correlation matrix with the optical functions of this parametrization as shown in Table \ref{compa_corr_matrix}.

\begin{table}[hbt!]
\centering
\begin{tabular}{@{}lllll@{}} \toprule
       Elements & & Lebedev \& Bogacz \cite{Lebedev} & & Wolski \cite{wolskiNormalFormAnalysis2004} \\
  \midrule
  $<x^2>$ &  & $\beta_{1x} \varepsilon_I + \beta_{2x} \varepsilon_{II}$ &  & $\beta_{1x} \varepsilon_I + |\zeta_y|^2 \varepsilon_{II}$ \\
  $<y^2>$ &  & $\beta_{1y} \varepsilon_I + \beta_{2y} \varepsilon_{II}$ &  & $|\zeta_x|^2 \varepsilon_I + \beta_{2y} \varepsilon_{II}$ \\
  $<xy>$ &  & $\sqrt{\beta_{1x}\beta_{1y}} cos(\nu_1) \varepsilon_I $ &  & $\sqrt{\beta_{1x}} Re(\zeta_x) \varepsilon_I $ \\
  &  & $+ \sqrt{\beta_{2x}\beta_{2y}} cos(\nu_2) \varepsilon_{II}$ &  & $+ \sqrt{\beta_y} Re(\zeta_y) \varepsilon_{II}$\\
  $<x p_x> $ &  & $-\alpha_{1x} \varepsilon_I -\alpha_{2x} \varepsilon_{II}$ &  & $- \alpha_{1x} \varepsilon_I + Re(\zeta_y \overset{\sim}{\zeta_x})\varepsilon_{II}$ \\
  $<yp_y>$ &  & $-\alpha_{1y}\varepsilon_I -\alpha_{2y} \varepsilon_{II}$ &  &  $Re(\zeta_x \overset{\sim}{\zeta_y})\varepsilon_I -\alpha_{2y} \varepsilon_{II}$\\
      \bottomrule
    \end{tabular}
    \caption{Expression of the correlation matrix elements with the parameters appearing in \cite{Lebedev} and \cite{wolskiNormalFormAnalysis2004}}.
    \label{compa_corr_matrix}
\end{table}

The expressions in Table \ref{compa_corr_matrix} give the horizontal and vertical beam sizes, which are always positive because the $\beta$-functions are always positive, and the beam tilt, which represents the orientation angle of the ellipse formed by the projection of the 4D ellipsoid in the plane $x-y$. A brief discussion on the 4D phase space ellipsoid can be found in Appendix \ref{appendix_ellipsoid}, along with the link between the correlation matrix $\Sigma$ and the bilinear form that describes the ellipsoid surface. It should be noted that whatever the parametrization in the MR category, it will always be possible to link the MR parameters to the physical parameters of the beam. Moreover, it is also possible to measure the parameters of the MR parametrization using the $\Sigma$ matrix. In Ref.~\cite{wolskiSimpleWayCharacterize2004}, Wolski presents an experimental method to  obtain the phase advances and ratios of lattice functions ($\beta$-functions and $\zeta$ modulus) from BPM measurements. Finally, the $\beta$ and $\alpha$ functions are related to the physical directions and are calculated from the eigenvectors associated with a tune. With the oscillation eigenmodes also associated with these tunes, we will no longer have problems with mode identification. The MR parametrization allows univocally determining the generalized Twiss parameters from the transfer matrix eigenvectors.

\section{Interpretation and clarification of the relationships between parametrization types\label{Lien_ET_MR}}
The ET parametrizations directly express the linear invariants in terms of the Twiss parameters in the decoupled space, and the MR parametrization provides direct expressions for the $\Sigma$ matrix in terms of the generalized Twiss parameters. Expressing the linear invariants with the MR parametrizations or expressing the $\Sigma$ matrix with the ET parametrization proves difficult. This section details the links between the parametrizations belonging to the ET or MR categories.

Instead of describing the motion in the decoupled axes, the MR parametrization directly parametrizes the principal and non-principal oscillations. These oscillations originate from the two oscillation eigenmodes and form the motion in one of the physical directions. The generalized Twiss parameters ($\alpha_{1x}$, $\beta_{1x}$, $\alpha_{2x}$, $\beta_{2x}$, $\alpha_{1y}$, $\beta_{1y}$, $\alpha_{2y}$, $\beta_{2y}$, $\mu_1$, $\mu_2$) do not allow to have an elegant expression for the linear invariants but are related to measurable parameters of the beam and make it possible to calculate the beam horizontal and vertical sizes. With the coupled and decoupled spaces being linked to each other by the symplectic rotation matrix $\widetilde{\mathbf{R}}$, one can find a relation between the parameters involved in the two types of parametrization, as first highlighted in Ref.~\cite{Lebedev}.

Section~\ref{ET_parametrization} shows that it is possible to go from the coupled transfer matrix $\mathbf{\hat{M}}$ to the decoupled transfer matrix $\mathbf{\hat{P}}$ using the symplectic rotation matrix $\widetilde{\mathbf{R}}$:
\begin{equation*}
\mathbf{\hat{P}}(s) = \widetilde{\mathbf{R}}^{-1} \mathbf{\hat{M}}(s) \widetilde{\mathbf{R}}\text{.}
\end{equation*}
It is also possible to express this transfer matrix $\mathbf{\hat{P}}$ as the product of a rotation matrix $\mathbf{R}(\mu_1, \mu_2)$ and the decoupled space normalization matrix $\mathbf{T}$, which depends on the lattice parameters of the ET parametrization:
\begin{equation}
    \mathbf{\hat{P}} = \mathbf{T} \mathbf{R}(\mu_1, \mu_2) \mathbf{T}^{-1}\text{, }
\end{equation}
\begin{align}
    \mathbf{T} = \begin{pmatrix}
    \sqrt{\beta_1(s)} & 0 & 0 & 0\\
    \frac{- \alpha_1(s)}{\sqrt{\beta_1(s)}} & \frac{1}{\sqrt{\beta_1(s)}} & 0 & 0 \\
    0 & 0 &\sqrt{\beta_2(s)} & 0 \\
    0 & 0 &\frac{- \alpha_2(s)}{\sqrt{\beta_2(s)}} & \frac{1}{\sqrt{\beta_2(s)}}
    \end{pmatrix}\text{.}
\end{align}\\

In addition, Sec.~\ref{MR_parametrization} shows that it is possible to transform the coupled transfer matrix into its normal form using the normalization matrix $\mathbf{N}$:
\begin{equation*}
    \mathbf{N^{-1} \hat{M} N} = \mathbf{R}(\mu_1, \mu_2)\text{.}
\end{equation*}
Putting all these expressions together:
\begin{eqnarray*}
    \mathbf{\hat{M}} &=& \mathbf{N} \mathbf{R}(\mu_1, \mu_2) \mathbf{N}^{-1} \\
               &=& \widetilde{\mathbf{R}} \mathbf{\hat{P}}(s) \widetilde{\mathbf{R}}^{-1}\\
               &=& \widetilde{\mathbf{R}} \mathbf{T} \mathbf{R}(\mu_1, \mu_2) \mathbf{T}^{-1} \widetilde{\mathbf{R}}^{-1}\text{.}
\end{eqnarray*}
We can therefore rewrite the normalization matrix $\mathbf{N}$, which depends on the MR parameters, as a product of the normalization matrix $\mathbf{T}$ and the symplectic rotation $\widetilde{\mathbf{R}}$, both of which depend on the ET parameters:\\

\begin{widetext}
\begin{equation}
    \mathbf{N} = \widetilde{\mathbf{R}} \mathbf{T}
\end{equation}
\begin{equation}
    \begin{split}
        & \begin{pmatrix}
        \sqrt{\beta_{1x}} & 0  & \sqrt{\beta_{2x}}\cos\nu_2 & -\sqrt{\beta_{2x}}\sin\nu_2 \\
        - \frac{\alpha_{1x}}{\sqrt{\beta_{1x}}}  & \frac{1-u}{\sqrt{\beta_{1x}}} & \frac{u \sin \nu_2 -\alpha_{2x}\cos\nu_2}{\sqrt{\beta_{2x}}} & \frac{u \cos \nu_2 +\alpha_{2x}\sin\nu_2}{\sqrt{\beta_{2x}}}\\
        \sqrt{\beta_{1y}}\cos\nu_1  & - \sqrt{\beta_{1y}}\sin\nu_1 & \sqrt{\beta_{2y}} &  0 \\
        \frac{u sin \nu_1 -\alpha_{1y}\cos\nu_1}{\sqrt{\beta_{1y}}} & \frac{u \cos\nu_1 +\alpha_{1y} \sin\nu_1}{\sqrt{\beta_{1y}}} & - \frac{\alpha_{2y}}{\sqrt{\beta_{2y}}} & \frac{1-u}{\sqrt{\beta_{2y}}}\\
        \end{pmatrix}\\
         = &\begin{pmatrix}
        \sqrt{\beta_1}\cos{\phi} & 0 & (d \sqrt{\beta_2} +\frac{b \alpha_2}{\sqrt{\beta_2}}) \sin{\phi} & -\frac{b}{\sqrt{\beta_2}} \sin{\phi}\\
         -\frac{\alpha_1}{\sqrt{\beta_1}}\cos{\phi} & \frac{\cos{\phi}}{\sqrt{\beta_1}}& (-c\sqrt{\beta_2} - \frac{a \alpha_2}{\sqrt{\beta_2}}) \sin{\phi} & \frac{a \sin{\phi}}{\sqrt{\beta_2}} \\
         (-a \sqrt{\beta_1} +\frac{b \alpha_1}{\sqrt{\beta_1}}) \sin{\phi} & -\frac{b}{\sqrt{\beta_1}} \sin{\phi} & \sqrt{\beta_2}\cos{\phi} & 0  \\
         (-c\sqrt{\beta_1} + \frac{d \alpha_1}{\sqrt{\beta_1}}) \sin{\phi} & -\frac{d \sin{\phi}}{\sqrt{\beta_1}}  &  -\frac{\alpha_2}{\sqrt{\beta_2}}\cos{\phi} & \frac{\cos{\phi}}{\sqrt{\beta_2}}
        \end{pmatrix}\text{.}
    \end{split}
\end{equation}
\end{widetext}
By comparing the blocks on the diagonal of these two matrices, we obtain directly:
\begin{align}
    1-u &=  \cos^2{\phi} \quad & \Rightarrow & \sin{\phi} = \pm \sqrt{u} \label{u_sin} \text{, }
\end{align}
\begin{align}
    \beta_{1x} &= \beta_1\cos^2{\phi} \quad & \Rightarrow & \beta_{1} = \frac{\beta_{1x}}{1-u} \text{, }\\
    \alpha_{1x} &= \alpha_1 \cos^2{\phi}\quad & \Rightarrow & \alpha_{1} = \frac{\alpha_{1x}}{1-u}\text{, }
\end{align}
\begin{align}
    \beta_{2y} &= \beta_2\cos^2{\phi}\quad & \Rightarrow & \beta_{2} = \frac{\beta_{2y}}{1-u}\text{, }\\
    \alpha_{2y} &= \alpha_2 \cos^2{\phi}\quad & \Rightarrow & \alpha_{2} = \frac{\alpha_{2y}}{1-u}\text{.}
\end{align}
We thus find the relations presented in Ref.~\cite{Lebedev} enabling us to link the different parametrizations' lattice functions. The parameter $u$, which reflects the coupling in the parametrization of LB, is related to the angle of rotation in the ET parametrization. We also see that if the parameter $u$ is negative, the rotation angle $\phi$ of the ET parametrization is complex. It is equivalent to the situation where $|\mathbf{B}+\mathbf{\bar{C}}| < 0$. There is then only one solution for the ET parametrization: when the parameter $u$ changes sign, a mode flip is forced. In addition, it is also possible to find the link between the decoupling matrix parameters (the four elements of the matrix $\mathcal{D}$ in the symplectic rotation) and those of LB. The elements $\mathbf{N}_{32}$ and $\mathbf{N}_{42}$ allow to find the parameters $b$ and $d$ of the matrix $\mathcal{D}$, while the elements $\mathbf{N}_{31}$ and $\mathbf{N}_{41}$ allow us to find the parameters $a$ and $ c $ of this same matrix:
\begin{widetext}
\begin{align}
    b \tan{\phi} &= \sqrt{\beta_{1x}\beta_{1y}} \frac{\sin{\nu_1}}{1-u} \text{, }\\
    d \tan{\phi} &= -\sqrt{\frac{\beta_{1x}}{\beta_{1y}}} \frac{u \cos{\nu_1} + \alpha_{1y}\sin{\nu_1}}{1-u} \text{, }\\
    a \tan{\phi} &=  \frac{\alpha_{1x}}{\beta_{1x}} b \tan{\phi} - \sqrt{\frac{\beta_{1y}}{\beta_{1x}}}\cos{\nu_1} \\
                 &= \sqrt{\frac{\beta_{1y}}{\beta_{1x}}} \frac{\alpha_{1x} \sin{\nu_1} - (1-u)\cos{\nu_1}}{1-u}\text{, }\\
    c \tan{\phi} &= \frac{\alpha_{1x}}{\beta_{1x}} d \tan{\phi} - \frac{u\sin{\nu_1} - \alpha_{1y}\cos{\nu_1}}{\sqrt{\beta_1y}}\\
                 &= \frac{\cos{\nu_1}[\alpha_{1y} (1-u)- \alpha_{1x} u] - \sin{\nu_1} [\alpha_{1x}\alpha_{1y} +u(1-u)]}{\sqrt{\beta_{1x}\beta_{1y}}(1-u)}\text{.}
\end{align}
\end{widetext}
From the above expressions, we can verify that $\mathcal{D}$ is symplectic, $|\mathcal{D}| = ad - bc = 1$. Moreover, we can also express the parameters of $\mathcal{D}$ in terms of the lattice functions linked to the second oscillation eigenmode. To do this, we should start from the elements of the upper right block of the matrix $\mathbf{N}$. Note also that, as mentioned by LB in Ref.~\cite{Lebedev}, the phase advances $\mu_1$ and $\mu_2$ are the same in the ET and the MR parametrizations. These phase advances are thus linked to the principal oscillation of the eigenmode, which is the oscillation in the direction primarily linked to the considered oscillation eigenmode. Moreover, starting from the definition of the normalization matrix $\mathbf{N}$ expressed with the eigenvectors (Eq.~\eqref{N_vecteurs_propres}) or in terms of the lattice parameters (Eq.~\eqref{param_lebedev}), we note that the knowledge of the coupled transfer matrix eigenvectors directly and uniquely provides the parameters of LB. Nevertheless, it is impossible to uniquely determine the eigenvectors if only the lattice functions are given. We can find four normalization matrices giving the same lattice functions $\alpha$ and $\beta$ so that it is impossible to univocally find the eigenvectors from the knowledge of these lattice functions \cite{Lebedev}.

Regarding the ET parametrization, knowing the eigenvectors of the coupled transfer matrix allows calculating the parameter $u$. However, there are four possible angles of rotation $\phi$ for a given parameter $u$ (see Eq.~\eqref{u_sin}). It is thus impossible to uniquely determine the ET parameters from the knowledge of the eigenvectors. To determine the Twiss parameters from the coupled transfer matrix, it is necessary to choose $\phi$, or equivalently, to choose one of the possible solutions for the decoupling matrix. Depending on the chosen solution, the generalized Twiss parameters of ET will be different. These different solutions for the angles $\phi$ correspond to distinct mode identifications. This problem of mode identification (linked to the mode flips) only appears when there is a rotation of the axes, as in the case of the ET parametrization. Finally, contrary to the MR parametrization, the knowledge of the ET generalized Twiss parameters allows uniquely determining the eigenvectors.

\section{Applications and interpretation on typical lattices \label{implementation}}
The different parametrizations have been implemented in Zgoubidoo \cite{Zgoubidoo}, a Python interface for the ray-tracing code Zgoubi (\cite{Zgoubi1999, Zgoubi, Zgoubi_url}) and validated by comparing with the coupled lattice functions obtained with MAD-X \cite{madx} and PTC. Ray-tracing codes, like Zgoubi, allow particles to be tracked in arbitrary electro-magnetic fields. The ability to perform step-by-step tracking makes Zgoubi a method of choice for (v)FFA studies (\cite{ZGOUBI_FFAG_1, ZGOUBI_FFAG, Zgoubi_Spiral_FFAG}). Another advantage of ray-tracing codes is the possibility to use them and extend the machine model at successive steps of the design process: from the optics study and lattice design to the simulations using complete models with computed or measured field maps, with the possibility to take into account magnetic field imperfections and fringe fields. At each integration step, the particle positions and velocities are calculated, and the field components and derivatives are evaluated. The positions and velocities are obtained with Taylor series truncated at the 5\textsuperscript{th} or 6\textsuperscript{th} order. The magnetic (and electric) fields are obtained either from field maps or from analytical models implemented in Zgoubi. Zgoubi can also compute other relevant quantities, such as transfer matrices or lattice functions. Zgoubidoo is a Python interface for the Zgoubi ray-tracing code. It provides a user-friendly Python interface and is capable of processing the tracking data to extract relevant quantities for beam dynamics studies. Zgoubidoo has also already been used to study beam dynamics in FFAs \cite{IPAC22_FFA_Zgoubidoo}. We chose to use this library to explore vFFAs, and therefore we have numerically implemented the different coupled parametrizations in Zgoubidoo.

The ET and LB parametrizations are implemented in Zgoubidoo. The ET parametrization was implemented using the method presented by Parzen in \cite{Parzen} and allows finding linear invariants. The LB parametrization was chosen among all the parametrizations of the MR category because it provides interesting additional quantities ($u, \nu_1, \nu_2$) together with the lattice functions of the MR parametrization. The LB lattice functions provide the evolution of the beam envelope in the laboratory axes along the lattice. The implementations were first tested on weakly coupled example lattices, then validated with more complicated strongly coupled lattices against MAD-X results \cite{madx}. The methods presented in the previous sections allow, on the one hand, to find the periodic conditions for periodic lattices and, on the other hand, to propagate initial lattice functions in a beamline. The examples presented below allow validating the coupled periodic lattice functions and the propagation of initial lattice functions. In addition, other concepts, such as forced mode flips, local coupling, and interpretation of lattice parameters, are analyzed in detail. Table~\ref{examples_table} summarizes the different example lattices discussed in this section. The lattice function computation method (periodic conditions or propagation of initial lattice functions) and the concepts the example illustrates are also indicated.

\begin{table}[hbt!]
\centering
\begin{tabular}{@{}lllll@{}} \toprule
       Examples & & Computation  & & Illustration \\
  \midrule
   FODO + & & Periodic & & - Global coupling and $u$\\
    \ skew quad  & &          & & - ET/MR functions to \\
                    & &          & & \ \ characterize decoupled \\
                    & &          & & \ \ and coupled phase spaces \\
   FODO + & & Periodic    & & - WR/LB param. (MR  \\
     \ solenoid            & &             & & \ \  category) characterizing\\
                 & &             & &  \ \ geometric and canonical\\
                 & &             & &  \ \ coupled phase spaces\\
   Snake lattice \cite{USPAS2013} & & Propagation  & & - Forced mode flip\\
                         & &               & & - Local coupling and $u$\\
   Spin rotator \cite{USPAS2013} & & Propagation & & - Local coupling and $u$\\

      \bottomrule
    \end{tabular}
    \caption{Examples used to validate the parametrization implementation and illustrate some concepts discussed previously in this work. ``Periodic'' stands for ``Periodic initial conditions'', while ``Propagation'' stands for ``Propagation of initial lattice functions''.}
    \label{examples_table}
\end{table}

\begin{figure}[h]
\begin{center}
    \includegraphics[width=1.0\linewidth]{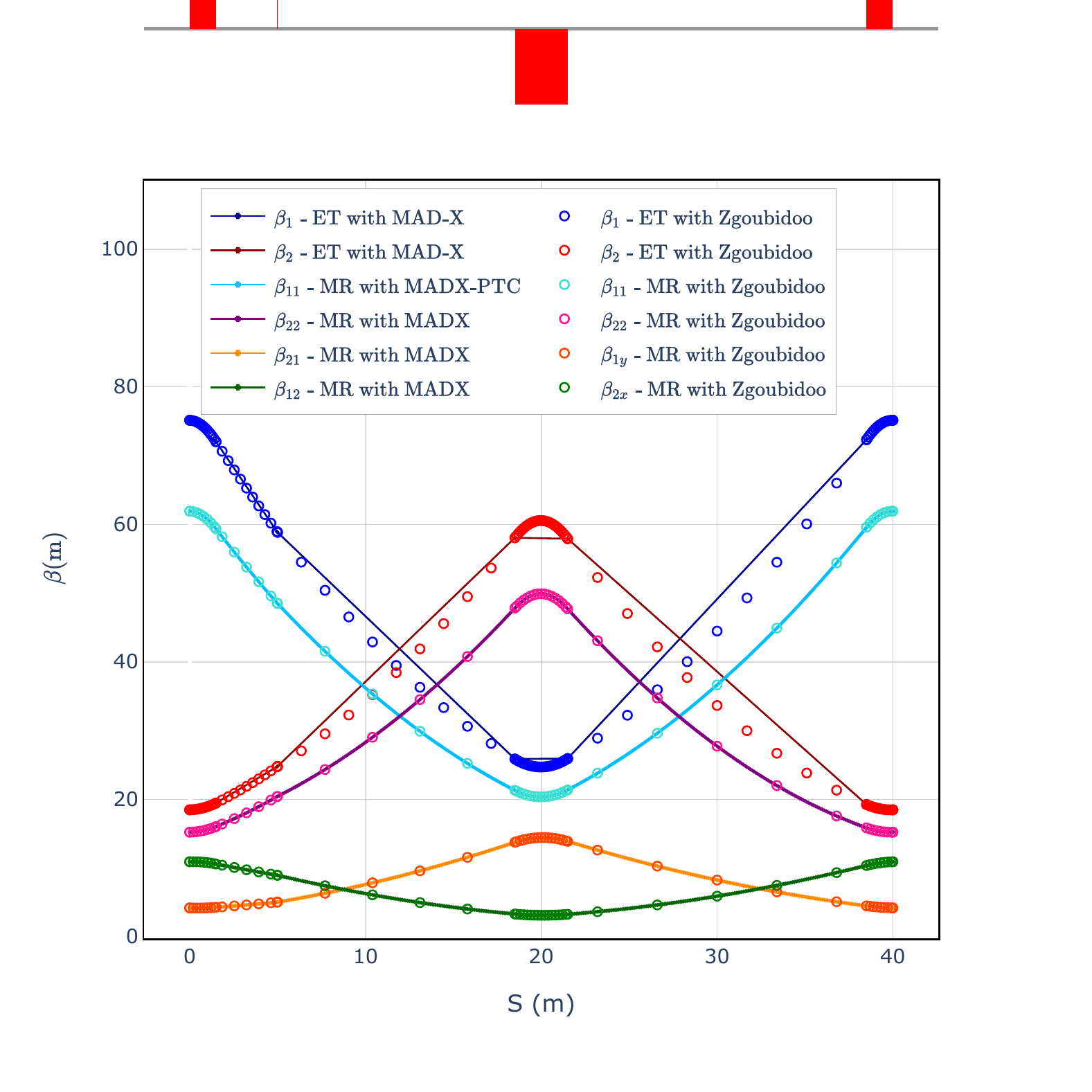}
    \includegraphics[width=1.0\linewidth]{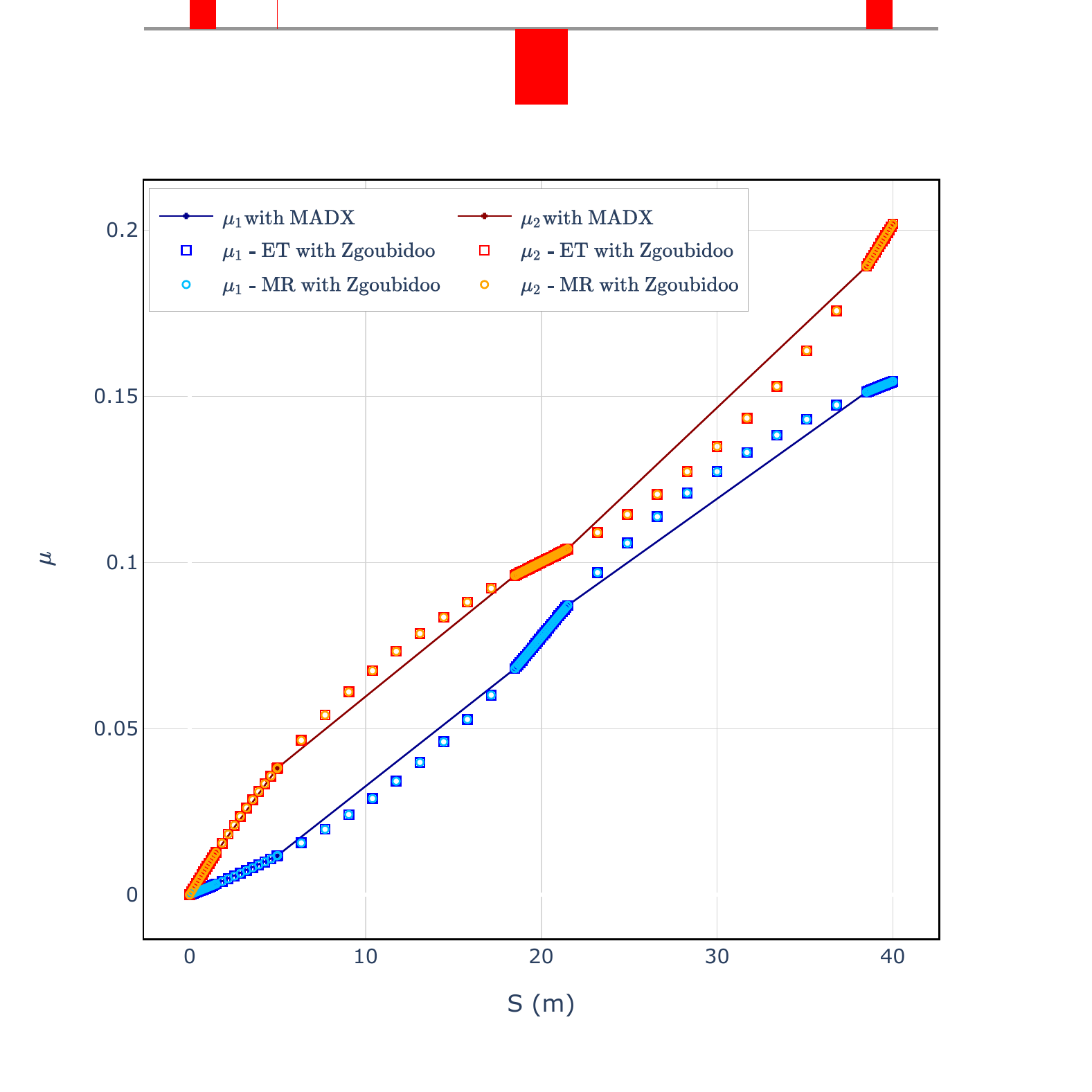}
\caption{Comparison between the coupled lattice functions (ET and MR parametrizations) obtained with Zgoubidoo and those obtained with MAD-X on a lattice consisting of a FODO with a small skew quadrupole.}
\label{FODO_skew}
\end{center}
\end{figure}

\begin{figure}[h]
\begin{center}
    \includegraphics[width=1.0\linewidth]{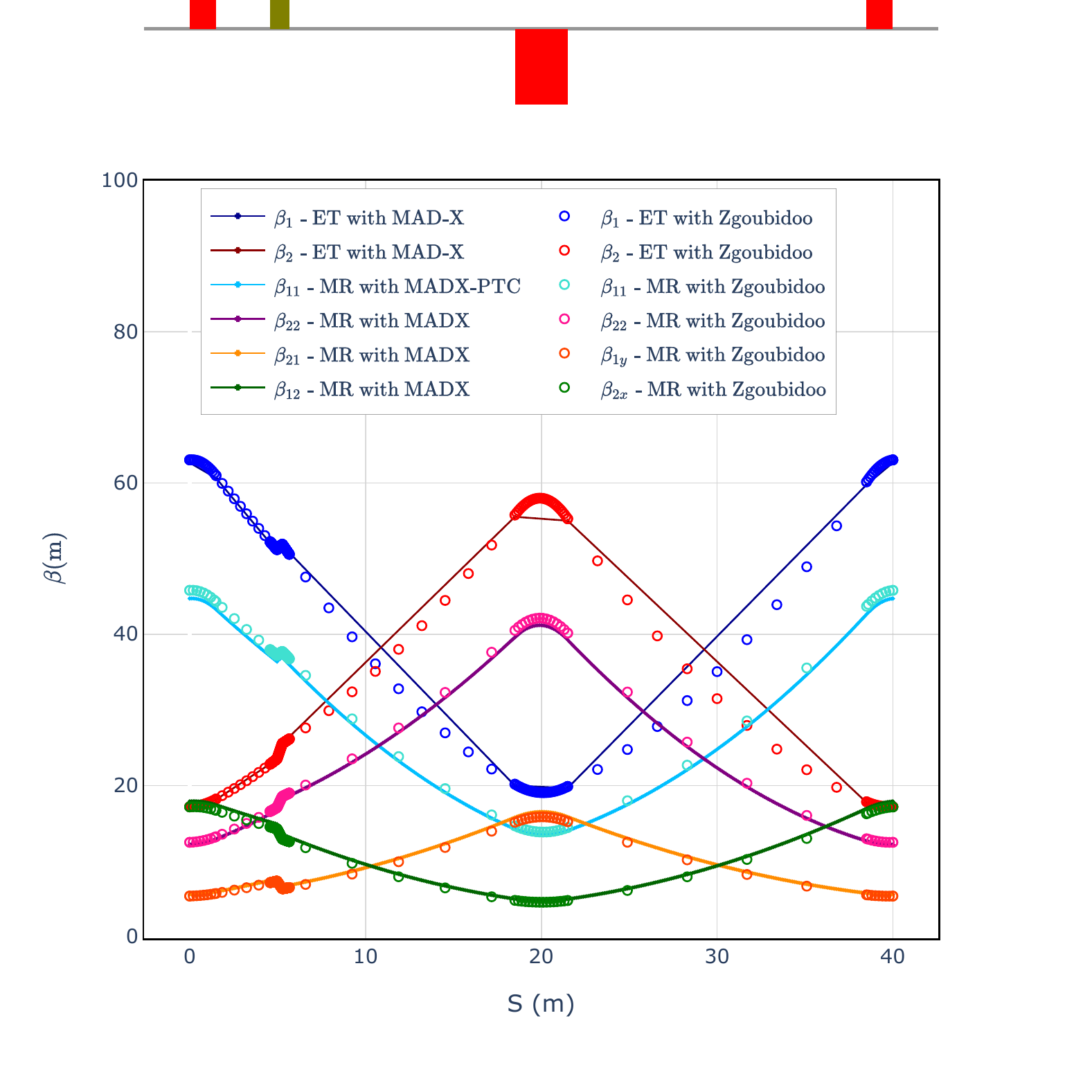}
    \includegraphics[width=1.0\linewidth]{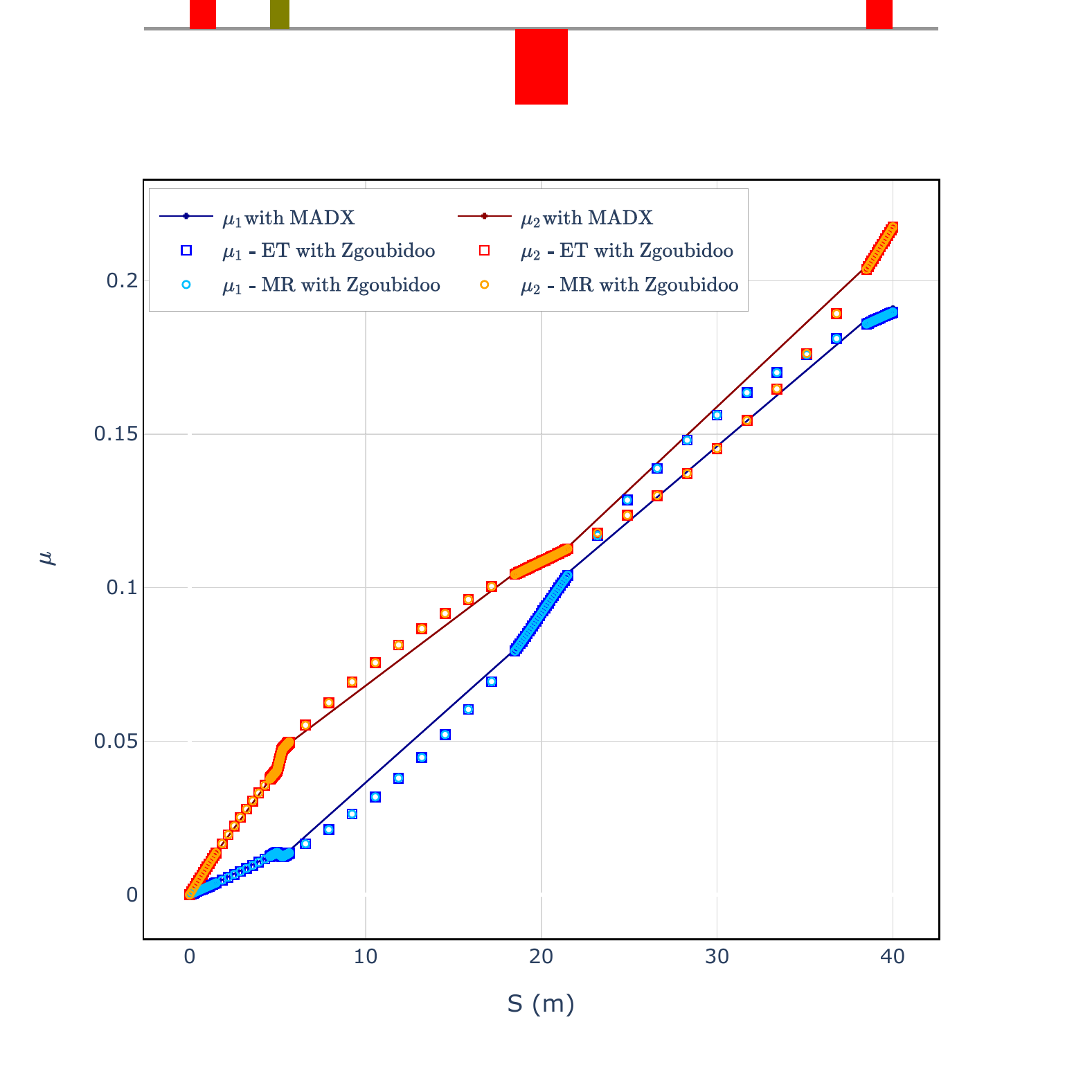}
\caption{Comparison between the coupled lattice functions (ET and MR parametrizations) obtained with Zgoubidoo and those obtained with MAD-X on a lattice consisting of a FODO with a small solenoid.}
\label{FODO_sole}
\end{center}
\end{figure}

The longitudinal and skew quadrupolar field components being the principal sources of coupling, the weakly coupled lattices are FODO lattices featuring short skew quadrupolar or solenoidal insertions. Zgoubidoo is used to calculate the lattice functions (in the ET and LB parametrizations), which are then compared with MAD-X results for the ET parametrization and with PTC results for the MR parametrization. One can observe in Figs.~\ref{FODO_skew} and \ref{FODO_sole} that an excellent agreement is found for the two weakly coupled example FODO lattices. In addition, the phase advances $\mu_1$ and $\mu_2$ obtained from the ET and MR parametrizations are identical, as expected.

It should be noted that the solenoid models in MAD-X or Zgoubi are different, which may introduce differences in the computed lattice functions. MAD-X models an ``ideal'' solenoid. The solenoid model in Zgoubi is more realistic and has fringe fields whose length depends on the (finite) radius of the solenoid. When computing this lattice, the initial conditions were obtained assuming a periodic transfer matrix. The lattice functions represented in Figs.~\ref{FODO_skew} and \ref{FODO_sole} are thus periodic optical functions and reflect a global coupling of the lattice. This global coupling can be understood by analyzing the lattice parameters of LB. First, we can observe that the non-principal lattice functions ($\beta_{1y}$ and $\beta_{2x}$) are non-zero at the beginning of the lattice. They are computed with periodic conditions and therefore take into account the coupling present in the whole lattice and not only the coupling at the location where they are calculated. The coupling is distributed over the entire line.

\begin{figure}
\begin{center}
    \includegraphics[width=1.0\linewidth]{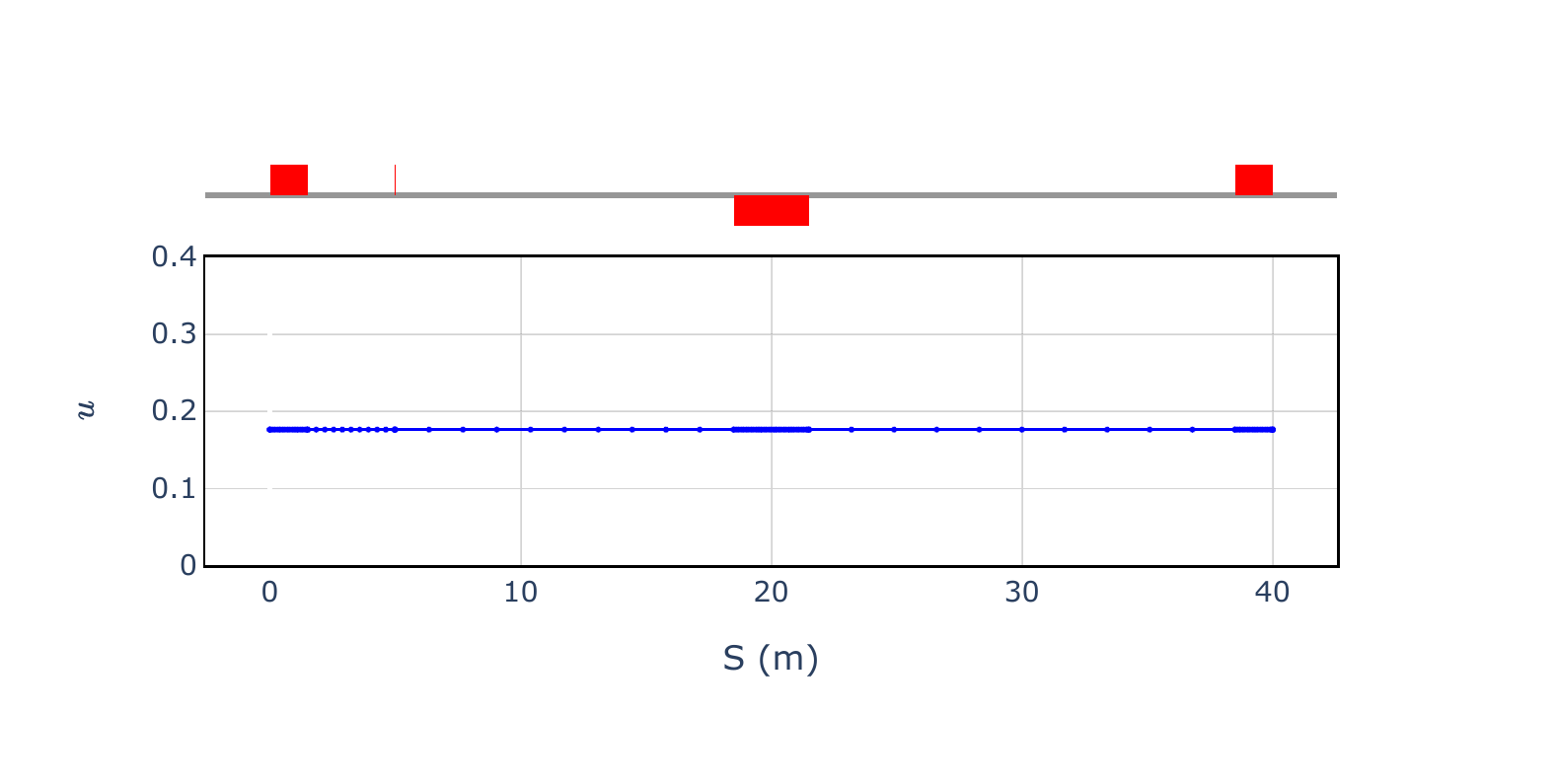}
\caption{Parameter $u$ of the LB parametrization, obtained by assuming periodic initial conditions, on a lattice consisting of a FODO with a small skew quadrupole.}
\label{FODO_skew_u}
\end{center}
\end{figure}

The real parameter $u$ on the full lattice (as shown in Fig.~\ref{FODO_skew_u}) remains constant on the entire line at a value of $0.176$. When calculated with periodic conditions, this parameter reflects an average of the coupling over the whole lattice. It provides insight into the weight of the non-principal lattice functions compared to the principal ones over the complete lattice. For the FODO lattice featuring a short skew quadrupole, in the absence of any longitudinal field, the ratio between the $\beta$-functions for the mode I can be written:

\begin{equation}
    \frac{\beta_{1y}}{\beta_{1x}} = \frac{u \ \mu_1^{'}}{(1-u)(\mu_1 - \nu_1)^{'}}\text{.}
\end{equation}

A fully coupled lattice would have principal functions equal to the non-principal ones and $u=0.5$: it is the case, for example, of a FODO cell in which all the elements are rotated by 45 degrees. When computed with the periodic conditions, the parameter $u$ thus gives a measure of the overall coupling of the lattice. This parameter has a finite value in the elements not introducing coupling and represents the average coupling of the lattice. Nevertheless, it varies in the elements introducing coupling and indicates whether the element couples more or less the motion than the lattice does globally.

\begin{figure}[h!]
\begin{center}
    \includegraphics[width=1.0\linewidth]{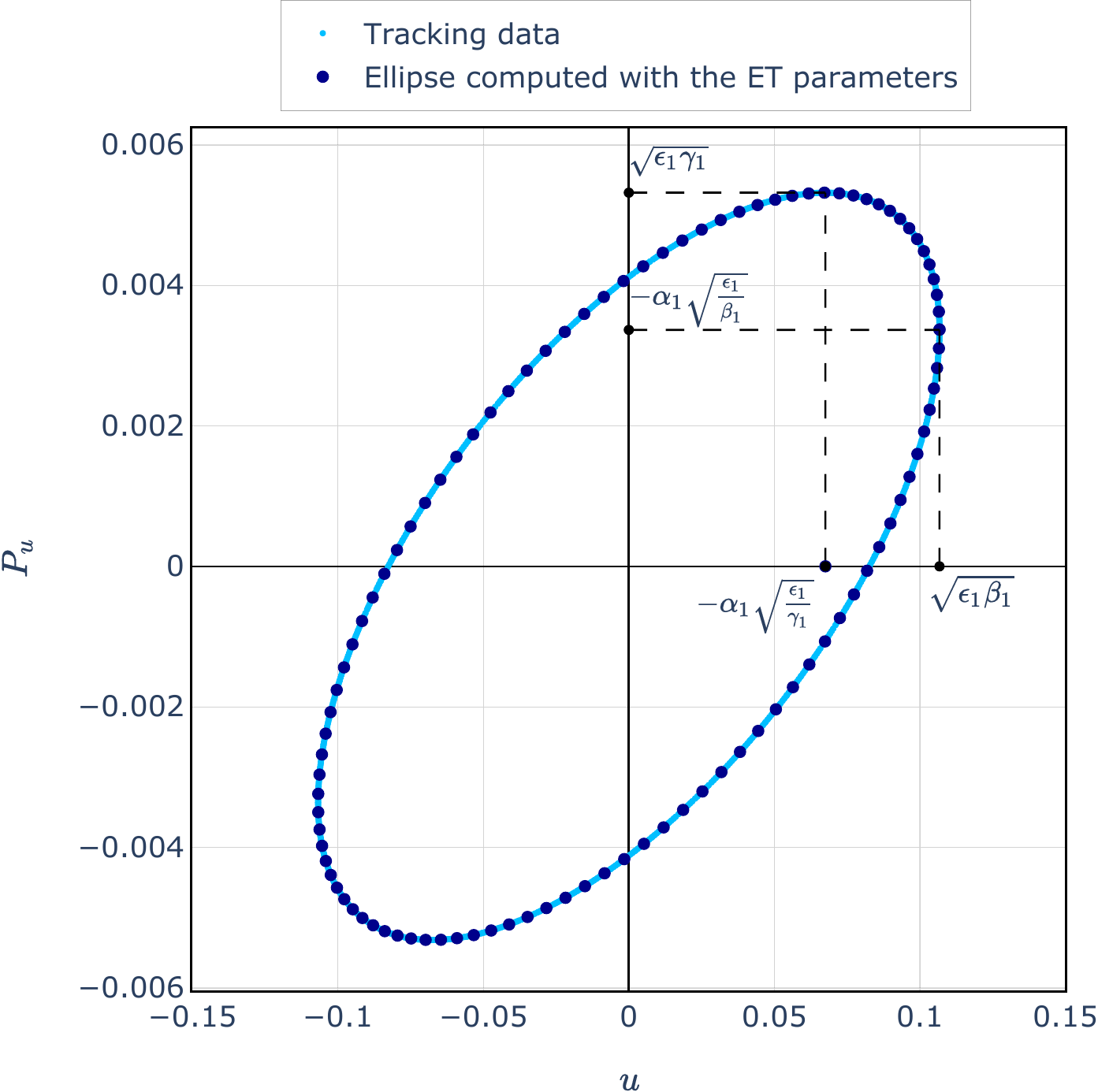}\\
\caption{Decoupled phase space $(u-p_u)$ obtained by tracking a particle 1000 times in a cell composed of a FODO with a skew quadrupole. The ET lattice functions describe the decoupled phase space ellipse whose area is $\pi\epsilon_1$.}
\label{ellipses_decoupled}
\end{center}
\end{figure}

\begin{figure}[h!]
\begin{center}
    \includegraphics[width=1.0\linewidth]{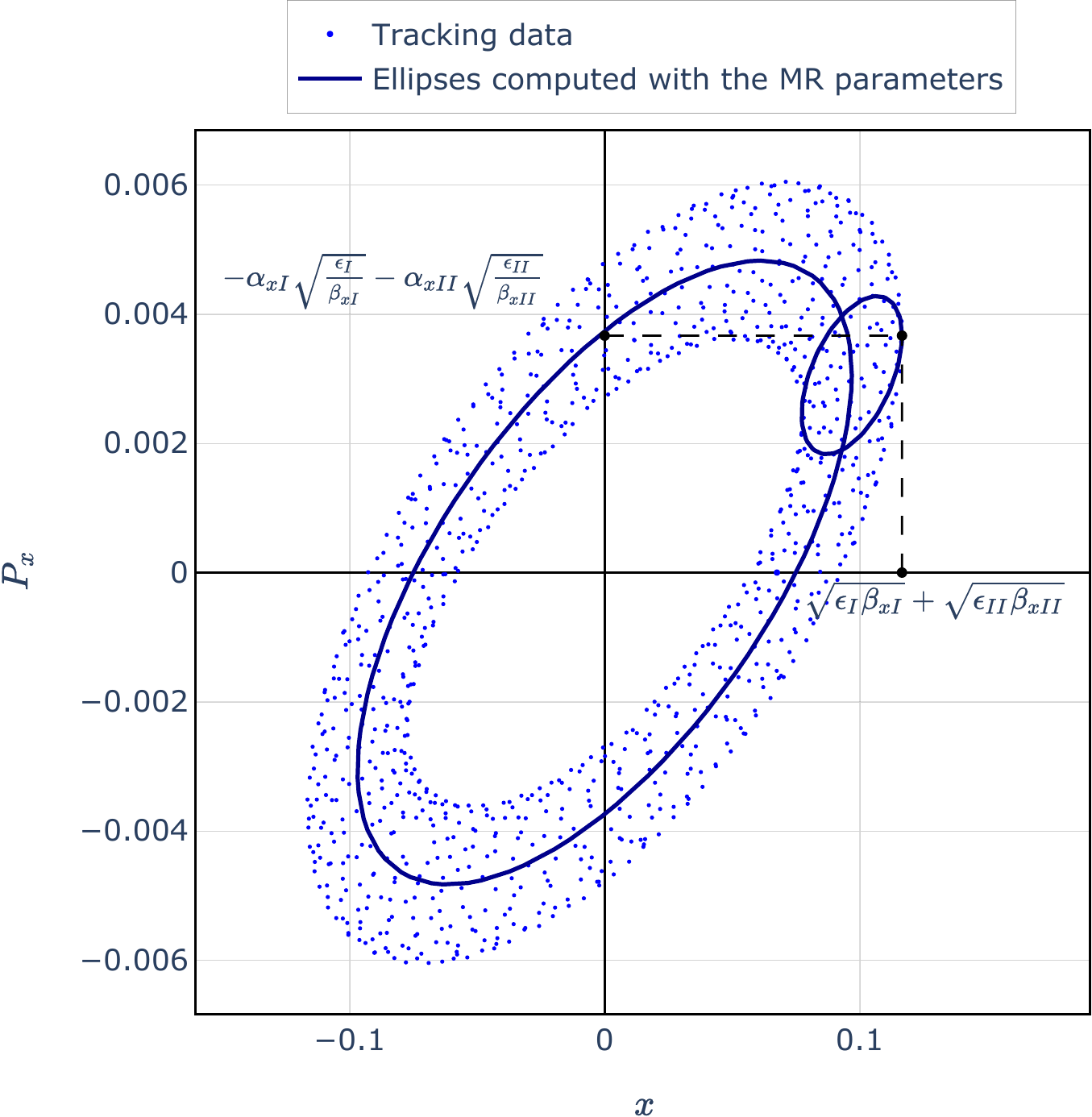}
\caption{Coupled phase spaces $(x-p_x)$ obtained by tracking a particle 1000 times in a cell composed of a FODO with a skew quadrupole. The MR lattice functions characterize the two ellipses appearing in the coupled phase space. These two ellipses are due to the two oscillation eigenmodes. The lattice functions used are those that connect the two eigenmodes to the observed transverse direction ($x$ in this case). The ellipse areas are given by $\pi\epsilon_1(1-u)$ and $\pi\epsilon_2u$.}
\label{ellipses_coupled}
\end{center}
\end{figure}

The parameter $u$ can also be linked to the area of the ellipses in the coupled phase spaces. The decoupled (resp. coupled) phase spaces can be related to the ET parameters (resp. MR parameters). Figures~\ref{ellipses_decoupled} and \ref{ellipses_coupled} show the decoupled phase space $(u - p_u)$ and the coupled phase space $(x - p_x)$, obtained by tracking a particle with an initial horizontal amplitude in the cell composed of the FODO and the additional skew quadrupole. The phase spaces are constructed by sampling the particle coordinates at a point in the lattice (in this case, just after the defocusing quadrupole) and tracking 1000 iterations. We observe that the lattice functions of the ET parametrization allow describing the ellipse in the decoupled phase space, while the lattice functions of the MR parametrization allow describing the two ellipses in the coupled phase space. The area of the ellipse in the phase space $(u-p_u)$ is given by $\pi \epsilon_1$ and is an invariant of the motion, while the areas of the ellipses corresponding to the two oscillation modes projected into the transverse phase plane $(x-p_x)$ can be calculated using the parameters of LB: $\Gamma_{1x} = \pi \epsilon_1 (1-u)$, $\Gamma_{2x } = \pi \epsilon_2 u$. The parameter $u$ gives the relative importance of the two ellipses coming from an oscillation eigenmode in the two transverse phase spaces $(x-p_x)$ and $(y-p_y)$.

When the phase space corresponds to geometric variables ($x-x'$), the chosen MR lattice functions are those of WR rather than those of LB. Indeed, we have seen that the major difference between the $\beta$- and $\alpha$-functions of WR and those of LB comes from taking into account or not the coupling due to the longitudinal magnetic field (see Table \ref {compa_params}). In the case where $R_{1,2} = 0$, $(1-u)$ becomes $\beta_{xI}\phi_{xI}'$ and $u$ becomes $\beta_{xII }\phi_{xII}'$ (see Eqs. \refeq{expressions_u_3} - \refeq{expressions_u_2}), which allows us to find the expressions of the ellipse areas in the coupled phase spaces in geometric variables presented in Ref.~\cite{ Willeke} (see section \ref{generating_vectors_parametrization}, Eqs. \refeq{area_1} and \refeq{area_2}).

\begin{figure}[h!]
\begin{center}
    \includegraphics[width=1.0\linewidth]{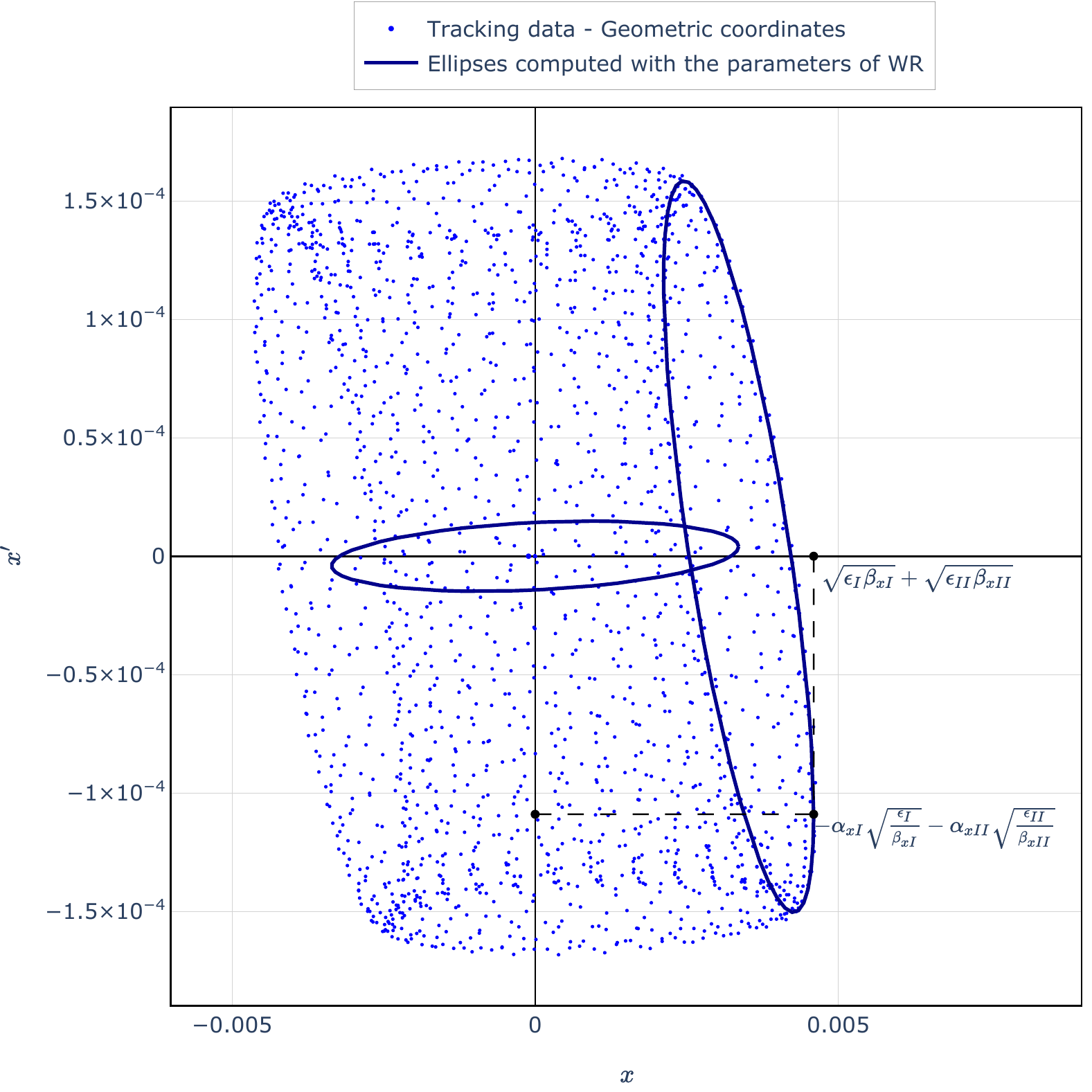}\\
    \includegraphics[width=1.0\linewidth]{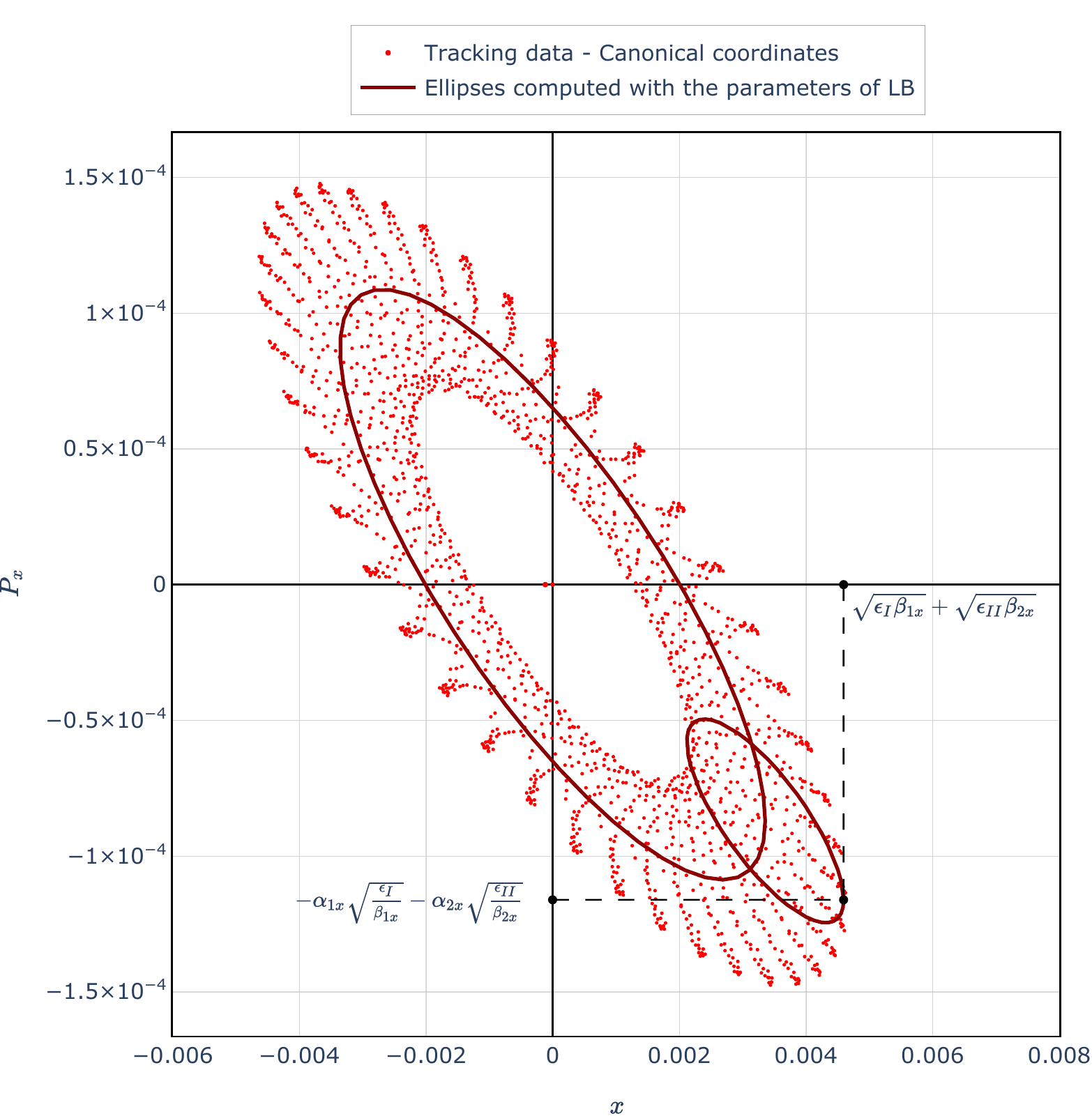}\\
\caption{Coupled phase spaces in geometric (top) or canonical (bottom) coordinates. The lattice functions of WR describe the geometric phase space, while the lattice functions of LB describe the canonical phase space and thus take into account the longitudinal field.}
\label{FODO_sole_canon_vs_geom}
\end{center}
\end{figure}

In the example of the FODO with a skew quadrupole, no element introduces coupling due to the longitudinal magnetic field in the lattice; there is no longitudinal field at the place where the coordinates are sampled. The lattice functions of WR and those of LB are thus equivalent ($R_{1,2} = 0$), and the phase spaces in canonical or geometric coordinates are the same.
However, if we study the example of the FODO with a solenoid, it is possible to sample at a place where the longitudinal field is non-zero. It is notably the case if we place our marker inside the solenoid. We then obtain geometric or canonical phase spaces that are very different (see Fig.~\ref{FODO_sole_canon_vs_geom}). The point cloud obtained in each case is characterized by the superposition of two ellipses. These ellipses are described by the parameters of WR in the geometric case and by the parameters of LB in the canonical case.

\begin{figure}[h]
\begin{center}
    \includegraphics[width=0.9\linewidth]{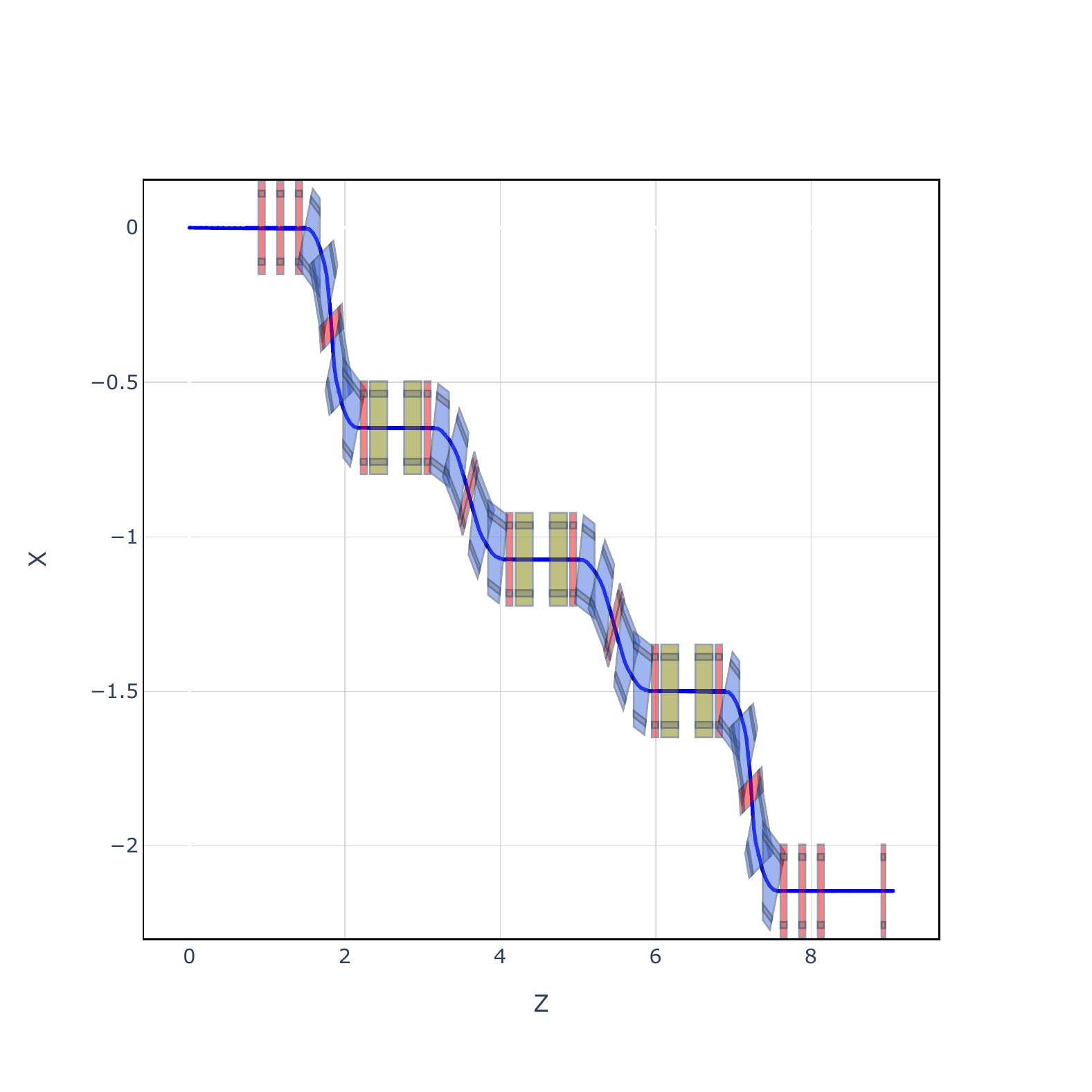}
\caption{Representation of the example Snake lattice. Quadrupoles are depicted in red, dipoles in blue, and solenoids in yellow.}
\label{snake}
\end{center}
\end{figure}

With weakly coupled lattices validated with an excellent agreement, more complex strongly coupled lattices were tackled. As examples we use a ``Snake'' lattice and a ``Spin Rotator'' beamline. The Snake lattice, as shown in Fig.~\ref{snake}, is a muon cooling channel, which contains two parametric resonance cooling cells as well as parts to match the dispersion at the entry and exit of these cells \cite{USPAS2013, Snake_ref}. The Spin Rotator lattice is a beamline of about 46 meters, designed for the ``Figure-8 Electron Collider Ring''. It allows transforming the vertical spin of the electrons in the arcs into a longitudinal spin at interaction points \cite{USPAS2013, SpinRotator_these}. It consists of solenoids, quadrupoles, and dipoles. Solenoids rotate the electron spin but also introduce coupling. The parts containing the solenoids have been designed to compensate for this coupling. Thus, the solenoids are split in two to introduce elements to locally compensate for the solenoid coupling\footnote{More details are provided in Ref.~\cite{SpinRotator_these}.}.

\begin{figure}
\begin{center}
    \includegraphics[width=1.0\linewidth]{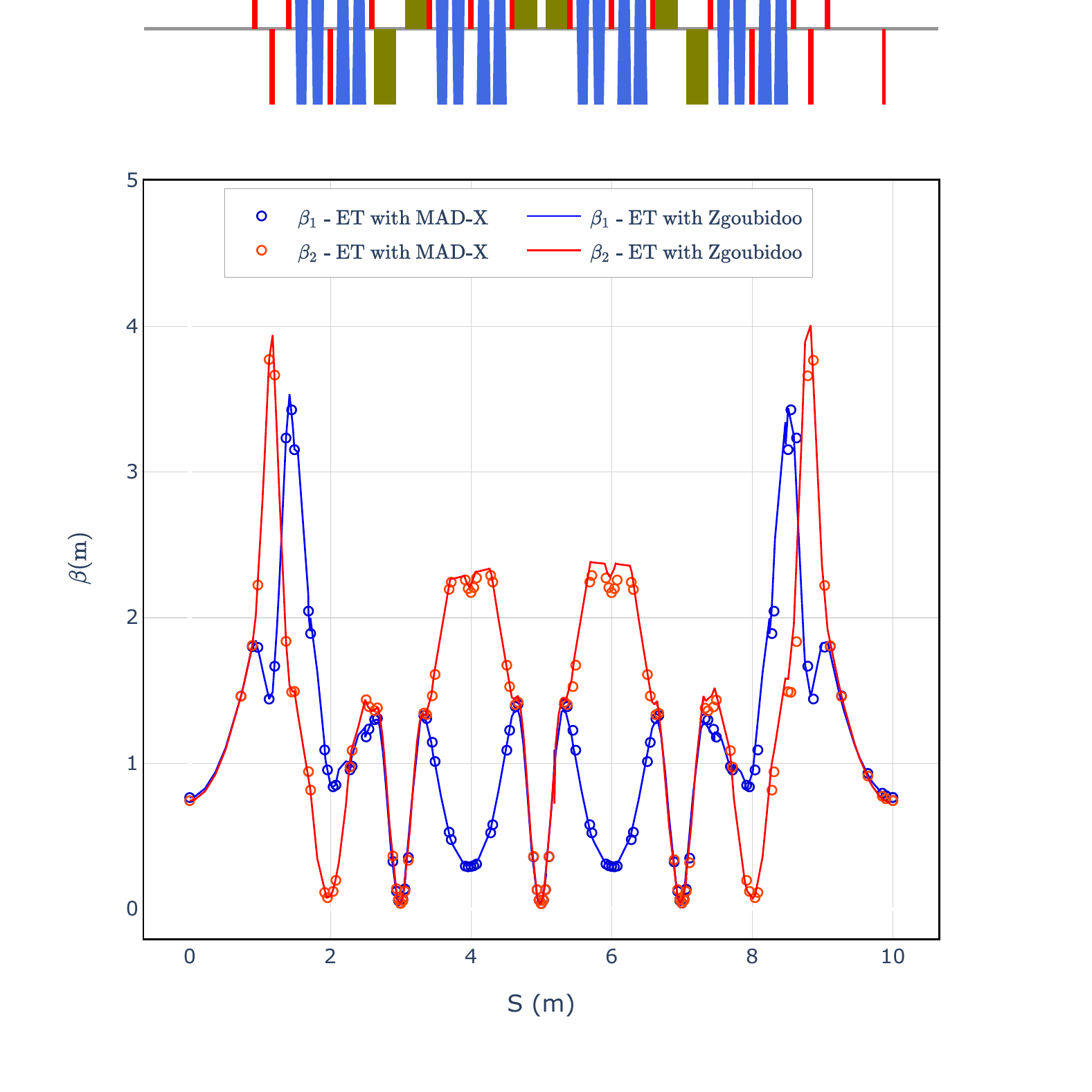}\\
    \includegraphics[width=1.0\linewidth]{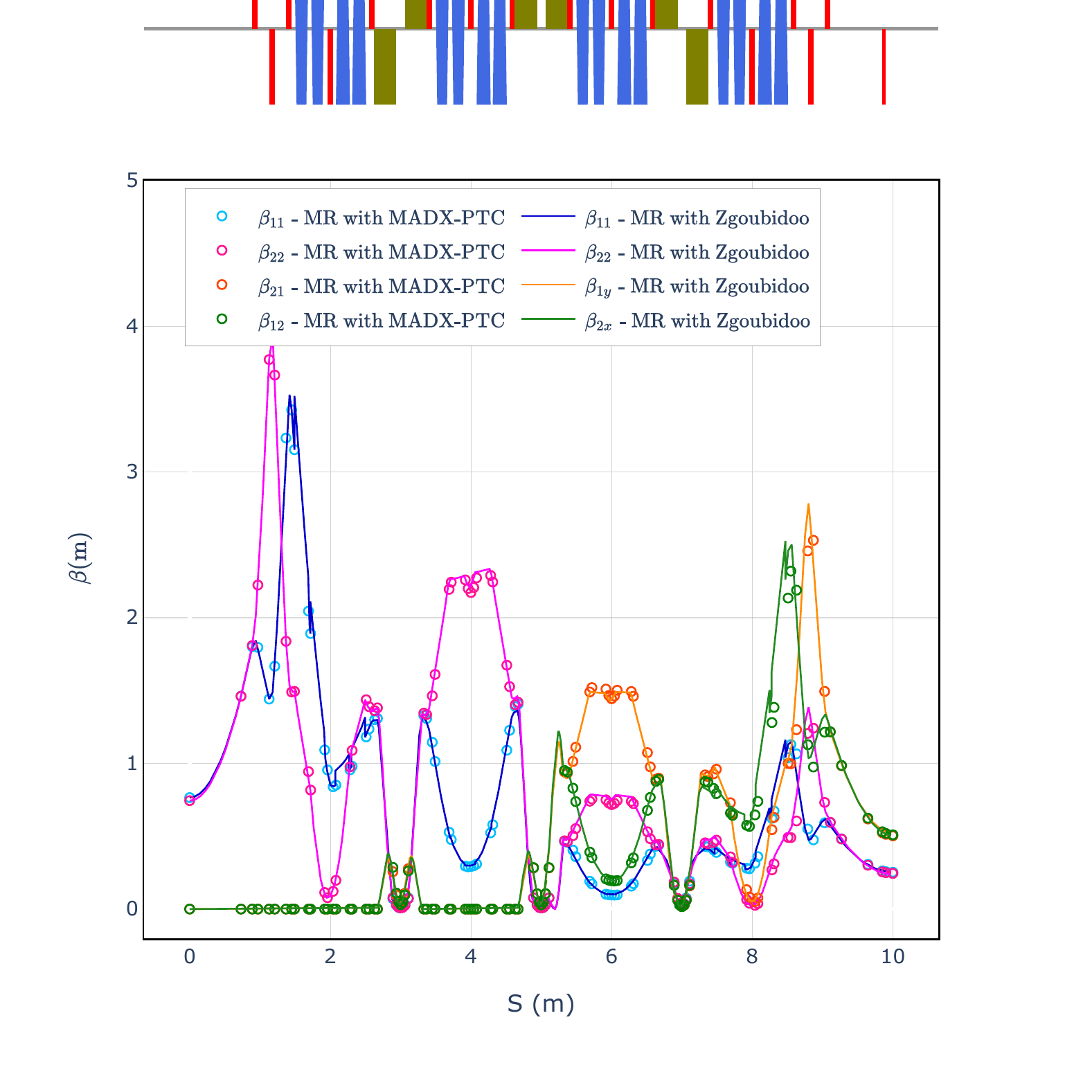}
\caption{Comparison between the coupled $\beta$-functions (ET and MR parametrization) obtained with Zgoubidoo and those obtained with MAD-X on the Snake lattice.}
\label{Snake_beta_functions}
\end{center}
\end{figure}

\begin{figure}
\begin{center}
    \includegraphics[width=1.0\linewidth]{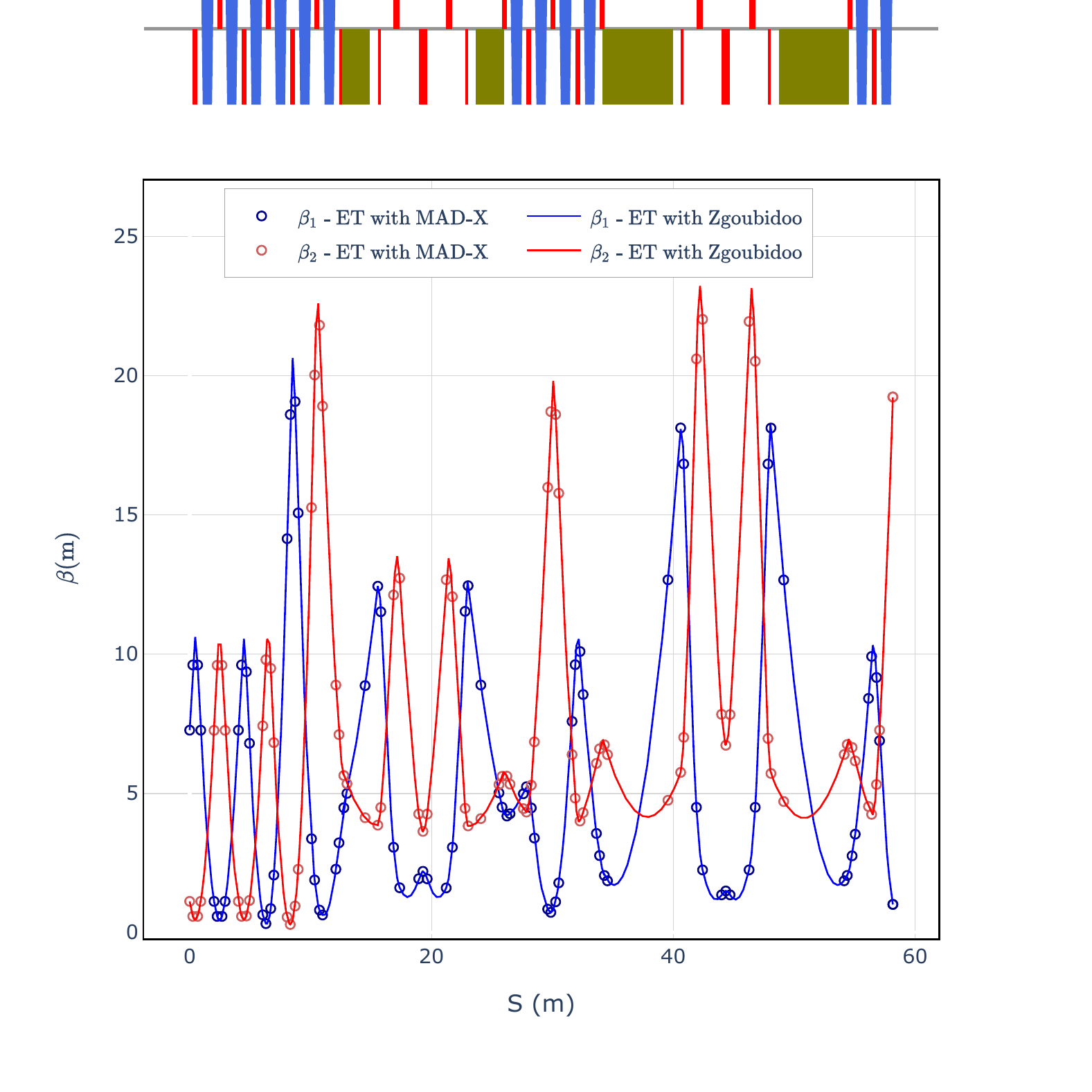}
    \includegraphics[width=1.0\linewidth]{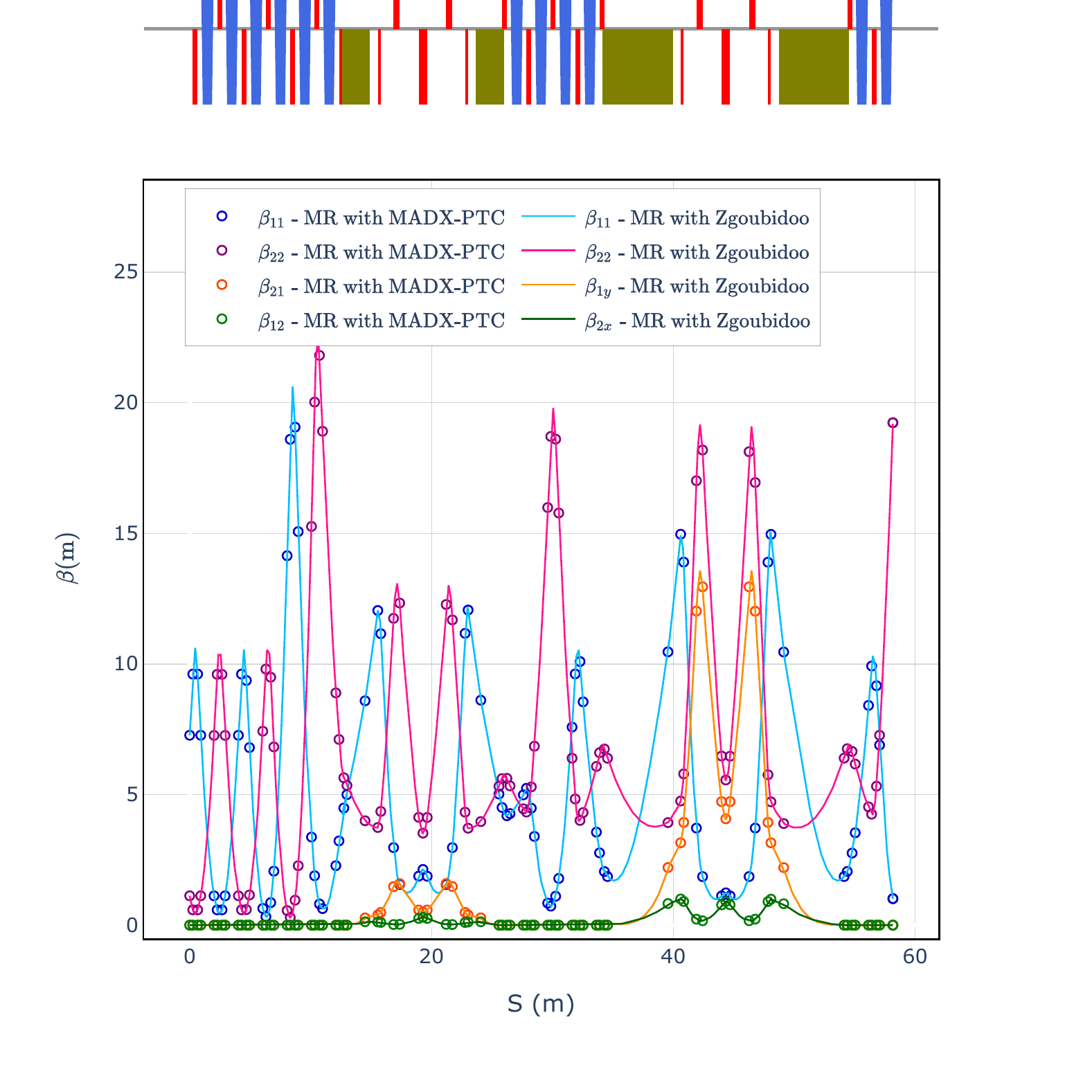}
\caption{Comparison between the coupled $\beta$-functions (ET and MR parametrization) obtained with Zgoubidoo and those obtained with MAD-X in the Spin Rotator lattice.}
\label{SpinRotator_beta_functions}
\end{center}
\end{figure}

\begin{figure}
\begin{center}
    \includegraphics[width=1.0\linewidth]{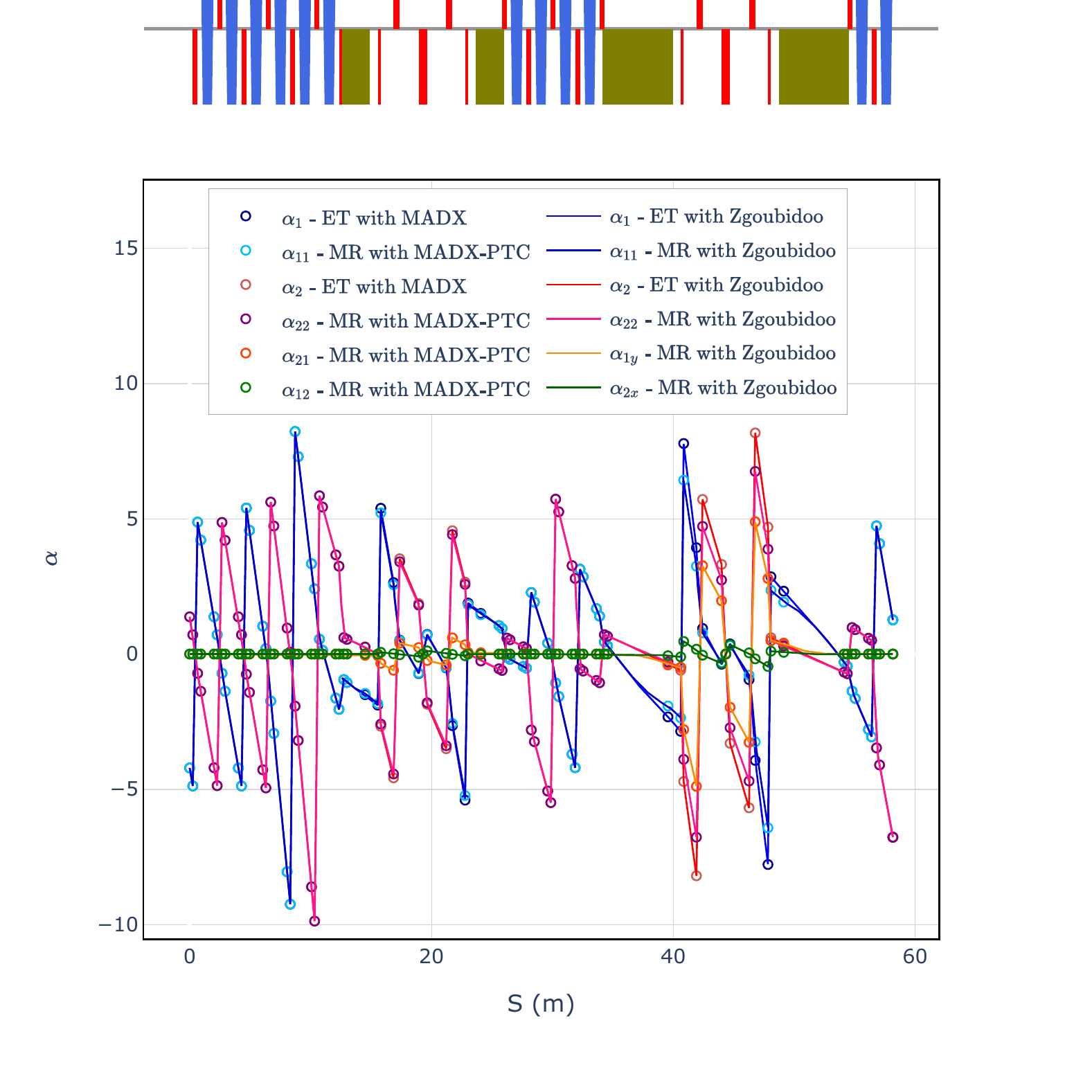}
    \includegraphics[width=1.0\linewidth]{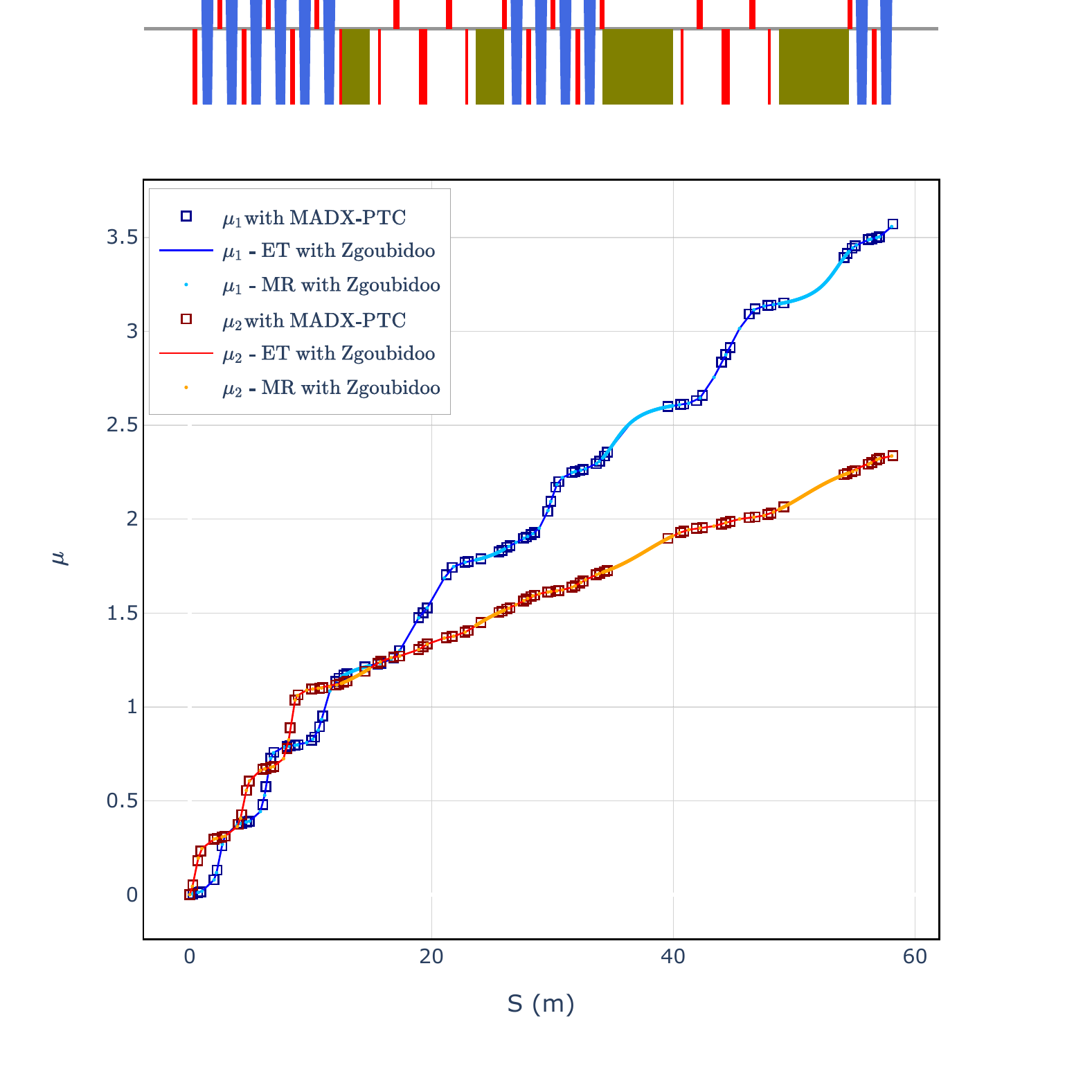}
\caption{Comparison between the coupled $\alpha$-functions (ET and MR parametrization) and the phase advances $\mu$ obtained with Zgoubidoo and those obtained with MAD-X in the Spin Rotator lattice.}
\label{SpinRotator_functions}
\end{center}
\end{figure}

The calculation of the lattice functions is done by the propagation of initial conditions. The initial lattice functions are known and are not coupled. The coupling parameters are thus zero: in the ET parametrization, the decoupling matrix equals the identity; in the MR parametrization, only the main lattice functions are initially non-zero. Non-main lattice functions and additional parameters ($u$, $\nu_1$, $\nu_2$) are zero at the beginning of the transfer line. The propagation of the initial ET lattice functions is discussed in section \ref{ET_parametrization}. For the MR parametrization, the relation used to propagate initial conditions is given in Ref \cite{Lebedev}:
\begin{equation}
     \mathbf{N_{2}} = \mathbf{M_{12}} \mathbf{N_{1}} \mathbf{R}(\Delta \mu_1, \Delta \mu_2) \text{, }
\end{equation}
where $\mathbf{N_{1}}$ and $\mathbf{N_{2}}$ are the normalization matrices at $s_1$ and $s_2$, $\mathbf{M_{12}} = \mathbf{M_{s_1 \rightarrow s_2}}$ is the coupled transfer matrix between $s_1$ and $s_2$ of the beamline, and $\Delta \mu_{1,2}$ are phase advances between $s_1$ and $s_2$. Figure~\ref{Snake_beta_functions} shows the comparison of the coupled $\beta$-functions (in the ET and MR parametrizations) computed with Zgoubidoo and those obtained with MAD-X and PTC on the Snake lattice. We see a good agreement, even if there are some discrepancies due to the difference in the solenoid model. Figure \ref{SpinRotator_beta_functions} shows the same comparison but on the Spin Rotator lattice, and Figure \ref{SpinRotator_functions} compares the other lattice functions ($\alpha$-functions, and phase advances $\mu$) on the same lattice. Again, an excellent agreement is found.

\begin{figure}[h]
\begin{center}
    \includegraphics[width=1.0\linewidth]{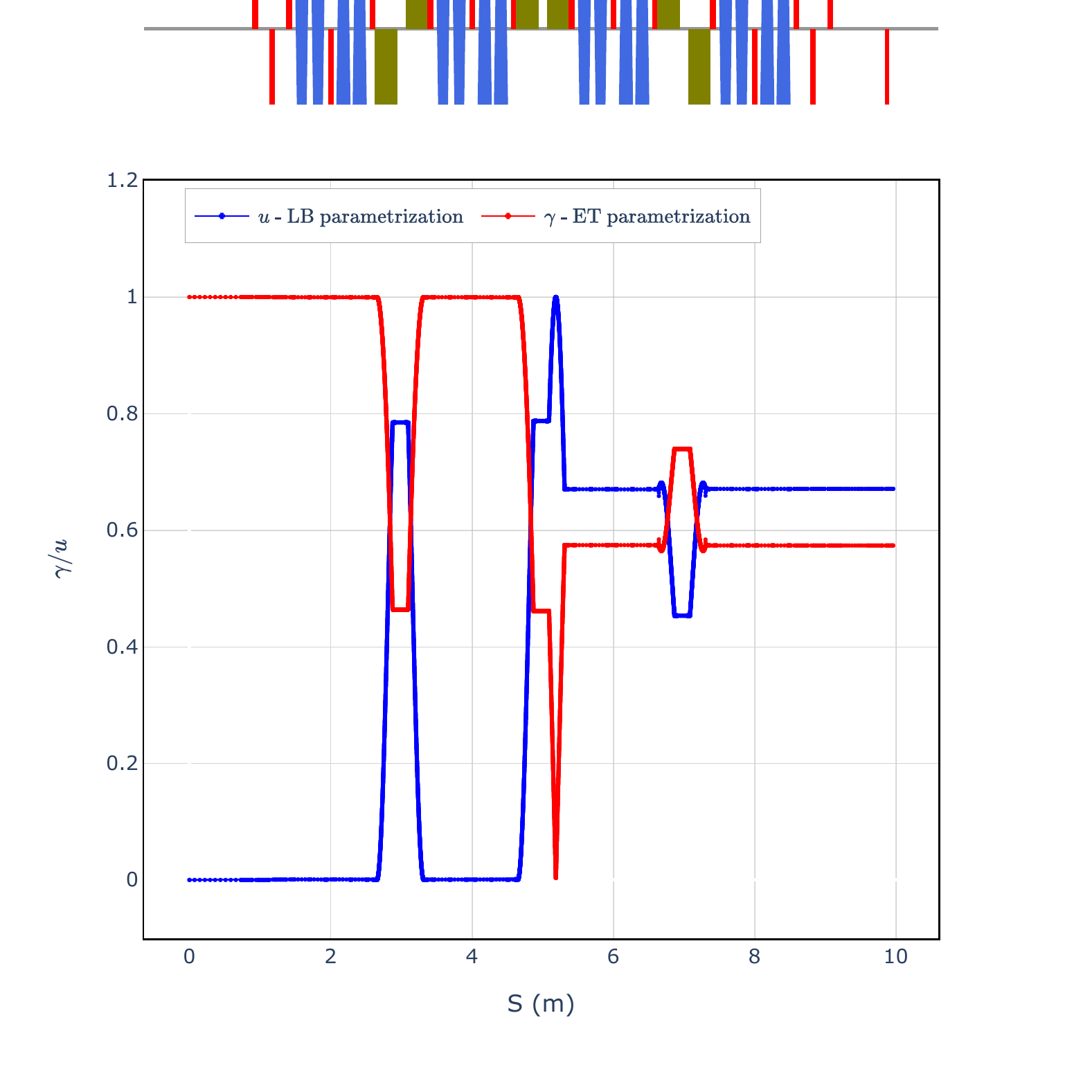}
\caption{Parameter $u$ of the LB parametrization and $\gamma$ of the ET parametrization, obtained by the propagation of initial lattice functions on the Snake lattice.}
\label{Snake_u}
\end{center}
\end{figure}

\begin{figure}[h]
\begin{center}
    \includegraphics[width=0.9\linewidth]{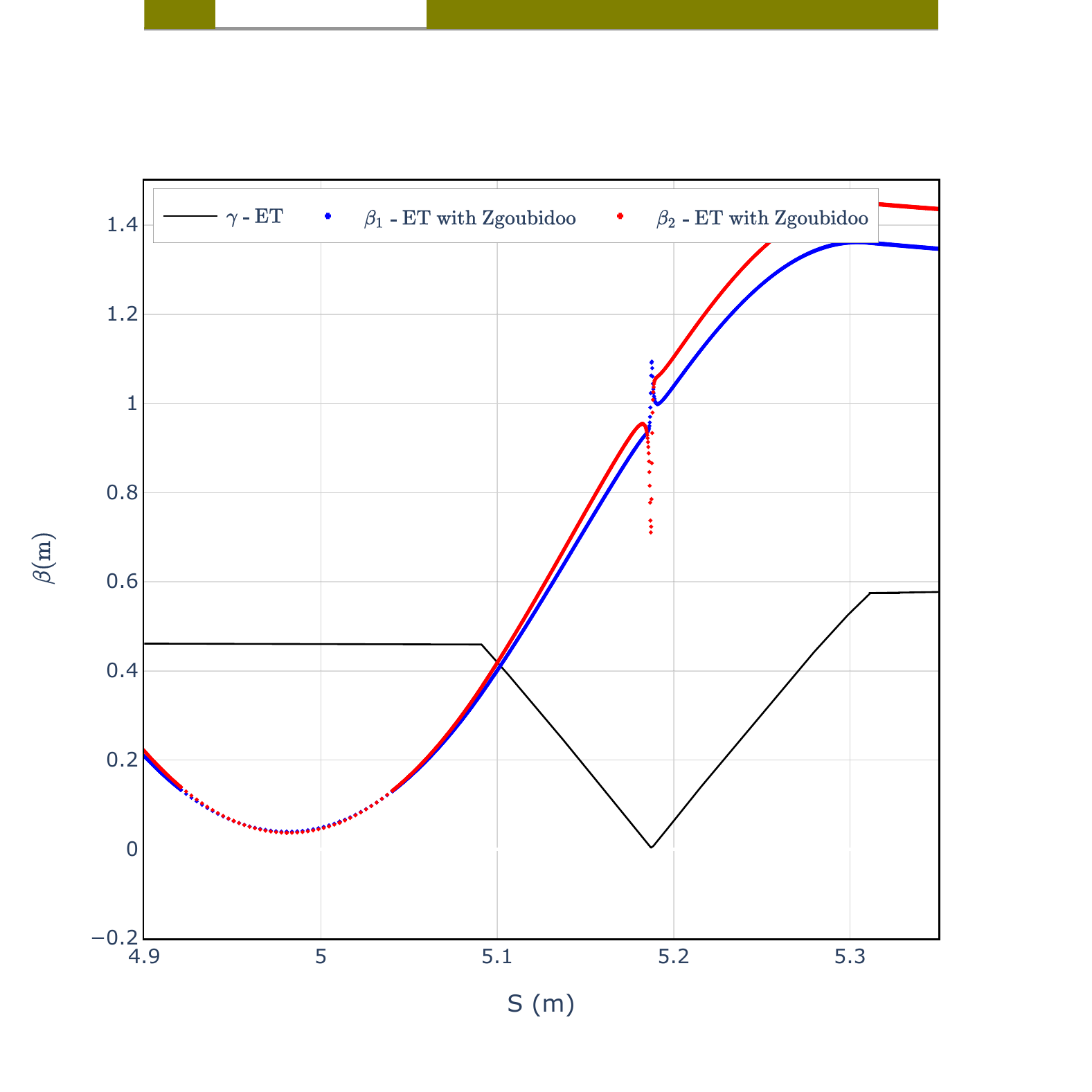}
\caption{$\beta$-functions and $\gamma$ of the ET parametrization on a part of the Snake line. This figure is a zoom on the lattice location that shows forced mode flip conditions ($\gamma \rightarrow 0$). At this location, the $\beta$-functions can diverge and can not anymore be related to beam sizes.}
\label{Snake_beta_ET_gamma}
\end{center}
\end{figure}

The propagation of the lattice functions which is used allows an in-depth study of key concepts. In the Snake lattice, we can highlight the problems related to forced mode flips by analyzing the propagation of the $\beta$ functions of the ET parametrization. In addition, by studying the propagated parameters of the MR parametrization, the interpretation of the parameter $u$ can be refined. In particular, the notion of ``local coupling'' might be clarified. To highlight the potential places where a forced mode flip can occur, we look at the evolution of the ET $\gamma$ parameter throughout the lattice (see Fig \ref{Snake_u}). By propagating initial lattice functions, a forced mode flip can occur when $\gamma \rightarrow 0$. Figure \ref{Snake_beta_ET_gamma} shows the $\beta$-functions of the ET parametrization and the $\gamma$ parameter on a transfer line part. We observe that when $\gamma \rightarrow 0$ (at a specific lattice location), the $\beta$-functions of ET seem to diverge; it illustrates that the ET lattice functions can sometimes be discontinuous or negative. Therefore, they cannot be related to the beam size.

To better understand what is happening, we looked at the behavior of transfer matrices in coupled and decoupled spaces. The transfer matrix in decoupled space is block-diagonal at all lattice points, which indicates that there is no mode flip between the start and the end of the lattice. The initial mode identification is kept throughout the transfer line. It is due to the method we have chosen to implement the ET parametrization, which is Parzen's method (refer to Sec.~\ref{subsection_second_method}). This method solves the problem of mode identification of the ET parametrization. Indeed, it is based on the eigenvectors of the transfer matrix. Each of these eigenvectors is associated with an eigenvalue. The eigenvalues of $\mathbf{M}$ being the same as those of $\mathbf{P}$ (matrices related by a similarity transformation), the oscillation eigenmodes can also be associated with these eigenvalues. It is then possible to identify the oscillation eigenmodes in the decoupled space with these eigenvalues. To keep a specific mode identification throughout the lattice, it is thus sufficient to calculate the optical parameters of each eigenmode with the eigenvectors of the coupled transfer matrix corresponding to the same eigenvalues. One can thus ensure that the Twiss parameters always correspond to the same oscillation eigenmode. This method allows removing the mode flips.

However, when a mode flip is ``forced'', it means that the mode identification is incorrect: it is not possible to correctly compute lattice functions with this mode identification. At the location of a forced mode flip, the planes are completely exchanged. The eigenaxes correspond to the horizontal and vertical axes (x and y), but the axes are switched: the two modes are identified with the perpendicular axes so that the $\beta$-functions that are computed by keeping the mode identification diverge totally. When $\gamma \rightarrow 0$, the transfer matrix in the coupled space tends towards an anti-diagonal matrix: any initial offset in x is transformed almost entirely into a motion in y and vice versa. It confirms the interpretation of the total axes exchange due to the strong coupling of the lattice. To summarize, when one is in the conditions of a forced mode flip at a place of the lattice ($\gamma \rightarrow 0$), either the mode identification is changed, which allows keeping finite $\beta$ functions but poses mode identification difficulties, or the mode identification is kept, which leads to lattice functions that can diverge and thus can no longer be associated with finite beam sizes.

\begin{figure}[h!]
\centering
\begin{subfigure}[h]{0.35\textwidth}
    \hspace{-55pt}
    \includegraphics[scale=0.35]{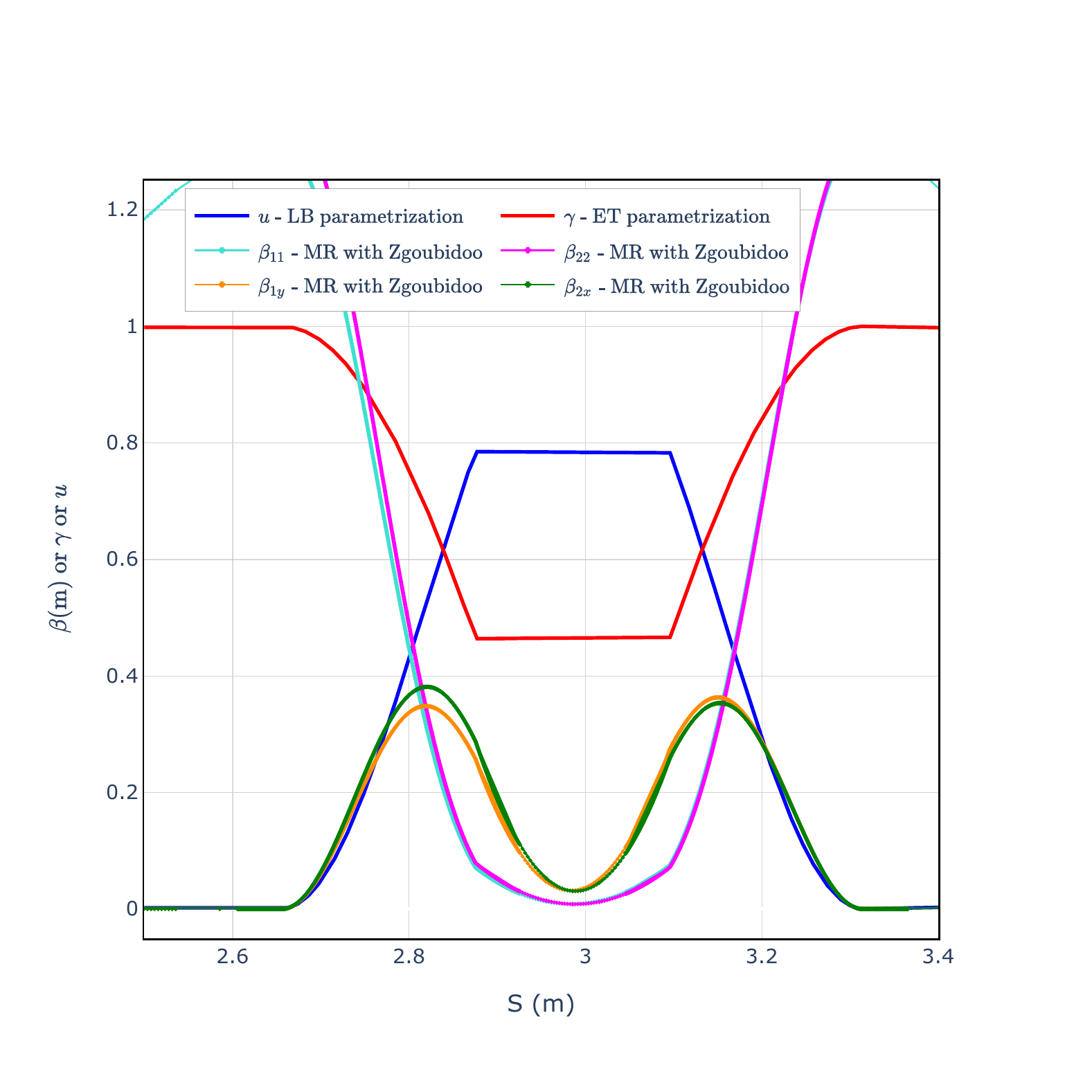}
\end{subfigure}%
\begin{subfigure}[h]{0.35\textwidth}
    \hspace{-135pt}
    \includegraphics[scale=0.35]{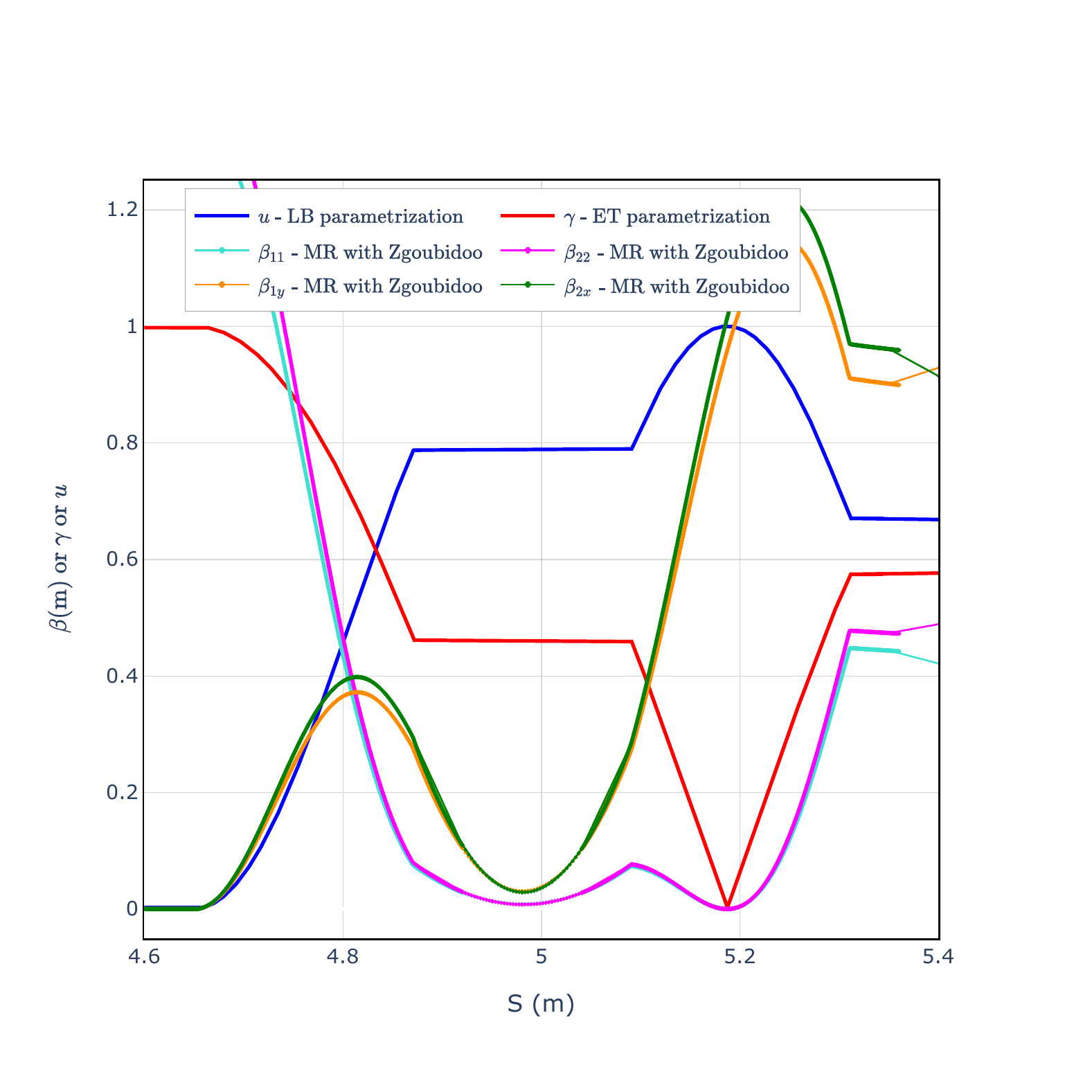}
\end{subfigure}
\caption{$\beta$-functions and $u$ parameter of the MR parametrization in addition to the $\gamma$ parameter of the ET parametrization. This figure is a zoom on some lattice parts where $u>0.5$, including the lattice location that shows forced mode flip conditions ($\gamma \rightarrow 0$). When $u>0.5$, the non-principal functions become more important than the principal ones.}
\label{Snake_beta_MR_u}
\end{figure}

By analyzing the $\beta$-functions of the MR parametrization (see Fig.~\ref{Snake_beta_MR_u}), we note that this phenomenon results in the fact that a mode is first more reflected on a plane and then more on the other plane. When $\gamma \rightarrow 0$ ($s\approx  5.19m$), the main $\beta$-functions ($\beta_{1x}$ and $\beta_{2y}$) are zero, and the eigenmode oscillations are reflected on the other axis, which translates into the non-principal lattice functions ($\beta_{1y}$ and $\beta_{2x}$).
Finally, it should be noted that the forced mode flip conditions appear in this lattice inside a solenoid. The potential problems are thus only fully detected when the tracking code allows step-by-step tracking inside the elements, with a well-chosen integration step. Zgoubidoo, with Zgoubi in the backend, allows obtaining the transfer matrices step by step inside the element and thus detecting any potential problem related to the ET parametrization.

In addition to the forced mode flips analysis, the propagation of the generalized Twiss parameters in the Snake lattice allows a better understanding of the ``local coupling'' (a term used previously in the paper and in many references). To that end, we can first analyze the evolution of the parameter $u$ propagated throughout the lattice. We can see in Fig.~\ref{Snake_u} that $u$ is initially zero because we imposed uncoupled initial conditions. Then, it remains zero in all the elements that do not introduce coupling until reaching the first solenoid. We observe that if an element does not introduce coupling, the parameter $u$ remains constant. From an eigenvector point of view, it means that the ratio between the x and y components of the eigenvector remains constant \cite{Lebedev} because the element does not introduce more coupling than the initial coupling at the element entry. In this beamline, the only elements that change the parameter $u$ are the solenoids, which have a longitudinal field that couples the transverse motion. In parallel with the parameter $u$, we can analyze the principal and non-principal $\beta$-functions of the MR parametrization (see Fig.~\ref{Snake_beta_MR_u}). When $u$ is greater than 0.5, the non-principal functions become more important than the principal ones. If $u$ remains greater than 0.5 at a solenoid output, which is the case at the end of the Snake line, the non-main functions remain dominant until the end of the line. Moreover, when $u \rightarrow 1$, the principal functions cancel each other out. It corresponds to $\gamma \rightarrow 0$: the transfer line is then so coupled that the planes have been totally inverted (forced mode flip). A strong enough coupling is necessary to have this mode inversion along the line; however, where $u$ tends to 1, the line is locally totally decoupled if we invert the mode identification.

\begin{figure}[h]
\begin{center}
    \includegraphics[width=1.0\linewidth]{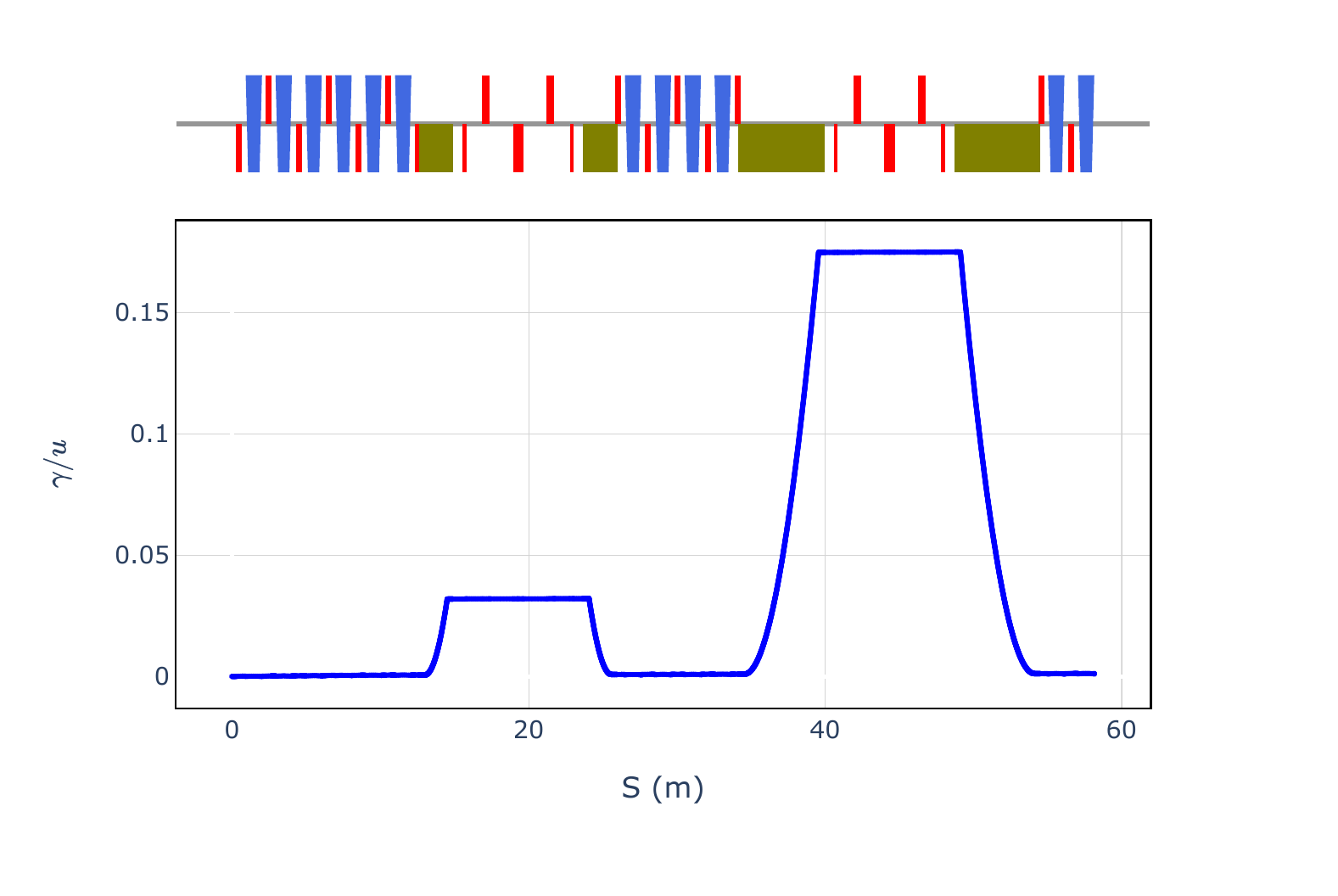}
\caption{Parameter $u$ of the LB parametrization, obtained by the propagation of initial lattice functions on the Spin Rotator lattice.}
\label{SpinRotator_u}
\end{center}
\end{figure}

When propagated in a lattice from initial conditions, the parameter $u$ thus gives a measure of the local coupling. If initially uncoupled lattice functions are propagated into an element where no local coupling is present, the non-principal $\beta$ and $\alpha$ functions (or equivalently, the complex $\zeta$ functions) remain zero. However, these parameters can be non-zero in elements without coupling if they follow lattice parts introducing coupling; the parameter $u$ will then have a finite value that will remain constant in these elements without local coupling. To support this interpretation, we can also examine $u$ in the case of the Spin rotator line as shown in Fig.~\ref{SpinRotator_u}. The various observations made for the Snake lattice also hold in this example. The parameter $u$ is constant except in the solenoids, which allow turning electron spin. We observe that the coupled insertion in this line is designed to cancel the coupling of the solenoids. In Fig.~\ref{SpinRotator_beta_functions}, we can see that the non-principal functions are non-zero only in the coupled insertions; when $u$ returns to 0, indicating, in this case, a zero local coupling, the non-principal lattice functions are also zero.

We have been able to observe in various examples that the parameter $u$ of LB, calculated with periodic conditions or by the propagation of initial lattice functions, gives an idea, respectively, of the lattice average coupling strength or the local coupling at a specific place in the lattice. Moreover, we have highlighted the link between this parameter $u$ and the ellipses in the physical coupled phase space. However, even if this parameter can be used qualitatively, it cannot be rigorously used in all cases to evaluate the coupling strength. Indeed, by analyzing Eqs. \eqref{expressions_u_1} and \eqref{expressions_u_2}, we see that this parameter includes different terms that can cancel each other out in some situations, in particular when a longitudinal field is present (resulting in the constants $R_{1,2 }$). The interpretation of the constant value of $u$ in the elements not introducing coupling remains nevertheless valid because the relative importance of the x and y components of the coupled transfer matrix eigenvectors does not change if the element does not introduce any additional coupling.

\section{Summary and conclusions\label{section_conclusion}}
Transverse betatron motion coupling is a frequent occurrence, whether originating from residual coupling that appears due to imperfections or being coupling ``by design'' from strong systematic coupling fields. Vertical excursion FFAs exhibit strong coupling due to their longitudinal and skew quadrupolar field components. The in-depth study of their linear optics must be studied using models adapted to strongly coupled optics. To support that effort, the available parametrization methods have been extensively reviewed. It has been shown that the ET parametrization allows readily finding the linear invariants of motion by exploring the motion in the decoupled axes. However, the generalized lattice functions of this parametrization are not easily interpretable in terms of beam $\Sigma$-matrix. The MR parametrization allows having a lattice function interpretation similar to that of the Courant-Snyder theory, allowing to link these lattice functions to measurable beam parameters. The ET and MR parametrizations are therefore complementary and are used for different purposes. The minute details and differences of variants of the ET and MR parametrizations have been pointed out. To prepare a detailed analysis of vFFA lattices and to benchmark the different parametrization methods, the different methods have been implemented in a Python interface to the Zgoubi ray-tracing code. The validation of these implementations was carried out on different example lattices with a remarkable agreement with MAD-X and PTC.

The ET parametrization is used to find linear invariants and to analyze the motion in linearly decoupled phase spaces, for example for the computation of the dynamic aperture. There are two main methods to find ET parameters: the generalized Twiss parameters in decoupled axes and the decoupling matrix parameters. The first method uses the analytical solution of a system of equations, while the second method uses the eigenvectors of the coupled and decoupled transfer matrices. We implemented the second method (presented by Parzen in Ref.~\cite{Parzen}) as it can easily be generalized in higher-dimensional phase space. We obtain an excellent agreement between our results and those obtained with MAD-X, which implements the first method (presented by Edwards and Teng in Ref.~\cite{Edwards_Teng} and extended by Sagan and Rubin \cite{Sagan_Rubin}). In parallel with the decoupled motion study, one can also use a parametrization of the MR category to link it with measurable quantities, such as the beam sizes. We have shown that the parametrizations of the MR category describe the quasi-harmonic motions in the coupled phase spaces, resulting from the eigen oscillations in the decoupled space. Depending on the chosen parametrization, one can describe the principal and non-principal oscillations either independently with parameter sets for each oscillation (WR) or with parameters describing the non-principal oscillations relative to the principal ones. The amplitudes and phase shifts are then either given explicitly (LB) or with phasors gathering amplitude and phase in the same quantity (Wolski). If the motion is studied using geometric coordinates, the phase space can be described by the WR parameters. If the motion is expressed in canonical coordinates, the phase space is described by parameters accounting for the longitudinal field coupling, such as the LB parameters.

For our study of vFFA lattices, the LB parametrization is encouraged as it provides relevant additional parameters compared to the WR description. The parameter $u$ qualitatively evaluates the local coupling strength. It characterizes the size of the two ellipses coming from an oscillation eigenmode in the two transverse phase spaces, can be related to the local coupling concept, and can indicate a forced mode flip because it is linked to the $\gamma$ parameter of the ET parametrization. The in-depth review, interpretation, implementation, and validation of the available parametrization methods pave the way for complete and detailed studies of the beam dynamics in strongly coupled vFFA lattices.

\begin{acknowledgments}
The authors would like to thank J-B. Lagrange for the fruitful discussions and comments on the manuscript.
Marion Vanwelde is a Research Fellow of the Fonds de la Recherche Scientifique - FNRS.
This work is supported by U.S. Department of Energy under DE-AC05-06OR23177.
\end{acknowledgments}

\appendix
\section{Parametrization of the betatron motion in the decoupled case\label{appendix_decoupled}}
The equations that describe the linear and uncoupled transverse motion are:
\begin{align}
    \label{Hills_equations}
	\left\lbrace
		\begin{aligned}
            \frac{d^2 x}{ds^2} &+ (\frac{1}{\rho^2} + K(s)) x = 0 \\
            \frac{d^2 y}{ds^2} &- K(s) y  = 0 \\
		\end{aligned}
	\right.
\end{align}
where $K(s)$ is a focusing parameter that is related to the field gradient. This coefficient depends on $s$ and gives the focusing strength throughout the machine, thus reflecting the organization of the magnets in the accelerator. If $K(s)$ is periodic ($K(s) = K(s + L)$ where $L$ is the length of a period), the equations of motion are called Hill's equations. The equations describing betatron oscillations are similar in both directions and can be written
\begin{equation}
    y'' + k(s) y  = 0 \text{,}
\end{equation}
where $y$ represents both vertical and horizontal coordinates. This equation corresponds to the equation of a harmonic oscillator with a $s$-dependent frequency. We assume a quasi-harmonic solution, the so-called Floquet solution which depends on the amplitude function $\beta(s)$, on the phase advance $\phi(s)$ and on the invariants $\epsilon$ and $\phi_0$:
\begin{equation}
    y(s) = \sqrt{\epsilon \beta(s)} \sin{(\phi(s)+\phi_0)} \text{.}
\label{solution_Floquet}
\end{equation}
The particles will oscillate inside the envelope $\sqrt{\epsilon \beta(s)}$ which depends on $s$ \textit{via} the amplitude function $\beta(s)$. By reinjecting the Floquet solution (Eq.~\eqref{solution_Floquet}) into the Hill's equations (Eq.~\eqref{Hills_equations}), we obtain the link between the beta-function and the phase advance, as well as a differential equation for $\beta$ (the ``envelope equation'') \cite{NF94}:
\begin{align}
\label{lien_beta_phi}
& \phi(s) = \int_0^s \frac{d\sigma}{\beta(\sigma)} \text{, }\\
& \frac{1}{2}\beta \beta' - 1/4 \beta '^2 +k(s) \beta^2 = 1\text{.}
\label{enveloppe_equation}
\end{align}

It is possible to find the $\beta$-function from equation \eqref{enveloppe_equation} with correct boundary conditions. However, this equation is not easily integrable, and using matrix formalism is preferable. The elements of this matrix can also be expressed with the Floquet parameters $\beta(s)$ and $\phi(s)$. By setting $w(s)=\sqrt{\beta(s)}$, the most general form of the transfer matrix is expressed as follows \cite{wilsonBeamDynamics2020}:

\begin{widetext}
\begin{equation}
\mathbf{M_{s_1 \rightarrow s_2}} = \begin{pmatrix}
\frac{w_2}{w_1} \cos{(\Delta \mu_{12})} - w_2 w_1' \sin{(\Delta \mu_{12})} & w_1 w_2 \sin{(\Delta \mu_{12})} \\
- \frac{1+w_1 w_1' w_2 w_2'}{w_1 w_2} \sin{(\Delta \mu_{12})} - (\frac{w_1'}{w_2'} - \frac{w_2}{w_1} \cos{(\Delta \mu_{12})} & (\frac{w_1}{w_2} \cos{(\Delta \mu_{12})} + w_1 w_2' sin(\Delta \mu_{12})
\end{pmatrix}\text{, }
\end{equation}
\end{widetext}
where $w_1 = w(s_1)$, $w_2 = w(s_2)$ and $\Delta \mu_{12} = \phi(s_2) - \phi(s_1)$ is the phase advance between $s_1$ and $s_2$. This matrix is further simplified when we consider that $k(s)$ is periodic. In this case, $w(s_1) = w(s_2) = w$, $w'(s_1) = w'(s_2)= w'$ and $\Delta \mu_{12} = \mu$ (phase advance of one cell). The transfer matrix of a period becomes:
\begin{equation}
    \mathbf{\hat{M}}=\begin{pmatrix}
    \cos{(\mu)} - w w' \sin{(\mu)} &  w^2 \sin{(\mu)} \\
    -\frac{1+w^2 w'^2}{w^2} \sin{(\mu)} & \cos{(\mu)} + w w' \sin{(\mu)}
    \end{pmatrix}\text{.}
\end{equation}

When we express this matrix with the periodic lattice functions $\beta(s) = w^2(s)$, $\alpha = -w(s)w'(s) = -\frac{\beta ' (s)}{2}$ and $\gamma(s) = \frac{1+\alpha^2(s)}{\beta(s)}$, we obtain the so-called Twiss matrix:
\begin{equation}
    \mathbf{\hat{M}} = \begin{pmatrix}
    \cos(\mu) + \alpha(s) \sin(\mu) &  \beta(s) \sin(\mu) \\
    -\gamma(s) \sin(\mu) & \cos(\mu) - \alpha(s) \sin(\mu)
    \end{pmatrix}\text{.}
    \label{Twiss_matrix}
\end{equation}
By comparing this parametrized matrix with the matrix obtained by simulation, it is possible to determine the optical functions $\beta(s)$, $\alpha(s)$ and $\mu(s)$ along the lattice. The linear tune of the periodic cell\textemdash the number of oscillations the particle performs in the cell\textemdash is given by the phase advance of the cell divided by $2\pi$: $Q = \frac{\mu}{2\pi}$.

\subsection{Parametrization of the normalization matrix}
We can express the one-turn transfer matrix $\mathbf{\hat{M}}$ as being the product of a rotation matrix $\mathbf{R}$, depending on the phase advance $\phi(s)$, and a matrix $\mathbf{T}$, depending on the lattice parameters $\beta(s)$ and $\alpha(s)$ \cite{NF94}:
\begin{equation}
    \mathbf{\hat{M}} = \mathbf{T R T^{-1}}\text{, }
    \label{M_with_T}
\end{equation}
\begin{align}
    \mathbf{R} &= \begin{pmatrix}
    \cos{(\phi(s))} & \sin{(\phi(s))} \\
    -\sin{(\phi(s))} & \cos{(\phi(s))}
    \end{pmatrix} \text{, } \label{Rotation_matrix_eq}\\
    \mathbf{T} &= \begin{pmatrix}
    \sqrt{\beta(s)} & 0 \\
    \frac{- \alpha(s)}{\sqrt{\beta(s)}} & \frac{1}{\sqrt{\beta(s)}}
    \end{pmatrix}\text{.}
    \label{T_matrix}
\end{align}\\
The matrix $\mathbf{T}$ is a normalization transformation which transforms the one-turn transfer matrix into its normal form $ \mathbf{T^{-1}} \mathbf{\hat{M}} \mathbf{T} = \mathbf{R} $. The matrix $\mathbf{T}$ contains the local focusing properties of the lattice, and will therefore depend on the section at which we compute it. The rotation matrix $ \mathbf{R}$ contains global properties (phase advance over one period) and will not depend on this section \cite{wolskiSimpleWayCharacterize2004}. The linear tunes of the cell are therefore independent of the point at which the period is started. The normalization matrix $\mathbf{T}$ transforms the phase space coordinates $\mathbf{x}$ into the Courant-Snyder coordinates $\mathbf{\tilde{x} = T^{-1} x}$. In these coordinates, the transfer matrix is a rotation $\mathbf{\tilde{x}}(s_2) = \mathbf{R \tilde{x}}(s_1)$, and the motion reduces to a harmonic solution \cite{NF94}.

It is interesting to note that there is some freedom in the choice of $\mathbf{T}$. Indeed if we choose $\mathbf{\bar{T} = TS}$, where $\mathbf{S}$ is a rotation that commutes with $\mathbf{R}$, the one-turn transfer matrix can also be expressed with this normalization transformation: $\mathbf{\hat{M}} = \mathbf{\bar{T} R \bar{T}^{-1}}$. In accelerator physics, the free parameters of $\mathbf{S}$ are commonly chosen to have a normalization matrix of the form given by Eq.~\eqref{T_matrix}, \textit{i.e.} a transformation which preserves the areas, and whose $T_{12}$ element is zero \cite{NF94}. If we consider a transfer line instead of a periodic accelerator, we can express the transfer matrix from $s_1$ to $s_2$ thanks to the normalization matrices at these points and the phase advance $\Delta \mu_{12}$: $$ \mathbf{M_{s_1 \rightarrow s_2}} = \mathbf{T}(s_2) \mathbf{R} (\Delta \mu_{12}) \mathbf{T}(s_1) ^{-1}\text{.}$$

\subsection{Parametrization of generating vectors}
Another way to study the transverse motion is to look at the trajectory of a particle in the phase space. For an uncoupled linear motion, if we plot the divergence $y'$ (vertical or horizontal) as a function of the position $y$ (vertical or horizontal) at a given location $s$ at each turn, we get an ellipse. This ellipse is described by the optical functions $\beta(s)$, $\alpha(s)$ et $\gamma(s)$ and is expressed by the Courant-Snyder invariant:
\begin{equation}
\gamma(s) y^2 + 2 \alpha(s) y y'+ \beta(s)y'^2 = \epsilon \text{. }
\label{Courant_Snyder_invariant}
\end{equation}

The ellipse area is equal to $\pi\epsilon$, where $\epsilon$ is the emittance of the beam. If we consider a beam emittance $\epsilon$, all the particles inside the beam are included in the phase space ellipse of area $\pi \epsilon$. The ellipse area is conserved during motion. However, the aspect ratio of the ellipse depends on the s-dependent optical functions and can therefore change along the lattice. The ellipse shape will thus depend on the section where the coordinates are sampled. In other words, under a linear transformation given by any unimodular transfer matrix $\mathbf{M_{s_1 \rightarrow s_2}}$ ($\mathbf{x_2}=\mathbf{M_{s_1 \rightarrow s_2} x_1}$), the phase space ellipse will be transformed into another ellipse with the same area \cite{USPAS}. In addition, if the linear transformation is periodic and is described by the one-turn Twiss matrix $\mathbf{\hat{M}}$ (Eq.~\eqref{Twiss_matrix}), the ellipse will be invariant.

In the normalized Courant-Snyder coordinates, the phase space ellipse becomes a circle with the same area. In these new coordinates, the radius of the circle is invariant $\epsilon = \tilde{y}^2 + \tilde{y}'^2$. A particle initially on the circle (or the ellipse in the case of physical coordinates) will remain on it but will rotate at a certain angle on this circle at each period. By looking at the rotation angle of the particle, we can find the fractional part of the tune (because at each turn, the phase advance of the particle will be $\mu = Q*2\pi$).

The particle motion on the ellipse can be expressed with two generating vectors of this ellipse $\mathbf{y_1}$ and $\mathbf{y_2}$. The particle coordinates at any point $s$ can be expressed as follows \cite{Willeke}:
\begin{equation}
    \mathbf{y}(s) = \sqrt{\epsilon} (\mathbf{y_1}(s)\cos{(\phi_0)} - \mathbf{y_2}(s) \sin{(\phi_0)}) \text{.}
    \label{coordinates_generating_vectors}
\end{equation}
In order to have the same description of the particle motion as before (and therefore to have an ellipse whose shape is characterized by $\beta(s)$, $\alpha(s)$ et $\gamma(s)$), the generating vectors must be parametrized with the optical functions:
\begin{align}
    y_1(s) &= \sqrt{\beta(s)} \cos{(\phi(s))} \text{, }\label{uncoupled_generating_vectors_1}\\
    y_2(s) &= \sqrt{\beta(s)} \sin{(\phi(s))} \text{, }\label{uncoupled_generating_vectors_2}\\
    y_1^\prime(s) &= \sqrt{\gamma(s)} \cos{(\overset{\sim}\phi(s))} \text{, } \label{uncoupled_generating_vectors_3}\\
    y_2^\prime(s) &= \sqrt{\gamma(s)} \sin{(\overset{\sim}\phi(s))}\text{.}
    \label{uncoupled_generating_vectors}
\end{align}
\noeqref{uncoupled_generating_vectors_1, uncoupled_generating_vectors_2, uncoupled_generating_vectors_3}
We clearly see in these equations that $\beta(s)$ and $\gamma(s)$ are respectively linked to the envelopes of the position and angle coordinates, while $\phi(s)$ and $\overset{\sim}\phi(s)$ are phase functions. The generating vectors are normalized ($\mathbf{y_1}^T \ \mathbf{S} \  \mathbf{y_2} = 1$) so that the ellipse area is equal to $\pi \epsilon$. We see that $\epsilon$ weights the expression of the particle coordinates (Eq.~\eqref{coordinates_generating_vectors}) and thus characterizes the size of the ellipse. The normalization condition on the generating vectors allows to find the link between $\beta$ and $\phi$ (Eq.~\eqref{lien_beta_phi}), the link between $\alpha$, $\beta$ and $\gamma$, and the link between $\phi(s)$ and $\overset{\sim}\phi(s)$: $\overset{\sim}\phi(s) = \phi(s) - \arctan{\alpha ^{-1}} \text{.}$

The maximum displacement of the particle is described by the beam envelope $\sqrt{\epsilon \beta(s)}$. The size of the beam will thus depend on the beam itself via its emittance $\epsilon$, and on the machine via the $\beta$-function. This envelope modulates the oscillation amplitude of the particles along $s$. All the particles in the beam are injected differently into the accelerator, and therefore do not have the same individual trajectory, but they will all be bounded by this envelope. In the uncoupled motion parametrization, the lattice functions $\beta(s)$, $\alpha(s)$ and $\mu(s)$ have a clear physical meaning and give us information about the focusing properties of the lattice: $\beta(s)$ limits the betatron oscillation amplitude of the particles and is therefore related to the beam size, while $\mu(s)$ represents the phase advance of the oscillation. The functions $\alpha(s)$ and $\gamma(s)$ are directly related to the $\beta$-function, while the linear tune is directly related to the phase advance on a period. This clear physical interpretation of the optical parameters is, among other things, what we are looking for in the parametrization of coupled motion.

\section{Floquet's theorem\label{appendeix_floquet}}
Floquet's theorem states that the eigenvectors $\mathbf{\hat{v}_j}(s)$ of the one-turn transfer matrix $\mathbf{\hat{M}}(s)$ (where $\mathbf{\hat{v}_j}(s) = \mathbf{M_{s_0 \rightarrow s}}\mathbf{\hat{v}_j}(s_0)$) are special solutions of the equations of motion and can be written as the product of a periodic function $f$ and a harmonic factor \cite{Willeke}:
\begin{equation}
    \mathbf{\hat{v}_j}(s) = \mathbf{f_j}(s) e^{i 2 \pi Q_j (\frac{s}{L})}\text{, }
    \label{Floquet_theorem}
\end{equation}
where $\mathbf{f_j}(s+L)=\mathbf{f_j}(s)$.

\section{4D phase space beam ellipsoid\label{appendix_ellipsoid}}
\begin{figure}[ht!]
    \includegraphics[width=0.8\linewidth]{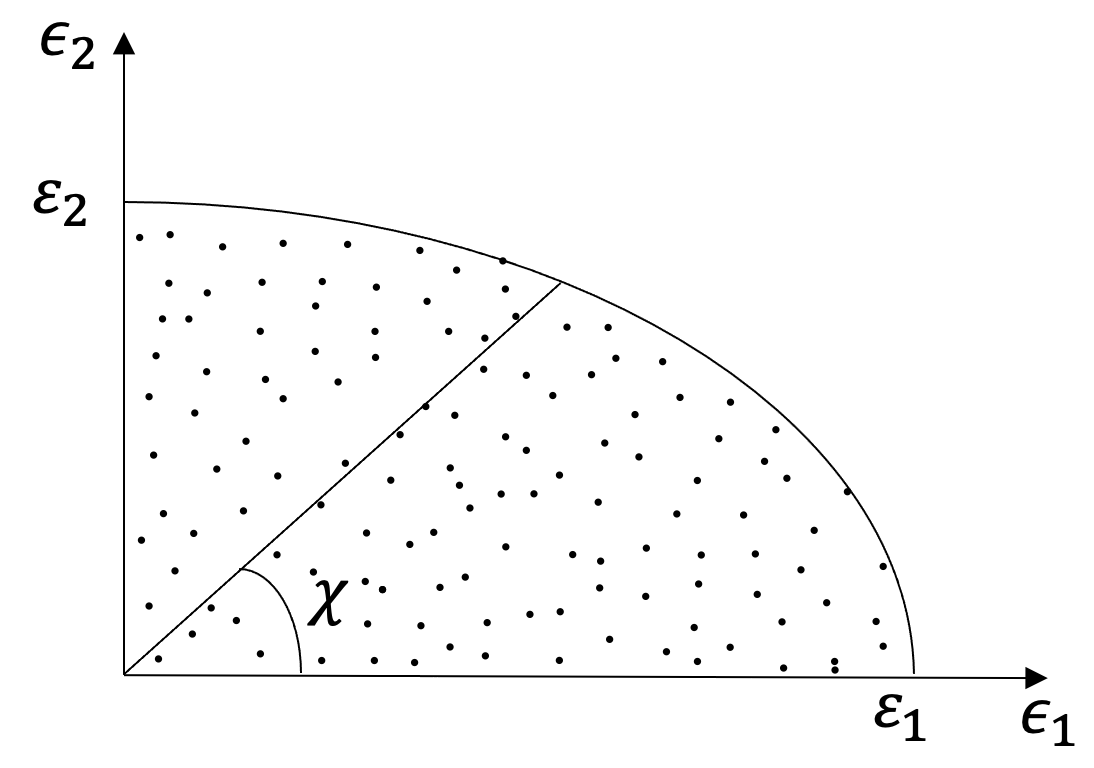}
\caption{Particle distribution with respect to their amplitudes. Reproduced from Ref.~\cite{Willeke}.}
\label{bunch_surface}
\end{figure}

We consider an ensemble of particles having different amplitudes and initial phases and confined within a 4D phase space ellipsoid. To describe the 3D surface which defines this ellipsoid, the motion of the particles whose betatron oscillation amplitude is maximum is investigated \cite{Lebedev, Willeke}. The particle distribution being characterized by the emittances $\varepsilon_1$ and $\varepsilon_2$, the particles which describe the 4D ellipsoid surface will have amplitude $\sqrt{\varepsilon_1} \cos{\chi}$ and $\sqrt{\varepsilon_2} \sin{\chi}$. Figure \ref{bunch_surface} shows the distribution of particles as a function of $\epsilon_1$ and $\epsilon_2$ \cite{Willeke}. The particles on the ellipsoid surface lie on the edge of the ellipse shown in this figure.

The trajectory of a particle on the ellipsoid surface can thus be written by replacing ($\sqrt{\epsilon_I}$, $\sqrt{\epsilon_{II}}$) by ($\sqrt{\varepsilon_I} \cos{\chi}$, $\sqrt{\varepsilon_{II}} \sin{\chi}$) \cite{Willeke}:
\begin{equation}
\label{particle_trajectory_bis}
\begin{aligned}
    \mathbf{z}(s) = & \sqrt{\varepsilon_{I}}\cos{\chi}[\mathbf{z_1}(s) \cos{\phi_{I,0}} - \mathbf{z_2}(s) \sin{\phi_{I,0}}] \\
                     + & \sqrt{\varepsilon_{II}} \sin{\chi}[\mathbf{z_3}(s) \cos{\phi_{II,0}} - \mathbf{z_4}(s) \sin{\phi_{II,0}}]\text{.}
\end{aligned}
\end{equation}
The projection of the 4D beam ellipsoid on a plane $z - z'$ (where $z = x, y$) is an ellipse whose maximum values are the same as what we obtained by superimposing the ellipses of the two modes in the case of a single-particle motion. These maximum values can be directly derived from equation \eqref{particle_trajectory_bis} and are given in \cite{Willeke}. For example, the maximum horizontal position is given by:
\begin{equation}
    x_{max} = \sqrt{\varepsilon_1 \beta_{x1} + \varepsilon_2 \beta_{x2}} \text{,}
\end{equation}
for particular initial phase values and angle value $\chi$ ($\sin{\chi} = \frac{\sqrt{\varepsilon_2 \beta_{x2}}}{\sqrt{\varepsilon_1 \beta_{x1} + \varepsilon_2 \beta_{x2}}}$, $\cos{\chi} = \frac{\sqrt{\varepsilon_1 \beta_{x1}}}{\sqrt{\varepsilon_1 \beta_{x1} + \varepsilon_2 \beta_{x2}}}$). We can define the horizontal and vertical emittances of the beam from the ellipses corresponding to the 4D ellipsoid projections on the $x-x'$ and $y-y'$ planes. The surface of the ellipse projected in the $x-x'$ (resp. $y-y'$) plane is referred to as the \textit{horizontal emittance} (resp. \textit{vertical emittance})\cite{Willeke}:
\begin{align}
\varepsilon_x &= \varepsilon_1\beta_{x1} \phi'_{x1} + \varepsilon_2\beta_{x2} \phi'_{x2}\text{, }\\
\varepsilon_y &= \varepsilon_1\beta_{y1} \phi'_{y1} + \varepsilon_2\beta_{y2} \phi'_{y2}\text{.}
\end{align}

We can directly see that the emittance $\varepsilon_x$ (resp. $\varepsilon_y$) corresponds to the sum (or difference according to the sign of $\phi'$) of the ellipse areas of modes I and II, appearing in the horizontal (resp. vertical) plane when looking at the single-particle motion. These horizontal and vertical emittances are not motion invariants, and therefore the sum of the two ellipse areas in the same plane is not a motion invariant. The ellipsoid surface can be described by the following bilinear form:
\begin{equation}
    \mathbf{\hat{x}^T} \mathbf{\Xi} \mathbf{\hat{x}} = 1 \text{,}
    \label{bilinear_form}
\end{equation}
which can be diagonalized by the normalization matrix $\mathbf{N}$:
\begin{equation*}
    \mathbf{\Xi^{'}} = \mathbf{N}^T \mathbf{\Xi} \mathbf{N}\text{, where} \quad
    \mathbf{\Xi^{'}} = \begin{pmatrix}
    \frac{1}{\varepsilon_1} & 0 & 0 & 0 \\
    0 & \frac{1}{\varepsilon_1} & 0 & 0 \\
    0 & 0 & \frac{1}{\varepsilon_2} & 0 \\
    0 & 0 & 0 & \frac{1}{\varepsilon_2}
    \end{pmatrix}\text{.}
\end{equation*}

The bilinear form can then be rewritten as follows:
\begin{equation}
\Xi_{11}^{'} x^{'2} + \Xi_{22}^{'} p_x^{'2} +\Xi_{33}^{'} y^{'2} + \Xi_{44}^{'} p_y^{'2} = 1 \text{.}
\end{equation}
The 4D beam emittance corresponds to the phase space volume occupied by the beam and is therefore equal to the product of the ellipsoid semi-axes \cite{Lebedev}: $\varepsilon_{4D} = \varepsilon_1 \varepsilon_2$. We can understand the bilinear form of equation \eqref{bilinear_form} from the analysis of the motion invariants. It is possible to write the particle action as a quadratic function of the phase space coordinates \cite{wolskiNormalFormAnalysis2004}:
\begin{equation}
J = \frac{1}{2} <\mathbf{\hat{x}}|\mathbf{A}|\mathbf{\hat{x}}>\text{, }
\end{equation}
where $J$ is invariant under the transformation $\mathbf{\hat{M}}$, which results in the following relation: $\mathbf{\hat{M}}^T \mathbf{A} \mathbf{\hat{M}} = \mathbf{A}$. We can define the emittances $\varepsilon_1$ and $\varepsilon_2$ as the average over all the particle actions: $\varepsilon = <J>$. It is possible to show \cite {wolskiNormalFormAnalysis2004} that there are two linearly independent solutions for the matrix $\mathbf{A}$ ($\mathbf{A_I}$ and $\mathbf{A_{II}}$), so two quadratic invariants $J_1$ and $J_2$ and two associated emittances. The quadratic form used to describe the 4D ellipsoid can be written using the matrices $\mathbf{A_I}$ and $\mathbf{A_ {II}}$ as follows:
\begin{equation}
    \mathbf{\Xi} = \biggl(\frac{\mathbf{A_I}}{\varepsilon_{I}} + \frac{\mathbf{A_{II}}}{\varepsilon_{II}}\biggr)\text{.}
\end{equation}
Finally, it is possible to find the correlation matrix $\mathbf{\Sigma}$ from $\mathbf{\Xi}$. The matrix $\mathbf{\Sigma}$ allows describing the matched beam distribution \textemdash a distribution invariant under the transformation $\mathbf{\hat{M}}$:
\begin{eqnarray*}
    \mathbf{\Sigma} &\rightarrow & \mathbf{\Sigma} \text{, }\\
    \mathbf{\hat{x}} &\rightarrow & \mathbf{\hat{M}}\mathbf{\hat{x}}\text{.}
\end{eqnarray*}
This distribution is invariant under the transformation $\mathbf{\hat{M}}$ so that it depends on the motion invariants, and therefore on $\frac{1}{2}\mathbf{\hat{x}^T} \mathbf{\Xi} \mathbf{\hat{x}}$. Considering a Gaussian distribution, we can write the distribution as follows \cite{Lebedev}:
\begin{equation}
    f(\mathbf{\hat{x}}) = \tilde{c} \exp{(-\frac{1}{2}\mathbf{\hat{x}^T} \mathbf{\Xi} \mathbf{\hat{x}}) }\text{, }
\end{equation}
where $\tilde{c}$ is a normalization coefficient. The correlation matrix $\mathbf{\Sigma}$, which contains the second-order moments of this distribution, is linked to the quadratic form $\mathbf{\Xi}$ by the following relation: $\mathbf{\Sigma} = \mathbf{\Xi^{-1}}$ \cite{Lebedev, wolskiNormalFormAnalysis2004}. This correlation matrix reads:
\begin{equation}
    \mathbf{\Sigma} = \mathbf{N}\begin{pmatrix}
    \varepsilon_1 & 0 & 0 & 0 \\
    0 & \varepsilon_1 & 0 & 0 \\
    0 & 0 & \varepsilon_2 & 0 \\
    0 & 0 & 0 & \varepsilon_2 \\
    \end{pmatrix} \mathbf{N}^T \text{.}
\end{equation}
The matrices $\mathbf{\Xi}$ and $\mathbf{\Sigma}$ are directly related to the beam emittances and to the eigenvectors of $\mathbf{\hat{M}}$, and thus to the generalized lattice functions. It is possible to calculate the correlation matrix elements with the MR parametrization lattice functions if we know the mode emittances (see Tab.~\ref{compa_corr_matrix}). Conversely, if we know $\mathbf{\Xi}$ or $\mathbf{\Sigma}$, we can find the mode emittances $\varepsilon_1$, $\varepsilon_2$ and the eigenvectors of $\mathbf{\hat{M}}$, and thus the normalization matrix. We can therefore determine the emittances as well as the eigenvectors from something measurable. Finally, it is possible to prove that $\mathbf{\Sigma S}$ and $\mathbf{\hat{M}}$ have the same eigenvectors, but different eigenvalues \cite{wolskiAlternativeApproachGeneral2006}. The matrix $\mathbf{\hat{M}}$ represents the propagation along the lattice and will therefore have eigenvalues related to the lattice tunes, while $\mathbf{\Sigma S}$ represents the phase space beam distribution, and its eigenvalues will thus be linked to the beam emittances.

\bibliography{bibliography.bib}

\end{document}